\documentclass[aps,prd,twocolumn,showpacs,10pt,superscriptaddress, footinbib]{revtex4-1}
\usepackage{amsmath,bm,amssymb}
\usepackage{autobreak}%%%
\usepackage{graphicx}
\usepackage{psfrag}
\usepackage{ulem}
\usepackage[caption=false]{subfig}
\usepackage{color,xcolor}
\definecolor{urlblue}{rgb}{0.2,0.4,0.7}
\definecolor{citegreen}{rgb}{0,0.6,0.2}
\definecolor{linkred}{rgb}{0.9,0.2,0.1}
\usepackage{hyperref}
\hypersetup{
colorlinks=true, citecolor=citegreen, linkcolor=blue,urlcolor=urlblue}

\usepackage{slashed}
\usepackage{array}
\newcolumntype{P}[1]{>{\centering\arraybackslash}p{#1}}

\allowdisplaybreaks

\def\be{\begin{equation}}
\def\ee{\end{equation}}
\def\nn{\nonumber\\}

\def\rmS{{\rm S}}
\def\rmJ{{\rm J}}
\def\rmP{{\rm P}}

\begin{document}

\def\FBo{\bm{\gamma}_{{\rm P},1}^{I}}
\def\FBtw{\bm{\gamma}_{{\rm P},2}^{I}}
\def\FBth{\bm{\gamma}_{{\rm P},3}^{I}}
\def\FBfo{\bm{\gamma}_{{\rm P},4}^{I}}

%\preprint{IMSc/2018/05/04}
%\begin{flushleft} 
%\mbox{IMSc/2017/08/07}
%\end{flushleft}

\title{Soft and Jet functions for SCET at four loops in QCD
}

\author{Saurav Goyal}
\email{sauravg@imsc.res.in}
\affiliation{The Institute of Mathematical Sciences,  Taramani, 600113 Chennai, India}
\affiliation{Homi Bhabha National Institute, Training School Complex, Anushakti Nagar, Mumbai 400094, India}
\author{Sven-Olaf Moch}
\email{sven-olaf.moch@desy.de}
\affiliation{II. Institute for Theoretical Physics, Hamburg University, D-22761 Hamburg, Germany} %Luruper Chaussee 149,
\author{Vaibhav Pathak}
\email{vaibhavp@imsc.res.in}
\affiliation{The Institute of Mathematical Sciences, Taramani, 600113 Chennai, India}
\affiliation{Homi Bhabha National Institute, Training School Complex, Anushakti Nagar, Mumbai 400094, India}
\author{V. Ravindran}
\email{ravindra@imsc.res.in}
\affiliation{The Institute of Mathematical Sciences, Taramani, 600113 Chennai, India}
\affiliation{Homi Bhabha National Institute, Training School Complex, Anushakti Nagar, Mumbai 400094, India}

\date{\today}

\begin{abstract}
Soft-Collinear Effective Theory is a framework for systematically organizing and resumming the logarithmic contributions that occur in high-energy reactions. 
It provides a factorized description of cross sections in terms of hard, jet, soft, and beam functions.
As the latter are universal, they can be obtained from the well-known perturbative results in quantum chromodynamics (QCD) for deep-inelastic scattering, Drell-Yan and Higgs boson productions.  
Using the recent results~\cite{Kniehl:2025ttz} on four-loop eikonal $(f^I)$ and collinear anomalous dimensions $(B^I)$ for quarks and gluons, $I=q,g$, as well as perturbative results from previous orders, we present four-loop predictions for the quark and gluon soft and jet functions. 
They constitute an important component of the $N$-jettiness subtraction method at $\rm{N^4LO}$ accuracy in QCD, which eventually may enable the calculation of fully-differential cross sections at higher orders.
\end{abstract}
%\pacs{12.38.Bx}

\maketitle
\section{I.\,INTRODUCTION}
%%%%%%%
The Standard Model (SM) of particle physics is currently being tested with unprecedented accuracy 
at the Large Hadron Collider (LHC). To achieve this level of accuracy, one requires highly 
precise measurements of various collider observables and their comparison with state-of-the-art theoretical predictions~\cite{Heinrich:2020ybq}.
These predictions rely on modern techniques and are essential not only for 
validating the SM but also for constraining potential signals resulting from models beyond the SM. 
For example, the processes, such as the production of lepton pairs, namely Drell-Yan (DY) process, the production of a Higgs boson in gluon-gluon fusion (ggF) or in bottom quark annihilation, 
the production of vector bosons, and of multi-jets are central to these efforts, as they serve as key probes of both electroweak and strong interaction dynamics. 
The Electron-Ion Collider, currently under construction at the Brookhaven National Laboratory (BNL), requires precise predictions from the SM in order to  probe the internal structure of hadrons through (semi-)inclusive deep-inelastic scatterings (DIS, SIDIS), see, e.g.~\cite{Vermaseren:2005qc,Goyal:2023zdi,Bonino:2024qbh}.
Such studies provide valuable insight into the underlying structure of the gauge theory of strong interaction, Quantum Chromodynamics (QCD).

In these high-momentum transfer processes, the hadronic cross sections factorize into hard and soft/jet components. 
The hard part encodes short-distance physics at the scale of the large momentum transfer $Q$, the characteristic scale of the process, while the soft and jet contribution capture the effects of soft and collinear radiations respectively. 
Soft-Collinear Effective Theory (SCET) \cite{Bauer:2000ew,Bauer:2000yr,Bauer:2001ct,
Bauer:2001yt,Bauer:2002nz,Beneke:2002ph} provides a systematic framework to describe this soft 
and collinear dynamics at high energies through two key ingredients: the soft and jet functions.
The soft functions in SCET parametrize the low-energy emissions of partons at wide angles and also mediate interactions between different jets.  
The jet functions, on the other hand, describe the dynamics of collinear partons propagating along energetic directions, effectively characterizing the internal structure of jets. 
They are universal and play important roles in precise predictions for collider observables. 
They are crucial for the resummation of logarithmically enhanced contributions in the threshold kinematic region and provide building blocks for subtraction methods like the $N$-jettiness formalism, which are vital for perturbative QCD predictions at next-to-next-to-leading order (NNLO) and beyond at the LHC as well as the EIC.

In the SCET framework, both quark and gluon soft and jet functions have been computed to high orders in perturbation theory. The quark and gluon soft functions are now known up to the three-loop level; one- and two-loop results are found in~\cite{Belitsky_1998}, while for three loops, see~\cite{Li:2014bfa,Ahmed:2014cla} and references therein.
For the quark jet function, results were initially available up to two loops~\cite{Bauer:2003pi,Bosch:2004th,Becher:2006qw}, and were later extended to three loops in~\cite{Bruser:2018rad}.
Similarly, the gluon jet function was first computed at one loop in~\cite{Becher:2009th}, followed by the two-loop results in~\cite{Becher:2010pd}, and most recently, the three-loop computation in~\cite{Banerjee:2018ozf}.

Due to their universality, both soft and jet functions act as essential building blocks for a broad spectrum of observables at hadron colliders. 
In SCET, the operator definition of the soft function is found in~\cite{Belitsky_1998}, see also~\cite{Li:2014bfa}, and for the jet function in~\cite{Becher:2006qw}. 
In the threshold region, the DY coefficient functions (CFs) are sensitive to the quark soft function, whereas Higgs production via ggF relies on the gluon soft function. 
Similarly, photon-mediated DIS processes depend on the quark jet function, whereas Higgs-mediated DIS relies on the gluon jet function. 
Since these functions are process-independent which means they do not depend on the specific nature of the hard scattering process, hard matching coefficients are required to relate them to the corresponding CFs for the respective cross sections. 
The hard matching coefficients are computed using the process dependent quark or gluon form factors (FFs).  

These hadron scattering processes  have been studied in full QCD for several decades and perturbative results for many important observables are known very precisely; see the latest results for DY~\cite{Duhr:2020seh}, for Higgs production in ggF~\cite{Anastasiou:2014vaa} and for DIS~\cite{Vermaseren:2005qc}. 
The results have revealed remarkable perturbative structures characterized by several process-independent quantities, namely the cusp anomalous dimension $(A)$, the collinear anomalous dimension $(B)$, and the eikonal anomalous dimension $(f)$, which are described through renormalization group (RG) equations in the infrared (IR) sector of QCD, see~\cite{Dixon:2008gr}.
Their study gives insight into the all-order structure of those contributions that are important in threshold kinematic regions of phase space. 
Systematic approaches have led to summing up dominant contributions resulting from these regions to all orders in perturbation theory  through threshold resummation \cite{Sterman:1986aj,Catani:1989ne}. 
The resummed results primarily depend on the process-independent cusp, collinear, and soft anomalous dimensions.

For DY and ggF Higgs production, it has been shown in \cite{Ravindran:2005vv,Ravindran:2006cg} that the CFs in the threshold limit factorize into a product involving the square of the ultraviolet (UV) renormalized FF, a soft distribution function, and the mass factorization kernels of incoming partons. 
A careful study of the soft distribution function reveals that its finite parts, as given in \cite{Ravindran:2005vv,Ravindran:2006cg}, correspond precisely to the soft functions defined in SCET.
The soft function for quarks was first computed up to two loops  in a seminal paper~\cite{Belitsky_1998} and the corresponding gluon soft function can be obtained by appropriately adjusting the color factors. 
Using the soft plus virtual QCD contributions to inclusive ggF Higgs boson production at N$^3$LO computed in \cite{Anastasiou_2014}, and following the framework in \cite{Ravindran:2005vv}, the gluon soft function was obtained at three loops for the first time in \cite{Ahmed:2014cla}.  
Applying the maximally non-Abelian property of soft functions, the three-loop quark soft function was obtained by multiplying the gluon soft function by the ratio of the color factors $C_F/C_A$.
The three-loop contributions to both gluon and quark soft functions obtained in this way agree with the results~\cite{Li:2014bfa} computed within SCET. 
These results have been employed to derive the soft plus virtual contributions to the production of lepton pairs in the DY process \cite{Ahmed:2014cla} and to Higgs boson production via bottom quark annihilation \cite{Ahmed:2014cha} up to N$^3$LO.

Similarly, it was found in~\cite{Ravindran:2006cg} that the DIS cross section in the threshold limit also factorizes into the square of the ultraviolet (UV) renormalized FF, a jet function, and the mass factorization kernel for the incoming parton.
In~\cite{Banerjee:2018ozf}, a novel connection between jet functions in SCET and the CFs in DIS cross sections computed in full QCD was established. 
This enabled the use of the three-loop CF for the Higgs-mediated DIS process, as calculated in~\cite{Soar:2009yh}, to derive the three-loop gluon jet function presented in~\cite{Banerjee:2018ozf}.
In addition, using the three-loop CFs of photon-mediated DIS \cite{Vermaseren:2005qc} process, the extraction of three-loop corrected quark jet function has been performed~\cite{Banerjee:2018ozf}, in agreement with the result obtained in SCET~\cite{Bruser:2018rad}.  
In summary, understanding the structure of these perturbative results provides the foundation for connecting the soft and soft-plus-jet distribution functions discussed in~\cite{Ravindran:2005vv,Ravindran:2006cg} in full QCD with the soft and jet functions in SCET.

In this article, utilizing the aforementioned formalism and the recent four-loop results on the eikonal $(f_4)$ and collinear $(B_4)$ anomalous dimensions \cite{Kniehl:2025ttz}, we derive the results for both the soft and jet functions at four loops, up to terms proportional to $\delta(1-z)$, where $z$ is the relevant scaling variable. 
Our approach relies on the application of the so-called $K$ plus $G$ $(KG)$ equation , RG invariance, 
the factorization theorem, and previously known three and four loop results. 
We work within the framework developed in \cite{Ravindran:2005vv,Ravindran:2006cg,Ravindran:2006bu,Banerjee:2017cfc,Banerjee:2018vvb,Banerjee:2018mkm,AH:2020qoa,Ahmed:2020amh,AH:2021vhf,Ravindran:2022aqr,Ravindran:2023qae} which systematically resums soft and collinear gluon effects to all orders in QCD perturbation theory.  
In \cite{Ravindran:2006bu}, an all order relation between CFs of inclusive cross sections and rapidity distributions of colorless particle production was established, see \cite{Ravindran:2007sv,Ahmed:2014uya,Ahmed:2014era,Banerjee:2017cfc} for further developments. 
Using these relations, we determine the soft function for the rapidity distribution of any colorless particle production at hadron colliders at the four-loop level.

\section{II.\,THEORETICAL FRAMEWORK AND RESULTS}
%%%%%%%%
Let us begin with the production of a color singlet state in hadron-hadron collision.  
This state can be a pair of leptons for the DY process, a scalar or pseudo-scalar Higgs boson or multiples of color-singlet states. 
The corresponding inclusive cross section in the QCD improved parton model is given by
\begin{eqnarray}
\label{eqdy}
\sigma^{I}_\rmS (\tau,Q^2)&\!=\!&\sigma^I_{\rmS,B}(\mu_R^2)\!\!\!\!\! \sum_{a,b=q,\overline q,g } 
\int_{\tau}^{1} \, \frac{d z_1}{z_1} 
\int_{\frac{\tau}{z_1}}^{1} \, \frac{d z_2}{z_2} 
f_{a}\left(z_1,\mu_{F}^2\right)  
\nonumber\\ &&
\!\!\!\times f_{b}\left(z_2,\mu_{F}^2\right)  
%\nonumber \\&&
 \Delta^{I}_{\rmS,ab}\left(a_s,\frac{\tau}{ z_1 z_2}, Q^2, \mu_{R}^2, \mu_{F}^2\right).
\nonumber\\
\end{eqnarray}
The hadronic scaling variable is defined as $\tau = \frac{Q^2}{S}$, where $S = (P_1 + P_2)^2$ is the square of the hadronic center-of-mass energy, with $P_1$ and $P_2$ denoting the momenta of the incoming hadrons. 
The variable $Q^2$ represents the invariant mass square of the final state: it corresponds to the dilepton invariant mass in DY production, or to the mass of a (pseudo-)scalar Higgs boson. 
For the production of multiple color-singlet states, $Q^2$ is the invariant mass squared of the entire final-state system.
The superscript $I=q, g$ distinguish the DY process $(q)$ and ggF Higgs production $(g)$.

The parton distribution function (PDF) for an incoming parton of type $a$ is denoted by $f_{a}(z_1, \mu_F^2)$, where $z_1$ is the momentum fraction carried by the parton relative to the hadron, and $\mu_F$ is the factorization scale. 
The PDFs for the incoming partons $a$ and $b$ are inherently non-perturbative and must be extracted from experimental data.
The CFs, denoted by $\Delta^{I}_{\rmS,ab}$, are perturbatively calculable in QCD and can be expanded as a series expansion in the strong coupling at the renormalization scale $\mu_R$, $a_s =a_s(\mu_R^2)$.
The CFs encode the short-distance dynamics of the process and depend on $\mu_R$, $\mu_F$ and the scaling variable $z$, which is defined as $z = \frac{\tau}{z_1 z_2}$ in terms of the momentum fractions of the incoming partons $z_1$ and $z_2$.

The Born contribution $\sigma^I_B$ is normalized so that $\Delta^I_{\rmS,ab}$ is equal to $\delta_{a \overline b}\delta(1-z)$ to lowest order
in perturbation theory. 
In dimensional regularization with $\varepsilon = n-4$ in $n$ space-time dimensions, the UV renormalized strong coupling is given in terms of bare coupling constant $\hat a_s$ through 
$$ \hat a_s S_{\varepsilon}= a_s (\mu_R^2)\Big(\frac{\mu^2}{\mu_R^2}\Big) ^{\frac{\varepsilon}{2}}  Z(a_s(\mu_R^2),\varepsilon)
\, , $$
where $S_{\varepsilon} = \exp\left(\frac{\varepsilon }{2}\left(\gamma_E-\ln(4 \pi)\right)\right)$ and $\gamma_E$ is the Euler-Mascheroni constant. 
$Z$ is the coupling constant renormalization term and the scale $\mu$ keeps $\hat a_s$ dimensionless in $n$ space-time dimensions.

For DIS, the inclusive cross section in QCD improved parton model is given by
\begin{eqnarray}
\label{eqdis}
\sigma^{I}_\rmJ (x_{Bj},Q^2)&=&\sigma^I_{\rmJ,B} (\mu_R^2) \sum_{a=q,\overline q,g }
\int_{x_{Bj}}^{1} \, \frac{d z_1}{z_1} 
f_{a}\left(z_1,\mu_{F}^2\right)  
\nonumber \\&&
\times \Delta^{I}_{\rmJ,a}\left(a_s,\frac{x_{Bj}}{z_1}, Q^2, \mu_{R}^2, \mu_{F}^2\right)\, ,
\end{eqnarray} 
where $x_{Bj}$ is the Bjorken scaling variable, $x_{Bj}=\frac{-q^2}{2 P.q}$, and $P$ and $q$ are the momenta of the incoming hadron and the intermediate off-shell particle with space-like momentum transfer $Q^2=-q^2 > 0$. 
The perturbatively calculable CFs are $\Delta^{I}_{\rmJ,a}$ and we consider two classes of DIS processes: (i) photon‑mediated scattering, denoted by $I=q$, and (ii) scalar‑mediated scattering, denoted by $I=g$. 
The partonic scaling variable is defined as $z = \frac{-q^2}{2p.q}$, with the momentum $p$ of the incoming parton. The variable $z_1$ denotes the momentum fraction of the hadron’s momentum carried by the parton.

In perturbative QCD, the CFs $\Delta_{{\rmS},ab}^I$ and $\Delta_{{\rmJ},a}^I$ receive contributions from several partonic subprocesses. 
The logarithmic structure that arises in the soft–collinear limit motivates a decomposition of the CFs. For clarity, we keep only the dependence on $z$, while suppressing the remaining arguments, and we omit the subscripts $ab$ for $S$ and $a$ for $J$.
\begin{align}
\label{eq3}
\Delta^I_{\rm{P}}(z)  =  
 \Delta^{I,\text{SV}}_{\rm{P}}(z) + 
  \Delta^{I,\text{NSV}}_{\rm P}(z)+  \cdots 
\end{align}
Here $\rmP=\rmS,\rmJ$, and the soft–virtual (SV) and next‑to‑\ soft–virtual (NSV) terms isolate the dominant contributions in the threshold region.
The SV terms collect all the leading power (LP) terms such as plus distributions ${\cal D}_i(z)=\Big(\frac{\log^i(1-z)}{1-z}\Big)_+$ and $\delta(1-z)$.  
These contributions come from partons whose momentum fractions are close to one, i.e., $z\rightarrow 1$.  
The NSV terms collect all $\log^j(1-z), j=0,1,\cdots$ contributions, also often referred to as next-to-LP (NLP) terms. 
The ellipsis in Eq.~\eqref{eq3} denotes the remaining hard part of the CFs.  
In the threshold region, the SV and NSV contributions dominate the hadronic cross section, provided that the parton distribution functions also receive their leading support near the partonic threshold. 

In a series of works (see \cite{Ravindran:2005vv} for SV and \cite{AH:2020iki,AH:2020xll} for NSV),
the sum of the SV and NSV parts of the CFs $\Delta^{I,{\rm SV+NSV}}_{{\rmP}}$ 
was shown to factorize into contributions from FFs and soft (jet) distribution functions appropriately convoluted with mass factorization kernels of incoming partons as 
\begin{align}
\label{master}
\Delta^{I,\rm {SV+NSV}}_{\rm{P}} = &  (Z^{I}_{\rm{UV}} (\hat a_s,\mu_R^2, \mu^2, \varepsilon))^2 \big| \hat {\cal F}^{I}_\rmP(\hat a_s, -q^2,\mu^2,\varepsilon)\big|^2
\nonumber \\ &
\times  {\cal S}^{I}_{\rm{P}}(\hat a_s, Q^2,z,\varepsilon) 
\nonumber \\ & 
\otimes \Big(\overline\Gamma_{II}(\hat a_s, \mu^2, \mu^2_{F}, z, \varepsilon)\Big)^{-2 m}\, ,
\end{align}
where $Q^2 = |-q^2|$ and $m=1,-q^2<0$ for $\rmP=\rmS$ and $m=1/2,-q^2 >0$ for $\rmP =\rmJ$.
The symbol $\otimes$ denotes the Mellin convolution in $z$, and the term $\overline\Gamma_{II}$ refers to the SV+NSV part of the Altarelli–Parisi (AP) kernel.
For DY production, the SV+NSV terms arise from quark–antiquark–initiated subprocesses ($a=q $). 
For ggF Higgs production, they originate from gluon‑initiated subprocesses ($a=g$).  
For DIS, $a=q$ or $a=\overline q$ in photon‑mediated channels, while $a = g$ in scalar (Higgs‑mediated) channels.

In the above equation, the contributions from unrenormalized FF $\hat {\cal F}_\rmP^{I}$, those from the AP kernels $\Gamma_{II} $ and the overall renormalisation constant $Z^{I}_{\rm UV}$ are factored out 
from the partonic cross section in such a way, that the remaining partonic contributions contain at least one real parton emission. 
In other words,  ${\cal S}^{I}_{\rm{P}}(\hat a_s=0,Q^2,z,\varepsilon)=\delta(1-z)$.

The UV renormalisation constant $(Z_{\text{UV}}^I)$, associated with the operator or amplitude satisfies an RG equation and hence admits an exponential form controlled by the UV anomalous dimension $(\gamma_{\text{UV}}^I)$,
\begin{eqnarray}
\label{ZUVsol}
  Z_{\text{UV}}^I(\mu_R^2,\varepsilon)=
   \exp\Bigg(\int_0^{\mu_R^2} {d \lambda^2 \over \lambda^2} \gamma_{\text{UV}}^I(a_s(\lambda^2))\Bigg)\, .
\end{eqnarray}
For the quark FF (\(I=q\)), \( \gamma_{\text{UV}}^{q} \) =0 and hence one finds that $Z^q_{\rm {UV}}=1$ to all orders, while for $I=g$, the constant $Z^g_{\rm UV}$ is non-zero. 
The latter is determined by matching the results from Higgs-gluon operator in the Higgs effective theory obtained in the large top-quark mass limit against the full theory. 
It is well-known that the anomalous dimension corresponding to $Z^g_{\rm UV}$ is related to the QCD $\beta$ function coefficients~\cite{Chetyrkin:1997un}, i.e., 
\begin{eqnarray}
\gamma^g_{\rm UV} = \sum_{i=1}^\infty a_s^i(\mu_R^2) \left(i~ \beta_{i-1}\right)\, .
\end{eqnarray}

The quark and gluon FFs satisfy the so‑called $K$ plus $G$ integro‑differential equation, which encodes their infrared singularity structure as a consequence of factorization, gauge invariance, and RG invariance\cite{Sudakov:1954sw,Mueller:1979ih,Collins:1980ih,Sen:1981sd}. 
Its universal applicability reflects the fact that soft and collinear divergences in gauge theories such as QCD not only factorize but also obey RG equations governed by well‑defined anomalous dimensions. This, in turn, ensures that the $KG$ equation exhibits a predictable and resummable structure for the FFs to all orders in the strong coupling. 
For a given parton type $I$, the UV renormalized FF,
\begin{align}
\label{factFhat}
{\cal F}^I_\rmP\big(-q^2,\mu_R^2,\varepsilon\big) =Z^I_{\rm{UV}} \big(\mu_R^2,\varepsilon\big)  \hat{{\cal F}}_{\rmP}^I \big(q^2,\varepsilon\big)
\end{align}
factorizes into an IR-divergent part and an IR-finite remainder:
\begin{equation}
\label{factF}
{{\cal F}}^I_\rmP\left(-q^2,\mu_R^2,\varepsilon\right) = Z_{{{\cal F_{\rmP}}}}^I\left(Q^2, \mu_s^2,\varepsilon\right) \, {\cal F}^I_{\rmP,\text{fin}}\left(Q^2, \mu_s^2,\mu_R^2,\varepsilon\right),
\end{equation}
where $Z_{{\cal F}_\rmP}^I$ contains all the IR singularities and ${\cal F}^{I}_{\rmP,\text{fin}}$ is finite as $\varepsilon \to 0$. 
The auxiliary scale $\mu_s$ acts as an IR factorization scale, separating the divergent and finite contributions. 
Taking the logarithmic derivative of ${\cal F}^I_\rmP$ with respect to $Q^2$ leads to the Sudakov differential equation:
\begin{equation}
\label{KplusGF}
Q^2 \frac{d}{d Q^2} \ln {\cal F}^I_{\rmP}  = \frac{1}{2} \Bigg[ K^I(\mu_s^2,\varepsilon) + G^I_{ {\cal F}_{\rmP}}\left({Q^2\over \mu_R^2}, \mu_s^2,\varepsilon\right) \Bigg],
\end{equation}
where
\begin{align}
\label{KandGF}
K^I  = 2Q^2\frac{d}{d Q^2} \ln Z_{{\cal F}_{\rmP}}^I ,  \quad \quad 
G^I_{ {\cal F}_{\rmP}} &= 2Q^2\frac{d}{d Q^2} \ln {\cal F}^{I}_{\rmP,\text{fin}} .
\end{align}
From the above equations it is clear that $K^I$ contains only divergent terms and $G^I_{\cal F_{\rmP}}$ is finite. 
The fact that ${\cal F}^I_{\rmP}$ is $\mu_s$ independent leads to the RG equation with respect to $\mu_s$, 
\begin{eqnarray}
\label{RGKplusG}
\mu_s^2 {d K^I\over d \mu_s^2} = - \mu_s^2 {d G^I_{\cal F_{\rmP}} \over d \mu_s^2} = -A^I(\mu_s^2)
\end{eqnarray}
where $A^I$ is the universal cusp anomalous dimension.
Since $K^I$ contains only divergent terms, the solution to $K^I$ from Eq.~\eqref{RGKplusG} will contain only QCD $\beta$ function coefficients and $A^I$, hence it is process independent.
However, $G^I_{\cal F_{\rmP}}$ is finite and  unlike $K^I$, $G^I_{ {\cal F_{\rmP}}} $ depends not only on universal anomalous dimensions, namely $B^I$ (collinear) and $f^I$ (eikonal), but also on process dependent constants.
The collinear anomalous dimension $B^I$ is also referred to as 
virtual anomalous dimension and the eikonal anomalous dimension $f^I$ is also known as soft anomalous dimension.

The solution to the $KG$ equation, valid to all orders in perturbation theory, takes the exponential form,
\begin{equation}
\label{Fsol}
{\cal F}^I_{\rmP}  = \exp\left( 
\Phi_{{\cal F_{\rmP}}}^I(-q^2,\mu_s^2,\varepsilon) \right)\, ,
\end{equation}
where we set $\mu_R=\mu_s$ and
\begin{equation}
\label{PhiF}
\Phi_{\cal F_{\rmP}}^{I} =\int_0^{-q^2} \frac{d\lambda^2}{\lambda^2} \, \Gamma^I_{\cal F_{\rmP}}(\lambda^2, \mu_s^2,\varepsilon),
\end{equation}
with the boundary condition ${\cal F}^I_{\rmP}(-q^2 = 0, \mu_s^2,\varepsilon) = 1$. 
The integrand kernel is defined as $\Gamma_{{\cal F_{\rmP}}}^{I} = (K^I + G^I_{ {\cal F_{\rmP}}})/2$. 
The solution in Eq.~\eqref{Fsol} resums large logarithms due to soft and collinear emissions and reflects the exponentiation of IR divergences in gauge theories. 
The universality and iterative structure of the anomalous dimensions entering $K^I$ and $G^I_{\cal F_{\rmP}}$ play a crucial role in this behavior.  
Comparing Eqs.~\eqref{factF} and \eqref{Fsol}, we find 
\begin{eqnarray}
\label{FFdiv}
Z^I_{{\cal F}_{\rmP}} (Q^2,\mu_s^2,\varepsilon) = \exp\Big(\Phi^{I,div}_{{\cal F_{\rmP}}}(-q^2,\mu_s^2,\varepsilon)\Big)\, ,
\end{eqnarray}
where we have split  $\Phi^{I}_{ {\cal F_{\rmP}}}$ as sum of terms containing IR divergences and the remaining finite terms: $\Phi^I_{ {\cal F_{\rmP}}} = \Phi^{I,div}_{ {\cal F_{\rmP}}}(\mu_s^2)+\Phi^{I,fin}_{ {\cal F_{\rmP}}}(\mu_s^2)$ at the scale $\mu_s$. 
The fact that $\Phi_{\cal F_{\rmP}}^I$ is independent of $\mu_s$ leads to renormalisation group equation for $\Phi_{\cal F_{\rmP}}^{I,fin}(\mu_s^2)$:
\begin{align}
    \mu_s^2 {d \over d\mu_s^2} \Phi_{\cal F_{\rmP}}^{I,fin}(Q^2,\mu_s^2) &=\gamma^{I}_{\cal F_{\rmP}}(Q^2,\mu_s^2)\, ,
\end{align}
where,
\begin{align}
\label{FFanom}
   \gamma^{I}_{\cal F_{\rmP}}&= 
 \frac{1}{2}\left(   A^I(\mu_s^2) \log\left({Q^2 \over \mu_s^2}\right) 
    - 2 B^I(\mu_s^2) - f^I(\mu_s^2)\right)\, .
\end{align}
The general solution to the $KG$ equation for FFs has been derived in~\cite{Moch:2005id,Ravindran:2005vv}. 
The cusp anomalous dimensions $A^I$, governing the behavior of soft-collinear emissions, are known up to four loops in QCD~\cite{Henn:2019swt,vonManteuffel:2020vjv,Agarwal:2021zft} (see~\cite{Moch:2004pa,Vogt:2004mw,Catani:1989ne,Catani:1990rp,Vogt:2000ci} for results at lower orders).

Similarly, the collinear anomalous dimension $B^I$ encodes contributions from purely collinear radiation~\cite{Moch:2004pa,Vogt:2004mw}, while the eikonal anomalous dimension $f^I$ accounts for wide‑angle soft‑gluon emission~\cite{Ravindran:2004mb,Vogt:2004mw}. 
Both quantities are known up to three loops in QCD.
The first four‑loop results for $B^I$ and $f^I$ were obtained in~\cite{Das:2019uvh,Das:2020adl} and subsequently refined in~\cite{Duhr:2022cob,Moch:2023tdj,Falcioni:2024qpd}.  
More recently, analytic expressions have become available~\cite{Kniehl:2025ttz}, determining  $B^I$ and $f^I$ up to a single constant, whose value is known numerically to high precision~\cite{Moch:2023tdj}.
Together, these quantities determine the all-order IR structure of the FFs, up to additive constants that depend on the specific processes.  

The AP kernels $\Gamma_{II} $ remove the collinear divergences from the incoming partons as their momenta are not integrated out at the parton level. 
They are known fully up to the three-loop level in QCD~\cite{Moch:2004pa,Vogt:2004mw}. 
Work at four loops is ongoing, see~\cite{Falcioni:2023luc,Falcioni:2023vqq,Falcioni:2024xyt,Falcioni:2024qpd} and references therein for fixed Mellin moments and \cite{Gracey:1994nn,Davies:2016jie,Moch:2017uml,Basdew-Sharma:2022vya,Gehrmann:2023iah,Davies:2017hyl,Kniehl:2025jfs} for partial analytic results. 
The AP kernels satisfy the standard AP evolution equation. 
The bar on the kernels, $\overline \Gamma_{II}$, indicates that only their SV and NSV contributions are retained. 
In the approximation relevant for our analysis-where only the diagonal part of the splitting function matrix is retained-the evolution simplifies
\begin{align}
  \mu_F^2{ d \over d\mu_F^2} \overline \Gamma_{II}(\mu_F^2,z,\varepsilon)
={1 \over 2 } \overline P_{II}(\mu_F^2,z) \otimes \overline \Gamma_{II}(\mu_F^2,z,\varepsilon)
\, .
\qquad
\end{align}
The diagonal AP splitting functions $\overline P_{II}$ capture the probability of a parton of flavor $I$ emitting radiation while retaining its identity. 
In the SV+NSV limit, they take the simple form,
\begin{eqnarray}
\label{APsplit}
\overline {P}_{II}(\mu_F^2,z)&=&{2 A^I(\mu_F^2)\over (1-z)_+}
+2 B^I(\mu_F^2) \delta(1-z) \nonumber
\\&&
+2 C^I(\mu_F^2)\ln(1-z) + 2 D^I(\mu_F^2)\, .
\end{eqnarray}
Note that the leading SV distributions are controlled by the universal cusp $A^I$ and collinear anomalous dimensions $B^I$, 
while the subleading NSV logarithms ($C^I,D^I$) are given in terms of expansion coefficients of $A^I$ and the QCD $\beta$ function from lower orders, see, e.g.~\cite{Dokshitzer:2005bf,Moch:2023tdj}.  
In this limit, the all-order solution to the AP evolution equations takes the form,
\begin{eqnarray}
\label{Gcsol}
\overline \Gamma_{II}(\mu_F^2,z,\varepsilon) = {\cal C} \exp\left( \frac{1}{2} \int_0^{\mu_F^2} 
\frac{d\lambda^2}{\lambda^2} \overline P_{II}(\lambda^2,z) \right)\, .
\end{eqnarray}   
The symbol ${\cal C}$ can be found in \cite{Ravindran:2005vv}: 
\begin{eqnarray*}
\label{eq5}
{\cal C} e^{\displaystyle f(z) }&=& \delta(1-z)  + {1 \over 1!} f(z)
 +{1 \over 2!} f(z) \otimes f(z) 
\nonumber\\
&& + {1 \over 3!} f(z) \otimes f(z) 
 \otimes f(z)
+ \cdot \cdot \cdot \,.
\end{eqnarray*}
While performing the convolutions, we drop all the terms that are neither distributions nor powers of the type $\log^j(1-z),j=0,1,\cdots$ as we are interested only in the $\text{SV+NSV}$ terms.

Our next task is to investigate the all order structure of the real emission component ${\cal S}^I_\rmP$ in Eq.~\eqref{master}.
In~\cite{Ravindran:2005vv,Ravindran:2006cg,AH:2020iki,AH:2020xll,AH:2022lpp}, it was demonstrated that the requirement of finiteness of $\Delta^{I,\rm SV+\rm NSV}_{\rm{P}}$ for inclusive cross sections imposes important constraints on the structure of ${\cal S}^I_{\rm{P}}$. 
Just as the FFs are governed by the $K$ plus $G$ differential RG equation due to the interplay of IR factorization and RG invariance, it was shown that the ${\cal S}^{I}_{\rm P}$ must satisfy a similar RG equation,
\begin{eqnarray}
\label{eq6}
 Q^2 \frac{d}{dQ^2} {\cal S}^I_{\rmP} = \Gamma_{\cal S_{\rmP}}^I(Q^2,z,\varepsilon) \otimes {\cal S}^I_{\rmP}(Q^2,z,\varepsilon)\, ,
\end{eqnarray}
where the kernel $\Gamma_{\cal S_{\rmP}}$ admits a $KG$-type decomposition:
\begin{eqnarray}
\Gamma_{\cal S_{\rm{P}}}^I=\frac{1}{2} \Bigg[ \overline{K}^I \Big(   {\mu_s^2} , z,
\varepsilon \Big) + \overline{G}^I_{\rm{P}} \Big(  {Q^2} , {\mu_s^2}, z, \varepsilon \Big) \Bigg]\, .
\end{eqnarray}
The separate terms are defined such that  $\overline{K}^I$ contains poles in $\varepsilon$ while $\overline{G}^I_{P}$ is finite.
Their expressions have the same structural form as the corresponding quantities in the FFs, enabling a systematic all‑order resummation of logarithmic enhancements in perturbative calculations, and ensuring the cancellation of IR divergences when combined with $Z^{I}_{\rm{UV}}$, FFs and AP kernels in Eq.~\eqref{master}.

This result extends the universality of the Sudakov resummation framework to real emission processes. 
The $KG$ equation for ${\cal S}^{I}_{\rm P}$ encodes the SV distributions ${\cal D}_j(z)=\Big(\frac{\log^j(1-z)}{1-z}\Big)_+$ and the $\delta(1-z)$ terms.
In addition, NSV terms of the form $\log^j(1-z)$ with $j=0,1,\cdots $ are accounted for as well as IR singularities arising from soft and collinear emissions in the $z \to 1$ limit, where $z$ denotes the scaling variable.
The structure of this equation mirrors the one satisfied by the FFs, with the anomalous dimensions $A^I, B^I$ and $f^I$ $(C^I,D^I)$ playing analogous roles in determination of the resummation coefficients. 
The solution to the above equation Eq.~\eqref{eq6} is found to be
\begin{eqnarray}
{\cal S}^I_{\rm{P}}(Q^2,z,\varepsilon)
&=&{\cal C} \exp\Bigg(\int_0^{Q^2} {d\lambda^2\over \lambda^2} \Gamma^I_{\cal S_{\rm{P}}} (\lambda^2,z,\varepsilon) \Bigg)
\nn 
&=&{\cal C} \exp\Big (2 ~\Phi^I_{\rm{P}} (Q^2,z,\varepsilon) \Big)
\end{eqnarray}
with the boundary condition 
${\cal S}^I_{\rm{P}} (Q^2=0,z,\varepsilon)=\delta(1-z)$ and $\Gamma_{\cal S_{\rm{P}}}^I = (\overline K^I + \overline G^I_{\rm{P}})/2$.  
Following~\cite{Ravindran:2005vv,Ravindran:2006cg,AH:2020iki,AH:2020xll,AH:2022lpp} the explicit expression reads
\begin{widetext}
\begin{align}
\label{eq7}
\Phi^I_{\rm{P}}(Q^2,z,\varepsilon) = \sum_{i=1}^\infty {\hat a}_s^i S_\varepsilon^i \left({Q^2 (1-z)^{2 m} \over \mu^2}\right)^{i{\varepsilon \over 2}}
 \Bigg[ {i m \varepsilon \over (1-z)} \hat \phi^{I,(i)}_{\rm{P}} (\varepsilon) + m~ \hat{\varphi}^{I,(i)}_{\rm{P}} (z,\varepsilon)\Bigg] \, .
 \end{align}
%csm \end{widetext} 
%
We find for the SV part 
$\hat \phi^{I,(i)}_{\rm{P}}(\varepsilon) = \Big[\overline{K}^{I,(i)}(\varepsilon) 
+ \overline{G}^{I,(i)}_{\rm{P},SV}(\varepsilon)\Big]\big/i \varepsilon$,  
and for the NSV part $\hat{\varphi}^{I,(i)}_{\rm{P}} (z,\varepsilon)$ = $\overline{G}^{I,(i)}_{\rm{P},L}(z,\varepsilon)\big/i \varepsilon$, where $\overline{G}^{I}_{\rm{P},L} = \overline{G}^{I}_{\rm{P}}-\overline{G}^{I}_{\rm{P},SV} $.
Expressing 
$\overline{K}^I = \sum_{i=1}^\infty \hat a_s^i \left({\mu_R^2 \over \mu^2}\right)^{i {\varepsilon \over 2} }
S_\varepsilon^i \overline{K}^{I,(i)}$,  
the coefficients $\overline{K}^{I,(i)}$ are given in terms of $A^I$ and the QCD $\beta$ function, while $\overline{G}^{I,(i)}_{\rm{P},SV}$ reads
%
%csm end widetext here for layout
\end{widetext}
\begin{align}
\label{eq8}
\overline{G}^{I}_{\rm{P},SV}(Q^2,z,\varepsilon) &= \sum_{i=1}^\infty \hat a_s^i \left( {Q^2 (1-z)^{2 m}  \over \mu^2}\right)^{i {\varepsilon \over 2}}
S_\varepsilon^i \overline{G}^{I,(i)}_{\rm{P},SV}(\varepsilon) 
\nonumber\\
&=
\sum_{i=1}^\infty a_s^i(Q^2 (1-z)^{2 m} ) 
\overline{{\cal G}}^I_{{\rm P},i}(\varepsilon)
\, .
\end{align}
Here, $\overline{{\cal G}}^I_{{\rm P},i}$ can be
expressed in terms of the $B^{I}, f^{I}$ and $\varepsilon$-dependent parts from lower orders
in the following way:
$\overline{{\cal G}}^I_{{\rm P},i} =  -( f_{i}^{I} + B_{i}^{I}\delta_{\rm{P},\rm{SJ}} ) + \overline{\chi}^{I}_{{\rm P},i} + \sum_{k=1}^{\infty} \varepsilon^k  
\overline{{\cal G}}^{I,k}_{{\rm P},i}\, .
$
The constants $\overline{\chi}_{{\rm P},i}^{I}$ up to four-loop order are given by,
\begin{align}
% \label{eq10}
\overline{\chi}_{{\rm P},1}^{I} &= 0\, , \, 
%\quad \quad
\nonumber\\
\overline{\chi}_{{\rm P},2}^{I} &= - 2 \beta_{0}~{\chi}_{{\rm P},1}^{I} \, ,\,
\nonumber\\
\overline{\chi}_{{\rm P},3}^{I} &= - 2 \beta_{1}~{\chi}_{{\rm P},1}^{I} - 2 \beta_{0}~ {\chi}_{{\rm P},2}^{I} \, ,\,
\nonumber\\
\overline\chi^{I}_{{\rm P},4} &= - 2 \beta_{2}~{\chi}_{{\rm P},1}^{I}  - 2
   \beta_{1}~{\chi}_{{\rm P},2}^{I} 
   - 2\beta_{0}~ 
    {\chi}_{{\rm P},3}^{I}    \,,
\end{align}
and ${\chi}_{{\rm P},i}^{I}$ contains a particular combination of $\overline{{\cal G}}^{I,k}_{{\rm P},i} $, 
\begin{align}
\label{ChiP}
{\chi}_{{\rm P},1}^{I} &=  \overline{{\cal G}}^{I,1}_{{\rm P},1} \, ,\,
\nonumber\\
{\chi}_{{\rm P},2}^{I} &=  \overline{{\cal G}}^{I,1}_{{\rm P},2} + 2 \beta_{0} ~\overline{{\cal G}}^{I,2}_{{\rm P},1}\, ,\,
\nonumber\\
\chi^{I}_{{\rm P},3} &= 
    \overline {\cal G}^{I,1}_{{\rm P},3}   + 2 \beta_{0}~\overline {\cal G}^{I,2}_{{\rm P},2} + 4 \beta_{0}^2~\overline {\cal G}^{I,3}_{{\rm P},1} + 2 \beta_{1}~\overline {\cal G}^{I,2}_{{\rm P},1}   \,.
\end{align}
The relevant constants $\overline {\cal G}^{I,i}_{{\rm P},j}$ for $P=S$ and $P=J$ can be found in \cite{Ravindran:2006cg,Ahmed:2014cla} and \cite{Ravindran:2006cg,Banerjee:2018ozf} respectively.
Similarly, for the NSV part we have
\begin{align}
    \overline{G}^{I}_{\rm{P},L}(Q^2,z,\varepsilon) &= \sum_{i=1}^\infty \hat a_s^i \left( {Q^2 (1-z)^{2 m}  \over \mu^2}\right)^{i {\varepsilon \over 2}}
S_\varepsilon^i \overline{G}^{I,(i)}_{\rm{P},L}(z,\varepsilon) \nonumber\\
&=
\sum_{i=1}^\infty a_s^i(Q^2 (1-z)^{2 m}) 
\overline{{\cal G}}^I_{{\rm P,L},i}(z,\varepsilon)\,.
\end{align}
Note, that the renormalized NSV quantities 
$\overline{{\cal G}}^I_{{\rm P,L},i}$ can be further decompose as 
$\overline{{\cal G}}^I_{{\rm P,L},i} =  2\big(C^{I}_{i}\log(1-z) + D^{I}_{i}\big) + \overline{\chi}^{I}_{{\rm P, L},i} + \sum_{k=1}^{\infty} \varepsilon^k  
\overline{{\cal G}}^{I,k}_{{\rm P,L},i} ,
$
where $\overline{\chi}^{I}_{{\rm P, L},i} = \overline{\chi}^{I}_{{\rm P},i}\Big|_{\overline{{\cal G}}^{I,k}_{{\rm P},i} \rightarrow\overline{{\cal G}}^{I,k}_{{\rm P,L},i}} $.
The expansion coefficients $C^I_i$ and $D^I_i$ are defined through $X^I=\sum_{i=1}^\infty a_s^i(\mu_R^2) X^I_i$ with $X^I=C^I,D^I$ and known at four loops, see, e.g.~\cite{Moch:2023tdj}.

We use the QCD results to third order in $a_s$ for the CFs of DIS~\cite{vanNeerven:1991nn,Zijlstra:1992kj,Zijlstra:1991qc,Zijlstra:1992qd,Hamberg:1990np,Vermaseren:2005qc,Soar:2009yh} and for the DY process, ggF Higgs boson production~\cite{Anastasiou:2015ema,Mistlberger:2018etf,Dulat:2018bfe,Duhr:2019kwi} and single-inclusive $e^+e^-$ annihilation~\cite{Rijken:1996ns,Mitov:2006wy, He:2025hin}
along with the corresponding QCD corrections to the splitting functions \cite{Moch:2004pa,Vogt:2004mw} and the quark and gluon FFs~\cite{Harlander_2003,Ravindran_2005,Kramer:1986sg,Matsuura:1987wt,Matsuura:1988sm,Gehrmann_2010,Baikov_2009,Lee_2010,Lee_2022,chak2022} to determine  $\Phi^I_{\rm{P}}$.
The CFs for the DY and ggF Higgs boson reactions determine the soft distribution functions $\Phi^q_{\rm{S}}$ and $\Phi^g_{\rm{S}}$ respectively, while the CFs of photon-mediated DIS are used for $\Phi^q_{\rm{J}}$ and of Higgs-mediated DIS for $\Phi^g_{\rm J}$.

Since the $\Phi^I_{\rm{P}}$ contain divergent terms as well as finite ones, we can separate them at the scale $\mu_s$ as
\begin{align}
\Phi^I_{\rm {P}}(Q^2,z,\varepsilon) = 
\Phi^{I,div}_{\rm {P}}\left(\mu_s^2,z,\varepsilon\right)
+\Phi^{I,fin}_{\rm {P}}\left({Q^2\over\mu_s^2},z,\varepsilon\right)\, .
\end{align}
In the above $\Phi^{I,fin}_{\rm {P}}$ is finite as $\varepsilon \rightarrow 0$.   
Defining the factor $Z^I_{{\cal S}_{\rm{P}}}= 
{\cal C} \exp(2 \Phi^{I,div}_{\rm P})$,
we obtain
\begin{eqnarray}
\label{Ssol}
    {\cal S}^I_{\rm{P}}(Q^2,z,\varepsilon)=
    Z^I_{{\cal S}_{\rm{P}}}\left({Q^2 \over \mu_s^2},z,\varepsilon\right)
    \otimes \mathbb{J}^I_{\rm P} \left({Q^2 \over \mu_s^2},z,\varepsilon\right),
\end{eqnarray}
where
\begin{align}
\label{JPhirel}
 \mathbb{J}^I_{\rm P} =   {\cal C} \exp{\left(2 \Phi^{I,fin}_{\rm{P}}\right)}.
\end{align}
The fact that ${\cal S}^I_P$ does not depend on $\mu_s$ leads
to the RG equation
\begin{align}
\label{eqRG}
\mu_s^2 \frac{d}{d \mu_s^2} \mathbb{J}^I_{\rm P}\left(\frac{Q^2}{\mu_s^2},z,\varepsilon\right) = \gamma_{\cal S_{\rm{P}}}^{I}\left(\frac{Q^2}{\mu_s^2},z,\varepsilon\right) \otimes \mathbb{J}^I_{\rm P}\left(\frac{Q^2}{\mu_s^2},z,\varepsilon\right).
\end{align}
The anomalous dimension $\gamma^I_{\cal S_{\rm P}}$ is obtained by demanding that $\Delta^{I,\rm SV+NSV}_{\rm P}$ is independent of $\mu_s$.  
Note that the IR finiteness of the CFs implies
\begin{align}
\label{Zprod}
\Big(Z_{{\cal F_{\rmP}}}^I\left(\mu_s^2\right) \!\! \Big)^2  Z^{I}_{{\cal S}_{P}}\left(\mu_s^2,z\right) \otimes
\!\Big(\overline{\Gamma}_{II}\left(\mu_s^2,z\right)\!\!\Big)^{-2 m} \!\!\!\!\!= \delta(1-z)
\end{align}
to all orders in perturbation theory.
This leads a consistency relation among the anomalous dimensions:
\begin{eqnarray}
2\gamma^I_{\cal F_{\rmP}}\delta(1-z)
+  \gamma_{\cal S_{\rm{P}}}^I + m \overline P_{II} =0 \,.
\end{eqnarray}
Using  Eqs.~\eqref{FFanom} and \eqref{APsplit} we obtain,
\begin{eqnarray}\gamma_{\cal S_{\rm{P}}}^{I} &=& 
\Bigg(B^I(\mu_s^2)\delta_{\rm{P},\rm{J}} + f^I(\mu_s^2)
\nonumber\\
&&
- A^I(\mu_s^2)
\ln \left( \frac{Q^2}{\mu_s^2}\right)\Bigg) \delta(1-z)
- 2 m\Big(A^I(\mu_s^2) {\cal D}^0_{z}
\nonumber\\
&&+ ~C^I (\mu_s^2)\log(1-z) + D^I(\mu_s^2)\Big) 
\, .
\end{eqnarray}
Substituting $Z^I_{\rm UV}$ (Eq.~\eqref{ZUVsol}), $\hat {\cal F}^I_\rmP$ (Eqs.~\eqref{factFhat},~\eqref{factF}), ${\cal S}^I_\rmP$ (Eq.~\eqref{Ssol}) and $\overline \Gamma_{II}$ (Eq.~\eqref{Gcsol}), in Eq.~\eqref{master}, setting $\mu_s=\mu_F=\mu_R$ and using Eq.~\eqref{Zprod}, we obtain
\begin{align}
\label{Deltaincl}
\Delta^{I,{\rm SV+NSV}}_{\rmP}(Q^2,z)\! =\left|{\cal F}_{\rmP,fin}^I\left({Q^2 },\mu_R^2\right)\right|^2 ~ \mathbb{J}^I_{\rm P}
\left({Q^2},\mu_R^2,z\right) \,.
\end{align}
Since $\mathbb{J}^I_{{\rm P}}$ gets contributions from SV as well as NSV terms, we expand as $\mathbb{J}^I_{{\rm P}}=
\mathbb{J}^I_{{\rm P},{\rm SV}} +\mathbb{J}^I_{{\rm P},\rm{NSV}}$. 
The soft functions  $\mathbb{J}^I_{{\rm S}}$ extracted from the CFs of the DY/ggF Higgs boson reactions and the jet functions $\mathbb{J}^I_{{\rm J}}$ from the DIS cross sections admit the following expansions for the SV part,
\begin{widetext}
\begin{align}\label{JPSVansatz}
\mathbb{J}_{{\rm P},{\rm SV}}^{I} &= \sum_{i=1}^{\infty}a_s^{i}(\mu_R^2)\Bigg[\text{J}^{I,i}_{{\rm P},-1}~\delta(1-z) 
+\sum_{k=0}^{2i-1} \text{J}^{I,i}_{{\rm P},k}~ {\cal D}^{k}_{Q}(z)\Bigg]
\, ,
\end{align} 
and for NSV part,
\begin{align}\label{JPNSVansatz}
\mathbb{J}_{{\rm P},{\rm NSV}}^{I} &=\sum_{i=0}^{\infty}a^{i}_{s}(\mu_R^2)
\Bigg[\sum_{k=0}^{i}\text{K}^{I,i}_{{\rm P},k} ~\mathcal{L}^k_{z} 
+\sum_{j=1}^{2i-1}~\sum_{k=0}^{2i-j-1} \text{R}^{I,i}_{{\rm P},j,k} ~\mathcal{L}^{k}_{z}
~{\cal L}^{j}_{Q}(z)
\Bigg]
\, .
\end{align}
Note, that all the $\left(\frac{Q^2(1-z)^{2m}}{\mu_R^2}\right)$ dependent terms are in the second term of both SV and NSV  expansions.  In the above equations, the distributions ${\cal D}^{k}_Q(z)$ and ${\cal L}^{k}_{Q,z}$ are related to standard distributions ${\cal D}^k_z$, $\delta(1-z)$ and logarithms, $\mathcal{L}_{z}^{k}$ as follows:
\begin{align}
 {\cal D}^{k}_{Q}(z)&=  \sum_{j=0}^{k}  \binom{k}{j}\log^{k-j}\left(\frac{Q^2}{\mu_R^2}\right) (2m)^{j}\mathcal{D}^{j}_{z}  +\frac{1}{2(k+1)m}\log^{k+1}\left(\frac{Q^2}{\mu_R^2}\right)\delta(1-z) \, , \label{calLdef}\\
\mathcal{ {L}}^k_{Q,z} &= \sum_{j=0}^{k}  \binom{k}{j}\log^{k-j}\left(\frac{Q^2}{\mu_R^2}\right) (2m)^{j}\mathcal{L}^{j}_{z} \, , \label{calLdef2}
\end{align}
\end{widetext}
where $\mathcal{D}_{z}^{j}=\left(\frac{\log^j(1-z)}{1-z}\right)_{+}$ and $\mathcal{L}_{z}^{j}=\log^j(1-z)$. 
Since the ${\cal D}^{k}_{Q}(z)$ and ${\cal L}^{k}_{Q,z}$ contain the complete $\mu_R^2$ dependence, we can use the RG Eq.~\eqref{eqRG} to express their coefficients at every order in $a_s(\mu_R^2)$ in terms of the anomalous dimension $A^I,B^I$ and $f^I$ $(C^I, D^I)$ as well as the process dependent constants $\text{J}_{{\rm P},-1}^{I,i}$ and $\text{K}_{{\rm P},k}^{I,i}$ of previous orders. 
The later are related to $\chi^I_{\rmP,i}$ and $\varphi_{\rmP,i}^{I,(j)}$.   

With the CFs for DY/ggF Higgs case as well those for DIS, all at third order $a_s$, and also the universal cusp, collinear and eikonal anomalous dimensions and the QCD $\beta$ functions to the required accuracy, we can determine $\Phi^I_\rmP$ exactly to order $a_s^3$ for $\rm P =J,S $ and $I=q,g$.
The finite part of $\Phi^I_\rmP$ can be expanded to order  $a_s^3$ to obtain the jet and soft functions ${\mathbb J}^I_{\rmP,\rm{SV}}$ and ${\mathbb J}^I_{\rmP,\rm{NSV}}$ to third order in $a_s$. 
The explicit results for jet functions can be found in~\cite{{Belitsky_1998},Li:2014bfa,Ahmed:2014cla} and for soft functions in~\cite{Bauer:2003pi,Bosch:2004th,Becher:2006qw,Bruser:2018rad,Becher:2009th,Becher:2010pd,Banerjee:2018ozf}.
 
At fourth order in $a_s$, we can use, for example, the RG Eq.~\eqref{eqRG} and the exact results to third order for ${\mathbb J}^I_{\rmP,\rm SV}$ to predict $\text{J}^{I,4}_{\rmP,k}$ for $I=q,g$ and $k=1,\cdots, 7$.  Thanks to the four-loop results in~\cite{Henn:2019swt,vonManteuffel:2020vjv,Agarwal:2021zft} for $A^I$ and, recently for $B^I$ and $f^I$ \cite{Kniehl:2025ttz}, we can predict the $k=0$ term, namely $\text{J}^{I,4}_{\rmP,0}$ for both $I=q,g$. 
Similarly using the results of ${\mathbb J}^I_{\rmP,\rm NSV}$ up to third order we can determine $\text{R}^{I,4}_{P,j,k}$ for $I=q,g$. 
The new fourth order results for ${\mathbb J}^I_{\rmP,{\rm SV}}$ are presented in App.~\ref{gensoftjet} and the corresponding expressions for ${\mathbb J}^I_{\rmP,{\rm NSV}}$ are included in an ancillary file.

With the help of these partial fourth order results for ${\mathbb J}^I_{\rmP}$ and those for the FF, ${\cal F}^I_{\rmP,fin}$, \cite{Lee:2022nhh}, we obtain fourth order SV results for the CFs (except for the $\delta(1-z)$ terms) for a variety of processes, confirming and extending the results of \cite{Kniehl:2025ttz}:
the DY process, Higgs boson production in ggF and bottom-quark annihilation, single-inclusive $e^+e^-$ annihilation as well as for DIS, both, with exchange of a photon or a scalar. 
Similarly, we obtain fourth order NSV results for these CFs except for the powers of logarithms ${\cal L}^k_z$ with $k=0,\cdots,3$.

Before concluding this section, a few comments are in order to relate our results for ${\mathbb J}^I_{\rmP}$ to the soft and jet functions of SCET~\cite{Bauer:2000ew,Bauer:2000yr,Bauer:2001ct,Bauer:2001yt,Bauer:2002nz,Beneke:2002ph}.  
In SCET, the inclusive production of a pair of leptons or of a $Z/W^\pm$ or a Higgs boson, near the partonic threshold, factorizes into hard and soft functions. 
Similarly, the DIS cross section is sensitive to jet function. 
The CFs of the inclusive DY (ggF Higgs) production cross section factorizes into quark (gluon) soft function times the corresponding hard part: for example, for the DY process,
$\Delta_{\rm DY}^{q,\rm{SV}} = H_{\rm DY}^q(Q^2,\mu_R^2) {\cal S}^q(Q^2,\mu_R^2,z)$, where $H^q_{\rm DY}$ is the hard function and ${\cal S}^q$ is the soft function.  Similarly, the perturbatively calculable CF of the structure function $F_2(Q^2,x_B)$ can be written as
$\Delta_2^{q,\rm{SV}}(Q^2,z) = H^q_2(Q^2,\mu_R^2) J^q(Q^2,\mu_R^2,z)$,
where $H^q_2$ is the hard function and $J_q$ is the quark jet function.  
Since hard functions in SCET are related to the finite part of the respective FFs, 
$|{\cal F}^I_{\rmP,fin}|^2$, we can identify jet and soft functions with our $\mathbb{J}_{\rm{P}}^I$ by setting $m=1/2$ for $\rmP = {\rm J}$ and $m=1$ for $\rmP = {\rm S}$, respectively, see Eq.~\eqref{Deltaincl}.

\section{Rapidity distributions}
In addition to the inclusive cross sections discussed above, we can study the impact of the four-loop anomalous dimensions $f_4$ and $B_4$ on the fourth‑order CFs of differential distributions. In particular, we are interested in the rapidity of the lepton pair in DY and of the Higgs boson in ggF or in the bottom‑quark annihilation process.
We follow the framework developed in a series of works  \cite{Ravindran:2006bu,Ravindran:2007sv,Ahmed:2014uya,Ahmed:2014era,Banerjee:2017cfc,Banerjee:2018vvb,Banerjee:2018mkm,AH:2020qoa,Ahmed:2020amh,AH:2021vhf,Ravindran:2022aqr,Ravindran:2023qae} to predict rapidity distributions for the DY and Higgs boson processes. 
Other alternative approaches that study the SV part of rapidity CFs can be found in 
\cite{Laenen:1992ey,Sterman:2000pt,Mukherjee:2006uu,Becher:2006nr,Becher:2007ty,Bonvini:2014qga,Ebert:2017uel,Lustermans:2019cau,Cacciari:2001cw}.

The rapidity distribution of a colorless final state $F$ in hadron-hadron collision is given by 
\begin{eqnarray}
\label{sighad}
{d \sigma^I\over dy } &=&
\sigma^I_{\rm B}(\mu_R^2) 
\sum_{a,b=q,\overline q,g}
\int_{x_1^0}^1 {dz_1 \over z_1}\int_{x_2^0}^1 {dz_2 \over z_2}~ 
f_{a}\left({x_1^0 \over z_1},\mu_F^2\right)
\nonumber \\ 
&& \times f_{b}\left({x_2^0\over z_2}, \mu_F^2\right)
\Delta^I_{d,ab} (z_1,z_2,q^2,\mu_F^2,\mu_R^2)\,,
\end{eqnarray}
where 
$\sigma^I_B(\mu_R^2)=\sigma^I_B(x_1^0,x_2^0,q^2,\mu_R^2)$ is the Born cross section,
and the indicies $I=q$ correspond to the DY case and $I=g(b)$ for Higgs boson production in ggF and bottom-quark annihilation. 
The hadronic scaling variables $x_l^0~(l=1,2)$ are defined through the rapidity $y$ of the lepton pair or the Higgs boson:~ $y={1 \over2} \ln(p_2.q/p_1.q)={1 \over 2 } \ln\left({x_1^0/x_2^0}\right)$ and $\tau=q^2/S=x_1^0 x_2^0$.
The observable in Eq.~\eqref{sighad} for the DY case is $\sigma^I = d\sigma^{q}(\tau,q^2,y)/dq^2$, i.e. the invariant mass distribution of lepton pair in the final state $F$, whereas for the Higgs production in ggF or in bottom quark annihilation $\sigma^I=\sigma^{g,b}(\tau,q^2,y)$ respectively. 
The parton level rapidity CFs, $\Delta_{d,ab}$, are functions of the scaling variables  $z_l=x_l^0/x_l~(l=1,2)$.  
In the threshold region $z_l=1$, the rapidity CFs contain SV terms such as $\delta(1-z_l)$ and ${\cal D}_j(z_l)\equiv\big(\frac{\ln^j(1-z_l)}{(1-z_l)}\big)_+$ and the NSV terms contain the combinations  ${\cal D}_i(z_l)\ln^k(1-z_j)$ and $\delta(1-z_l) \ln^k(1-z_j)$ with ($l,j=1,2),~(i,k=0,1,\cdots$).

As demonstrated in \cite {Ravindran:2006bu,Ravindran:2007sv} for the SV and in \cite{AH:2020qoa,AH:2020xll} for the NSV parts of the CFs $\Delta^{I,{\rm SV+NSV}}_{d,a \overline a }$, 
their sum factorizes into contributions from the FFs and soft distribution functions, 
appropriately convoluted with mass factorization kernels of the incoming partons as 
\begin{align}
\label{masterR}
\Delta^{I,\rm {SV+NSV}}_{d} = &  (Z^{I}_{\rm{UV}} (\hat a_s,\mu_R^2, \mu^2, \varepsilon))^2 \big| \hat {\cal F}^{I}_S(\hat a_s, -q^2,\mu^2,\varepsilon)\big|^2
\nonumber \\  &
\hspace{-0.8cm}\times  {\cal S}^{I}_{d}(\hat a_s, Q^2,z_1,z_2,\varepsilon) 
%\nonumber \\ & 
\otimes \Big(\overline\Gamma_{II}(\hat a_s, \mu^2, \mu^2_{F}, z_1, \varepsilon) 
\nonumber \\ &
\hspace{-0.8cm} 
\times \overline\Gamma_{II}(\hat a_s, \mu^2, \mu^2_{F}, z_2, \varepsilon)\Big)^{-1}\,.  
\end{align}
The soft distribution ${\cal S}^I_d$ satisfies a $K$ plus $G$ equation, see \cite{Ravindran:2006bu}  and hence accounts for the soft enhancements associated with the real emissions in the production channel. 
They are found to be universal and their solution takes the following form:
\begin{eqnarray}
\label{eq:SsolR}
{\cal S}^I_d = {\cal C} \exp\Big(2~\Phi^I_d(Q^2,z_1,z_2,\varepsilon)\Big)
\, ,
\end{eqnarray}
where 
\begin{align}
\label{eq:Phid}
\Phi^I_d =& \sum_{i=1}^\infty \hat a_s^i \left(Q^2 \overline z_1 \overline z_2 \over \mu^2\right)^{i{\varepsilon \over 2}} S_\varepsilon^i
%\nonumber \\&
\Bigg[ {(i \varepsilon)^2 \over 4  \overline z_1 \overline z_2 } \hat \phi_d^{I,(i)}(\varepsilon)
\nonumber \\&
+ {i \varepsilon \over 4  \overline z_1 } \varphi_{d,I}^{(i)} (\overline z_2,\varepsilon)
%\nonumber \\&
+ {i \varepsilon \over 4 \overline z_2 } \varphi_{d,I}^{(i)} (\overline z_1,\varepsilon)\Bigg] \,.
\end{align}
%
%}
The first term corresponds to the soft contributions while the remaining ones denote the next-to-soft contributions, with $\overline z_1 = (1-z_1)$ and $\overline z_2 = (1-z_2)$.  
The SV and NSV constants, $\hat \phi^{I,(i)}_d(\varepsilon)$ and $\varphi_{d,I}^{(i)}(z_l,\varepsilon)$ can be obtained using  $\hat \phi_{\rmP}^{I,(i)}(\varepsilon)$ and 
$\hat \varphi^{I,(i)}_{\rmP}(z,\varepsilon)$ (see Eq.~\eqref{eq7}), 
respectively, with the help of the following relation in the large $N$ limit~\cite{Ravindran:2006bu}:
%
%\begin{small}
\begin{align}
\label{eq:RapInc}
\int_0^1 
dx_1^0 \int_0^1 
dx_2^0 \left(x_1^0 x_2^0\right)^{N-1}
{d \sigma^I \over d y}
=\int_0^1 d\tau~ \tau^{N-1} ~\sigma^I\,,
\end{align}
%\end{small}
%
where $\sigma^I$ is the inclusive cross section.
The results up to third order for the DY process and Higgs boson production can be found in \cite{Ravindran:2006bu,Ahmed:2014uya,Ahmed:2014era}. 
Note that $\Phi^I_d$ contains IR singularities which cancel against those coming from the FF in Eq.~\eqref{FFdiv}, and the AP kernels in Eq.~\eqref{Gcsol}. 
We decompose $\Phi^I_d=\Phi^{I,div}_d+\Phi_d^{I,fin}$ to write ${\cal S}^I_d$ as
\begin{align}
{\cal S}^I_d = Z^I_d(Q^2,\mu_s^2,z_1,z_2)\tilde \otimes {\mathbb{J}}^{I}_d(Q^2,\mu_s^2,z_1,z_2)¸, ,
\end{align}
where $\tilde \otimes$ denotes the convolution in the scaling variables $z_1, z_2$. 
Here, we have chosen an arbitrary scale $\mu_s$ to factorize the divergent part of ${\cal S}^I_d$, 
while the finite part of rapidity soft function is given by
\begin{align}
\label{JPhirelR}
 \mathbb{J}^I_{d} =   \tilde {\cal C} \exp{\left(2 \Phi^{I,fin}_{d}\right)}
\end{align}
and $Z^I_d = \tilde {\cal C} \exp(2 \Phi^{I,div}_d)$. 
Substituting $Z^I_{\rm UV}$ (Eq.~\eqref{ZUVsol}), $\hat {\cal F}^I_\rmP$ (Eqs.~\eqref{factFhat},~\eqref{factF}), ${\cal S}^I_d$ (Eqs.~\eqref{eq:SsolR}, \eqref{eq:Phid}) and $\overline \Gamma_{II}$ (Eq.~\eqref{Gcsol}), in Eq.~\eqref{masterR}, setting $\mu_s=\mu_F=\mu_R$ and using Eq.~\eqref{Zprod}, we obtain 
\begin{align}
\label{DeltaFFJet}
\Delta^{I,{\rm SV+NSV}}_{d} =\left|{\cal F}_{\rmP,fin}^I\left({Q^2 },\mu_R^2\right)\right|^2 ~ \mathbb{J}^I_{d}
\left({Q^2},\mu_R^2,z_1,z_2\right) \,.
\end{align}
The rapidity soft function ${\mathbb {J}}_d^I$ can be related to the inclusive one ${\mathbb {J}}_S^I $ thanks to Eq.~\eqref{eq:RapInc}.
In analogy to the inclusive case, we can expand ${\mathbb J}^I_d$ as 
\begin{widetext}
\begin{align}
\label{prob}
{\mathbb J}^{I,(i)}_{d}&=
{\mathbb J}^{I,(i)}_{d,\delta \delta} ~ \delta(1-z_1) \delta(1-z_2)
+ \sum_{j=0}^{2 i-1} {\mathbb J}^{I,(i,j)}_{d,\delta { {\cal L}}}~ 
\delta(1-z_1) {\cal L}^j_{z_2}
 + \sum_{j=0}^{2 i -1} {\mathbb J}^{I,(i,j)}_{d, {\cal L}\delta}~ 
 \delta(1-z_2) {\cal L}^j_{z_1}
 \nonumber\\ &
+ \sum_{j=0}^{2 i-1} \sum_{k=0}^{2 i -1 -j}{\mathbb J}^{I,(i,j,k)}_{d,{\cal D}  {\cal D}}
 {\cal D}^{j}_{Q}(z_1)  {\cal D}^{k}_{Q}(z_2) 
 + \sum_{j=0}^{2 i-1} {\mathbb J}^{I,(i,j)}_{d,\delta {\cal D}}~ 
\delta(1-z_1) {\cal D}^{j}_{Q}(z_2)
 + \sum_{j=0}^{2 i-1} {\mathbb J}^{I,(i,j)}_{d, {\cal D}\delta}~ 
 \delta(1-z_2) {\cal D}^{j}_{Q}(z_1)
 \nonumber \\ &
  + \sum_{j=0}^{2 i-1}\sum_{k=0}^{2 i -1 -j} {\mathbb J}^{I,(i,j,k)}_{d,{\cal L} {\cal D}}~ 
{\cal L}^j_{z_1} {\cal D}^{k}_{Q}(z_2)
 + \sum_{j=0}^{2 i -1} \sum_{k=0}^{2 i -1 -j}{\mathbb J}^{I,(i,j,k)}_{d, {\cal D} {\cal L}}~ 
  {\cal D}^{k}_{Q}(z_1) {\cal L}^j_{z_2}\, .
% \nonumber \\ & 
% +  \sum_{j=0}^{2 i-1} {\mathbb J}^{I,(j)}_{d,%\delta{\cal L}}
%\delta(1-z_1)  {\cal L}^j (z_2) + 
%\sum_{j=0}^{2i -1} {\mathbb J}^{I,(j)}_{d,{\cal %L}\delta }
% {\cal L}^j(z_1)\delta(1-z_2)
\end{align}
\end{widetext}
Substituting  Eqs.~\eqref{Deltaincl} and~\eqref{DeltaFFJet} along with Eqs.~\eqref{JPSVansatz}, \eqref{JPNSVansatz}, \eqref{eq:SsolR},\eqref{eq:Phid} and \eqref{prob} in Eq.~\eqref{eq:RapInc}
and taking the large $N$ limit, we can determine all the coefficients ${\mathbb J}^I_{d,st}$, where $s,t=\delta,{\cal L}, {\cal D}$, in terms of the corresponding coefficients of ${\mathbb J}^I_S$.  
The results for the SV part of ${\mathbb J}^I_{d,st}$ up to fourth order are collected in the Appendix while those for the NSV  are provided them in ancillary files (as described in the Appendix). 
Since soft function for the rapidity distribution is obtained from the corresponding one for the inclusive cross section, we can determine the complete SV terms of ${\mathbb J}^I_{d,st}$ up to third order. 
However, at fourth order, we are missing the coefficients of terms proportional to $\delta(1-z_1) \delta(1-z_2), \delta(1-z_1){\cal L}^{j}_{z_2}$ and $\delta(1-z_2){\cal L}^{j}_{z_1}$ for $j=0,1,2,3$. 
With the available information on ${\mathbb J}^I_{d,st}$, we thus predict the CFs of the DY process and Higgs boson production at fourth order to the same accuracy, with explicit result deferred to the Appendix and, again the predictions for the NSV contributions to ${\mathbb J}^{I,(i)}_{d}$ and $\Delta^{I,{\rm SV+NSV}}_{d}$ provided in ancillary files. 

%csm up to here
%{\bf
%$----------------------------------------------$
%}
 
%%%%%%%
\section{CONCLUSION}

In this work, we have presented a systematic derivation of the soft and jet functions for the kinematics of DY and DIS processes, respectively, at four-loop order for both quark and gluon. These functions constitute the universal building blocks of hadronic cross sections in the threshold region. 
Our analysis is based on the universality of soft and collinear dynamics and establishes a direct connection between perturbative results obtained in full QCD and their effective-theory counterparts in SCET. 
By utilizing recent four-loop results for the collinear and eikonal anomalous dimensions, together with RG invariance and the factorization properties of QCD cross sections in the threshold region, we have extended the known results for soft and jet functions to the highest perturbative order currently accessible.

The four-loop soft and jet functions derived here represent essential building blocks for precision phenomenology at hadron colliders and future facilities such as the Electron–Ion Collider \cite{Goyal:2025bzf}.  In particular, they constitute a crucial ingredient for extending the $N$-jettiness subtraction formalism to $\mathrm{N}^4\mathrm{LO}$ accuracy, thereby opening the possibility of computing fully differential cross sections at unprecedented precision. Beyond fixed-order applications, these results also provide key input for threshold resummation at higher logarithmic accuracy.
Our results  demonstrate that high-order perturbative information encoded in inclusive observables can be systematically exploited to extract universal ingredients relevant for a wide class of collider processes.

In summary, we have provided four-loop results for the quark and gluon soft and jet functions, derived from known perturbative information in full QCD. 
These results supply essential ingredients for higher-order fixed-order and resummed calculations.

\section*{Acknowledgements}
This work has been supported through a joint Indo-German research grant by
the Department of Science and Technology (DST/INT/DFG/P-03/2021/dtd.12.11.21). 
S.M. acknowledges the ERC Advanced Grant 101095857 {\it Conformal-EIC}.

%%%%%%%%%%%%%%%%%%%%%%%%%%%%%%%%%%%%%%%%%
\appendix

\begin{widetext}

\section{Description of the ancillary files}
We provide the results of the present paper in \texttt{FORM} format~\cite{Kuipers:2012rf,Ruijl:2017dtg} in the following the files:
\begin{itemize}
    \item  \verb|Coeff_inc.list| -- results of inclusive cross sections for the following processes:
    \begin{itemize}      
\item Inclusive DIS, Higgs boson-mediated: $[C_{\rm DIS}^g]$,
\item Inclusive DIS, photon-mediated (${\cal F}_2$): $[C_{\rm DIS}^q]$,
\item Inclusive Higgs boson production through ggF: $[C_{\rm ggH}^g]$,
\item Inclusive DY-process: $[C_{\rm DY}^q]$,
\item Inclusive Higgs boson production through bottom-quark annihilation: $[C_{\rm bBH}^b]$,
\item Single-inclusive annihilation of $e^+e^-$ (SIA), photon-mediated $({\cal F}_T)$: $[C_{\rm SIA}^q]$.
    \end{itemize}
\item  \verb|Coeff_rap.list| -- results of rapidity cross sections for the following processes:
\begin{itemize} 
\item Rapidity distribution for Higgs boson production through ggF: $[C_{d,\rm ggF}^g]$,
\item Rapidity distribution for the DY process: $[C_{d,\rm DY}^q]$,
\item Rapidity distribution for Higgs boson production through bottom-quark annihilation: $[C_{d,\rm bBH}^b]$.
\end{itemize}
\item \verb|SoftJet_inc.list| -- results of the soft and jet functions for quarks and gluons (inclusive case) in terms of $\mathcal{D}^{k}_{Q}(z)$ and $\mathcal{ {L}}^k_{Q,z}$ defined in Eq.~\eqref{calLdef}, and Eq.~\eqref{calLdef2}, respectively, with $m = 1$ for the soft function and $m = \frac{1}{2}$ for the jet function.
\item \verb|SoftJet_rap.list| -- results of the soft function for quarks and gluons (rapidity case) in terms of $\mathcal{D}^{k}_{Q}(z_{i})$ with $m = 1$ and $\mathcal{L}^{j}_{z_i}$ = $\log^j(1-z_i)$ for $i = 1,2$.
\end{itemize}

\section{Soft and Jet functions}
\label{gensoftjet}
In the following, we present the general structure of  $\mathbb{J}_{{\rm P}, {\rm SV}}^{I}$  up to fourth order in $a_s$. 
The corresponding NSV terms, i.e., $\mathbb{J}_{{\rm P}, {\rm NSV}}^{I}$ are provided in the ancillary file.
\begin{small}
\begin{align}
  \mathbb{J}^{I,1}_{{\rm P},{\rm SV}} &=
%%%%%%%%%%%%%%%%%%%%%%%%%%
%           SV
%%%%%%%%%%%%%%%%%%%%%%%%%%
        \bm{\delta_{x}}   \bigg[ \text{J}^{I,1}_{{\rm P},-1} \bigg]
       + 2  m~ \bm{\mathcal{D}^{1}_{Q}}   \bigg[ A_{1}^{I} \bigg]
         - 2  m~ \bm{\mathcal{D}^{0}_{Q}}   \bigg[ \FBo  \bigg]
%%%%%%%%%%%%%%%%%%%%%%%%%%%%
%           NSV 
%%%%%%%%%%%%%%%%%%%%%%%%%%%%
       % + \bm{\mathcal{L}^{1}_{z}}   \bigg( \text{K}^{I,1}_{{\rm P},1} \bigg)
       % + \bm{\mathcal{L}^{0}_{z}}   \bigg( \text{K}^{I,1}_{{\rm P},0} \bigg)
       % + \bm{\mathcal{L}^{0}_{z}} \bm{\mathcal{\overline{L}}_{1}}   \bigg( 2  m~ D_{1}^{I} \bigg)

\end{align}
\begin{align}
   \mathbb{J}^{I,2}_{{\rm P},{\rm SV}} &=
        \bm{\delta_{x}}   \bigg[ \text{J}^{I,2}_{{\rm P},-1} \bigg]
       
       +  m~\bm{\mathcal{D}^{3}_{Q}}   \bigg[  {A_{1}^{I}}^2 \bigg]
       
       - m A_{1}^{I}~ \bm{\mathcal{D}^{2}_{Q}}   \bigg[ \beta_{0}  +  3\FBo   \bigg]
       
       + 2 m~ \bm{\mathcal{D}^{1}_{Q}}   \bigg[
        A_{2}^{I}  + \beta_{0} \FBo  - 4  m^2 \zeta_{2} {A_{1}^{I}}^2 + (\FBo)^2 + A_{1}^{I} \text{J}^{I,1}_{{\rm P},-1}\bigg]
\nonumber\\&       
       -2 m~ \bm{\mathcal{D}^{0}_{Q}}   \bigg[  \FBtw   - 4  m^2 \zeta_{2} A_{1}^{I} \FBo  - 8  m^3 \zeta_{3} {A_{1}^{I}}^2 + \FBo  \text{J}^{I,1}_{{\rm P},-1} + \beta_{0} \text{J}^{I,1}_{{\rm P},-1} \bigg]
%%%%%%%%%%%%%%%%%%%%%%%%%%%
%           NSV
%%%%%%%%%%%%%%%%%%%%%%%%%%%%
       % + \bm{\mathcal{L}^{2}_{z}}   \bigg( \text{K}^{I,2}_{{\rm P},2} \bigg)

       % + \bm{\mathcal{L}^{1}_{z}}   \bigg( \text{K}^{I,2}_{{\rm P},1} \bigg)

       % + \bm{\mathcal{L}^{1}_{z}} \bm{\mathcal{\overline{L}}_{2}}   \bigg( \frac{1}{2} A_{1}^{I} \text{K}^{I,1}_{{\rm P},1} \bigg)

       % + \bm{\mathcal{L}^{1}_{z}} \bm{\mathcal{\overline{L}}_{1}}   \bigg(  - \FBo  \text{K}^{I,1}_{{\rm P},1} - \beta_{0} \text{K}^{I,1}_{{\rm P},1} + 2  m~ C_{2}^{I} \bigg)

       % + \bm{\mathcal{L}^{0}_{z}}   \bigg( \text{K}^{I,2}_{{\rm P},0} \bigg)

       % + \bm{\mathcal{L}^{0}_{z}} \bm{\mathcal{\overline{L}}_{3}}   \bigg(  m~ A_{1}^{I} D_{1}^{I} \bigg)

       % + \bm{\mathcal{L}^{0}_{z}} \bm{\mathcal{\overline{L}}_{2}}   \bigg( \frac{1}{2} A_{1}^{I} \text{K}^{I,1}_{{\rm P},0} - 2  m~ D_{1}^{I} \FBo  - m~ \beta_{0} D_{1}^{I} + 2    m^2
       %   {A_{1}^{I}}^2 \bigg)

       % + \bm{\mathcal{L}^{0}_{z}} \bm{\mathcal{\overline{L}}_{1}}   \bigg(  - \FBo  \text{K}^{I,1}_{{\rm P},0} - \beta_{0} \text{K}^{I,1}_{{\rm P},0} + 2  m~ D_{2}^{I} + 2  m~     D_{1}^{I}
       %   \text{J}^{I,1}_{{\rm P},-1} - 2  m~ \zeta_{2} A_{1}^{I} \text{K}^{I,1}_{{\rm P},1} - 4  m^2 A_{1}^{I} \FBo  - 8  m^3 \zeta_{2}
       %   A_{1}^{I} D_{1}^{I} \bigg)

\end{align}
\begin{align}
   \mathbb{J}^{I,3}_{{\rm P},{\rm SV}} &=
        \bm{\delta_{x}}   \bigg[ \text{J}^{I,3}_{{\rm P},-1} \bigg]
       
       + \frac{1}{4}  m~ \bm{\mathcal{D}^{5}_{Q}}   \bigg[ {A_{1}^{I}}^3 \bigg]
       - \frac{5}{4}  m   {A_{1}^{I}}^2~ \bm{\mathcal{D}^{4}_{Q}}   \bigg[  \FBo  + \frac{2}{3} \beta_{0} \bigg]
       
       +2 m A_{1}^{I}~ \bm{\mathcal{D}^{3}_{Q}}   \bigg[  A_{2}^{I}   - 4  m^2 \zeta_{2} {A_{1}^{I}}^2 + (\FBo)^2 
     + \frac{5}{3} \beta_{0}  \FBo  
     + \frac{1}{3} \beta_{0}^2  
     \nonumber\\&   
       + \frac{1}{2} {A_{1}^{I}} \text{J}^{I,1}_{{\rm P},-1} 
         \bigg]
          
       - m~ \bm{\mathcal{D}^{2}_{Q}}   \bigg[A_{2}^{I} \Big(3  \FBo + 2 \beta_{0} \Big)     + (\FBo )^3 + 3 \beta_{0} (\FBo)^2  + 2 \beta_{0}^2 \FBo  
      - 24  m^2 \zeta_{2} {A_{1}^{I}}^2 \FBo 
    - 12  m^2 \zeta_{2} \beta_{0} {A_{1}^{I}}^2 
\nonumber\\&         
         - 40  m^3 \zeta_{3} {A_{1}^{I}}^3 + 3 A_{1}^{I} \FBtw + \beta_{1} A_{1}^{I} + 4 \beta_{0} A_{1}^{I} \text{J}^{I,1}_{{\rm P},-1} + 3 A_{1}^{I} \FBo  \text{J}^{I,1}_{{\rm P},-1} \bigg]
          + 2 m~ \bm{\mathcal{D}^{1}_{Q}}   \bigg[ 2 \FBo  \FBtw             
            + A_{3}^{I} + \beta_{1} \FBo  + 2 \beta_{0} \FBtw          
\nonumber\\&        
       - 8  m^2 \zeta_{2} A_{1}^{I} (\FBo)^2 - 8  m^2 \zeta_{2} A_{1}^{I} A_{2}^{I} - 12  m^2 \zeta_{2} \beta_{0} A_{1}^{I} \FBo  - 32  m^3 \zeta_{3} {A_{1}^{I}}^2 \FBo  - 24 m^3 \zeta_{3} \beta_{0} {A_{1}^{I}}^2 - \frac{16}{5}  m^4 \zeta_{2}^2 {A_{1}^{I}}^3 + 
        A_{2}^{I} \text{J}^{I,1}_{{\rm P},-1} 
\nonumber\\&         
        +(\FBo)^2 \text{J}^{I,1}_{{\rm P},-1}   + A_{1}^{I} \text{J}^{I,2}_{{\rm P},-1}+ 3 \beta_{0} \FBo  \text{J}^{I,1}_{{\rm P},-1}    + 2 \beta_{0}^2 \text{J}^{I,1}_{{\rm P},-1}          - 4  m^2 \zeta_{2} {A_{1}^{I}}^2 \text{J}^{I,1}_{{\rm P},-1}\bigg]
          
       - 2 m~ \bm{\mathcal{D}^{0}_{Q}}   \bigg[ \FBth - 4
          m^2 \zeta_{2} A_{2}^{I} \FBo  - 4  m^2 \zeta_{2} A_{1}^{I} \FBtw
\nonumber\\&  
   - 8  m^3 \zeta_{3} A_{1}^{I}
         (\FBo )^2 - 16  m^3 \zeta_{3} A_{1}^{I} A_{2}^{I} - 8
          m^3 \zeta_{3} \beta_{0} A_{1}^{I} \FBo  - \frac{16}{5}  m^4 \zeta_{2}^2 {A_{1}^{I}}^2 \FBo  - \frac{32}{5}
          m^4 \zeta_{2}^2 \beta_{0} {A_{1}^{I}}^2 - 96  m^5 \zeta_{5} {A_{1}^{I}}^3 + 64  m^5 \zeta_{2}
         \zeta_{3} {A_{1}^{I}}^3
\nonumber\\&          
+ \FBtw \text{J}^{I,1}_{{\rm P},-1} 
     + \FBo  \text{J}^{I,2}_{{\rm P},-1}     
     + \beta_{1} \text{J}^{I,1}_{{\rm P},-1} + 2 \beta_{0} \text{J}^{I,2}_{{\rm P},-1}  - 4  m^2 \zeta_{2} A_{1}^{I}
         \FBo  \text{J}^{I,1}_{{\rm P},-1} - 4  m^2 \zeta_{2} \beta_{0} A_{1}^{I} \text{J}^{I,1}_{{\rm P},-1} 
%\nonumber\\&         
 - 8  m^3 \zeta_{3} {A_{1}^{I}}^2 \text{J}^{I,1}_{{\rm P},-1} \bigg]

\end{align}

%\begin{widetezt}
%\begin{small}
\begin{align}
%\begin{autobreak}
\mathbb{J}^{I,4}_{{\rm P},{\rm SV}}&=  
        \bm{\delta_{x}}   \bigg[ \text{J}^{I,4}_{{\rm P},-1} \bigg]
       +\frac{1}{24}  m~ \bm{\mathcal{D}^{7}_{Q}}   \bigg[  {A_{1}^{I}}^4 \bigg]   
       - \frac{7}{24}  m   {A_{1}^{I}}^3~\bm{\mathcal{D}^{6}_{Q}}   \bigg[ \FBo  + \beta_{0} \bigg]
       +\frac{3}{4}m~ \bm{\mathcal{D}^{5}_{Q}}   \bigg[{A_{1}^{I}}^2 A_{2}^{I} +  {A_{1}^{I}}^2 (\FBo )^2   
        + \frac{7}{3}  \beta_{0} {A_{1}^{I}}^2 \FBo  + \frac{8}{9}  \beta_{0}^2 {A_{1}^{I}}^2 
        \nonumber\\&           
         - 4 m^2  \zeta_{2} {A_{1}^{I}}^4 
         + \frac{1}{3} {A_{1}^{I}}^3 
         \text{J}^{I,1}_{{\rm P},-1}\bigg]
       -\frac{5}{6} m~ \bm{\mathcal{D}^{4}_{Q}}   \bigg[3 A_{1}^{I} A_{2}^{I} \FBo + A_{1}^{I} (\FBo )^3   + \frac{3}{2} {A_{1}^{I}}^2 \FBtw  +
         \beta_{1} {A_{1}^{I}}^2 
         + 4 \beta_{0} A_{1}^{I} (\FBo )^2 + 3 \beta_{0} A_{1}^{I} A_{2}^{I} 
\nonumber\\&         
         + 4 \beta_{0}^2 A_{1}^{I} \FBo  + \frac{3}{5} \beta_{0}^3 A_{1}^{I} - 18  m^2 \zeta_{2} {A_{1}^{I}}^3 \FBo  - 16  m^2
         \zeta_{2} \beta_{0} {A_{1}^{I}}^3 - 28  m^3 \zeta_{3} {A_{1}^{I}}^4 + \frac{3}{2}{A_{1}^{I}}^2 \FBo  \text{J}^{I,1}_{{\rm P},-1} + \frac{5}{2}\beta_{0} {A_{1}^{I}}^2 \text{J}^{I,1}_{{\rm P},-1}  \bigg]
         + m~\bm{\mathcal{D}^{3}_{Q}}   \bigg[ 
\nonumber\\&        
      {A_{2}^{I}}^2 + 4 A_{1}^{I} \FBo \FBtw  + \frac{1}{3} (\FBo )^4 + 2 A_{2}^{I} (\FBo )^2   + 2 A_{1}^{I} A_{3}^{I} 
         + \frac{10}{3} \beta_{1} A_{1}^{I} \FBo         
        + 2 \beta_{0} (\FBo )^3 + \frac{14}{3} \beta_{0} A_{2}^{I} \FBo  + \frac{16}{3} \beta_{0} A_{1}^{I} \FBtw  
\nonumber\\&   
 + \frac{5}{3} \beta_{0} \beta_{1} A_{1}^{I} + \frac{11}{3} \beta_{0}^2 (\FBo   )^2     + 2 \beta_{0}^2 A_{2}^{I} 
         + 2 \beta_{0}^3 \FBo  - 24  m^2 \zeta_{2} {A_{1}^{I}}^2 (\FBo  )^2 - 24
          m^2 \zeta_{2} {A_{1}^{I}}^2 A_{2}^{I} 
%          \nonumber\\&  
          - \frac{160}{3}  m^2\zeta_{2} \beta_{0} {A_{1}^{I}}^2 \FBo  
           \nonumber\\&             
          - \frac{16}{5}  m^4 \zeta_{2}^2{A_{1}^{I}}^4  
          - \frac{44}{3}  m^2 \zeta_{2} \beta_{0}^2 {A_{1}^{I}}^2           
          - \frac{256}{3}  m^3
         \zeta_{3} {A_{1}^{I}}^3 \FBo  - \frac{272}{3}  m^3 \zeta_{3} \beta_{0} {A_{1}^{I}}^3   +  {A_{1}^{I}}^2 \text{J}^{I,2}_{{\rm P},-1} 
         + 2 A_{1}^{I} A_{2}^{I} \text{J}^{I,1}_{{\rm P},-1} 
%   \nonumber\\&        
         + \frac{22}{3} \beta_{0} A_{1}^{I}
         \FBo  \text{J}^{I,1}_{{\rm P},-1} 
 \nonumber\\&          
         + 2 A_{1}^{I} (\FBo )^2 \text{J}^{I,1}_{{\rm P},-1}    + 6 \beta_{0}^2 A_{1}^{I} \text{J}^{I,1}_{{\rm P},-1} - 8  m^2 \zeta_{2} {A_{1}^{I}}^3 \text{J}^{I,1}_{{\rm P},-1}  \bigg]
%\nonumber\\&          
       - 3m~ \bm{\mathcal{D}^{2}_{Q}}   \bigg[ A_{3}^{I} \FBo  + A_{2}^{I}    \FBtw + (\FBo )^2 \FBtw + A_{1}^{I} \FBth + \frac{1}{3} \beta_{2} A_{1}^{I}   
 \nonumber\\& 
       + \beta_{1} (\FBo )^2 
+ \frac{2}{3} \beta_{1} A_{2}^{I}+ 3 \beta_{0} \FBo  \FBtw + \beta_{0} A_{3}^{I}  + \frac{5}{3} \beta_{0} \beta_{1} \FBo    + 2 \beta_{0}^2 \FBtw   - 4  m^2 \zeta_{2} A_{1}^{I} (\FBo )^3 - 16  m^2 \zeta_{2} A_{1}^{I} A_{2}^{I} \FBo 
 \nonumber\\& 
- 8  m^2 \zeta_{2} {A_{1}^{I}}^2 \FBtw - 4
          m^2 \zeta_{2} \beta_{1} {A_{1}^{I}}^2 - 16  m^2 \zeta_{2} \beta_{0} A_{1}^{I} (\FBo )^2    - 12  m^2
         \zeta_{2} \beta_{0} A_{1}^{I} A_{2}^{I} -    \frac{44}{3}  m^2
         \zeta_{2} \beta_{0}^2 A_{1}^{I} \FBo  - 32  m^3 \zeta_{3} {A_{1}^{I}}^2 (\FBo )^2 
  \nonumber\\&        
         - 40  m^3 \zeta_{3} {A_{1}^{I}}^2 A_{2}^{I} - 72  m^3 \zeta_{3}  \beta_{0} {A_{1}^{I}}^2
         \FBo  - \frac{88}{3}  m^3 \zeta_{3} \beta_{0}^2 {A_{1}^{I}}^2 - \frac{16}{5}  m^4 \zeta_{2}^2 {A_{1}^{I}}^3 \FBo
          - \frac{64}{5}  m^4 \zeta_{2}^2 \beta_{0} {A_{1}^{I}}^3 - 224  m^5 \zeta_{5} {A_{1}^{I}}^4 
   \nonumber\\&          
          + 160  m^5 \zeta_{2} \zeta_{3} {A_{1}^{I}}^4
+ \frac{1}{3} (\FBo )^3 \text{J}^{I,1}_{{\rm P},-1} 
 + A_{2}^{I} \FBo \text{J}^{I,1}_{{\rm P},-1}  + A_{1}^{I} \FBtw \text{J}^{I,1}_{{\rm P},-1} + A_{1}^{I} \FBo  \text{J}^{I,2}_{{\rm P},-1} + \frac{4}{3} \beta_{1} A_{1}^{I} \text{J}^{I,1}_{{\rm P},-1} 
 + 2 \beta_{0} (\FBo )^2 \text{J}^{I,1}_{{\rm P},-1} 
\nonumber\\&         
         + \frac{5}{3} \beta_{0} A_{2}^{I} \text{J}^{I,1}_{{\rm P},-1} + \frac{7}{3} \beta_{0} A_{1}^{I} \text{J}^{I,2}_{{\rm P},-1}
 + \frac{11}{3} \beta_{0}^2 \FBo  \text{J}^{I,1}_{{\rm P},-1} + 2
         \beta_{0}^3 \text{J}^{I,1}_{{\rm P},-1} 
%\nonumber\\&         
 - 8  m^2 \zeta_{2} {A_{1}^{I}}^2 \FBo  \text{J}^{I,1}_{{\rm P},-1} 
%\nonumber\\&          
 - 12  m^2 \zeta_{2} \beta_{0} {A_{1}^{I}}^2 \text{J}^{I,1}_{{\rm P},-1}  
\nonumber\\& 
 - \frac{40}{3}  m^3 \zeta_{3} {A_{1}^{I}}^3 \text{J}^{I,1}_{{\rm P},-1}  \bigg]
%\nonumber\\&
       + 2 m~\bm{\mathcal{D}^{1}_{Q}}   \bigg[ A_{4}^{I} + 2 \beta_{1} \FBtw + (\FBtw)^2 + 2 \FBo  \FBth  + 3 \beta_{0} \FBth           - 4 m^2 \zeta_{2} {A_{2}^{I}}^2 - 16  m^2 \zeta_{2} A_{1}^{I} \FBo  \FBtw 
\nonumber\\&        
       + \beta_{2} \FBo  - 8  m^2 \zeta_{2}  A_{2}^{I} (\FBo )^2  - 8  m^2 \zeta_{2} A_{1}^{I} A_{3}^{I}  - 12  m^2 \zeta_{2} \beta_{1} A_{1}^{I} \FBo  - 16
          m^2 \zeta_{2} \beta_{0} A_{2}^{I} \FBo  - 20  m^2 \zeta_{2} \beta_{0} A_{1}^{I} \FBtw  - 16 m^3 \zeta_{3} A_{1}^{I} (\FBo )^3 
 \nonumber\\&          
          - 64  m^3 \zeta_{3} A_{1}^{I} A_{2}^{I} \FBo  - 32  m^3 \zeta_{3}
          {A_{1}^{I}}^2 \FBtw  - 24  m^3 \zeta_{3}   \beta_{1} {A_{1}^{I}}^2 - 72  m^3 \zeta_{3} \beta_{0} A_{1}^{I}   A_{2}^{I} - 56  m^3 \zeta_{3} \beta_{0} A_{1}^{I} (\FBo )^2 - 40  m^3 \zeta_{3} \beta_{0}^2 A_{1}^{I} \FBo 
 \nonumber\\&           
          - \frac{48}{5} m^4 \zeta_{2}^2 {A_{1}^{I}}^2 (\FBo )^2 - \frac{48}{5}  m^4 \zeta_{2}^2 {A_{1}^{I}}^2 A_{2}^{I} 
         - 32 m^4 \zeta_{2}^2 \beta_{0}^2 {A_{1}^{I}}^2
 - 32  m^4 \zeta_{2}^2 \beta_{0} {A_{1}^{I}}^2 \FBo   - 576  m^5 \zeta_{5} {A_{1}^{I}}^3 \FBo  - 608  m^5
         \zeta_{5} \beta_{0} {A_{1}^{I}}^3 
 \nonumber\\&          
         + 384  m^5 \zeta_{2} \zeta_{3} {A_{1}^{I}}^3 \FBo 
         + 384  m^5 \zeta_{2}
         \zeta_{3} \beta_{0} {A_{1}^{I}}^3 
         + 320  m^6 \zeta_{3}^2 {A_{1}^{I}}^4 - \frac{1856}{35}  m^6 \zeta_{2}^3 {A_{1}^{I}}^4
          + 2 \FBo  \FBtw \text{J}^{I,1}_{{\rm P},-1} + (\FBo )^2 \text{J}^{I,2}_{{\rm P},-1} 
         
          + A_{3}^{I} \text{J}^{I,1}_{{\rm P},-1} 
\nonumber\\&           
          + A_{2}^{I} \text{J}^{I,2}_{{\rm P},-1} + A_{1}^{I} \text{J}^{I,3}_{{\rm P},-1}   
%\nonumber\\&          
 + 3 \beta_{1} \FBo  \text{J}^{I,1}_{{\rm P},-1}  + 4 \beta_{0} \FBtw \text{J}^{I,1}_{{\rm P},-1} + 5 \beta_{0}  \FBo  \text{J}^{I,2}_{{\rm P},-1} + 5 \beta_{0} \beta_{1} \text{J}^{I,1}_{{\rm P},-1} + 6 \beta_{0}^2 \text{J}^{I,2}_{{\rm P},-1} 
- 8  m^2 \zeta_{2} A_{1}^{I}
          (\FBo )^2 \text{J}^{I,1}_{{\rm P},-1}  
\nonumber\\&           
          - 8  m^2 \zeta_{2} A_{1}^{I}   A_{2}^{I} \text{J}^{I,1}_{{\rm P},-1}
          - 4  m^2 \zeta_{2} {A_{1}^{I}}^2 \text{J}^{I,2}_{{\rm P},-1} - 28  m^2 \zeta_{2} \beta_{0} A_{1}^{I} \FBo  \text{J}^{I,1}_{{\rm P},-1} - 20  m^2 \zeta_{2} \beta_{0}^2 A_{1}^{I}    \text{J}^{I,1}_{{\rm P},-1}         
 - 32  m^3 \zeta_{3} {A_{1}^{I}}^2 \FBo  \text{J}^{I,1}_{{\rm P},-1}
\nonumber\\&         
  - 56 m^3 \zeta_{3} \beta_{0} {A_{1}^{I}}^2 \text{J}^{I,1}_{{\rm P},-1}    
  -   \frac{16}{5} m^4 \zeta_{2}^2 {A_{1}^{I}}^3 \text{J}^{I,1}_{{\rm P},-1}  \bigg] 
  
       - 2 m~\bm{\mathcal{D}^{0}_{Q}}   \bigg[ \FBfo - 4  m^2 \zeta_{2} A_{3}^{I} \FBo  - 4  m^2 \zeta_{2} A_{2}^{I} \FBtw - 4  m^2 \zeta_{2} A_{1}^{I} \FBth - 8  m^3 \zeta_{3} {A_{2}^{I}}^2
\nonumber\\&       
       - 8  m^3 \zeta_{3} A_{2}^{I} (\FBo )^2   - 16  m^3 \zeta_{3} A_{1}^{I} \FBo  \FBtw - 16  m^3 \zeta_{3} A_{1}^{I} A_{3}^{I}    - \frac{32}{5}  m^4 \zeta_{2}^2 A_{1}^{I} A_{2}^{I} \FBo  - \frac{16}{5}  m^4
         \zeta_{2}^2 {A_{1}^{I}}^2 \FBtw  - \frac{32}{5}
          m^4 \zeta_{2}^2 \beta_{1} {A_{1}^{I}}^2 
 \nonumber\\&        
          - \frac{96}{5}  m^4 \zeta_{2}^2 \beta_{0} A_{1}^{I} (\FBo)^2 
          - \frac{96}{5} m^4 \zeta_{2}^2 \beta_{0} A_{1}^{I} A_{2}^{I}  - \frac{64}{5}  m^4 \zeta_{2}^2 \beta_{0}^2 A_{1}^{I} \FBo  - 192  m^5 \zeta_{5} {A_{1}^{I}}^2 (\FBo )^2 - 288
          m^5 \zeta_{5} {A_{1}^{I}}^2 A_{2}^{I} 
          - 64  m^5 \zeta_{5} \beta_{0}^2 {A_{1}^{I}}^2
\nonumber\\&          
        - 320  m^5 \zeta_{5} \beta_{0} {A_{1}^{I}}^2 \FBo   + 96  m^5 \zeta_{2}
         \zeta_{3} {A_{1}^{I}}^2 (\FBo )^2 + 192  m^5 \zeta_{2} \zeta_{3} {A_{1}^{I}}^2 A_{2}^{I}  
         + 128  m^5 \zeta_{2} \zeta_{3} \beta_{0} {A_{1}^{I}}^2 \FBo  + 256 m^6 \zeta_{3}^2 {A_{1}^{I}}^3 \FBo 
 \nonumber\\&         
         - 1920 m^7 \zeta_{7} {A_{1}^{I}}^4  + 192  m^6 \zeta_{3}^2 \beta_{0} {A_{1}^{I}}^3 - \frac{1856}{35}
          m^6 \zeta_{2}^3 {A_{1}^{I}}^3 \FBo  - \frac{1536}{35}  m^6 \zeta_{2}^3 \beta_{0} {A_{1}^{I}}^3 
         
          + 1152  m^7 \zeta_{2} \zeta_{5} {A_{1}^{I}}^4 + \frac{128}{5}  m^7
         \zeta_{2}^2 \zeta_{3} {A_{1}^{I}}^4 
\nonumber\\&         
       + \FBth \text{J}^{I,1}_{{\rm P},-1} 
       + \FBtw \text{J}^{I,2}_{{\rm P},-1} + \FBo  \text{J}^{I,3}_{{\rm P},-1} + \beta_{2} \text{J}^{I,1}_{{\rm   P},-1}  + 2\beta_{1} \text{J}^{I,2}_{{\rm P},-1} + 3 \beta_{0} \text{J}^{I,3}_{{\rm P},-1}  - 4  m^2 \zeta_{2} A_{2}^{I} \FBo  \text{J}^{I,1}_{{\rm P},-1}  -
         4  m^2 \zeta_{2} A_{1}^{I} \FBtw \text{J}^{I,1}_{{\rm P},-1} 
\nonumber\\&         
         - 4  m^2 \zeta_{2} A_{1}^{I} \FBo  \text{J}^{I,2}_{{\rm P},-1} - 4 m^2 \zeta_{2} \beta_{1} A_{1}^{I} \text{J}^{I,1}_{{\rm P},-1} - 4  m^2 \zeta_{2} \beta_{0} A_{2}^{I} \text{J}^{I,1}_{{\rm P},-1} - 8
          m^2 \zeta_{2} \beta_{0} A_{1}^{I} \text{J}^{I,2}_{{\rm P},-1}  - 8  m^3 \zeta_{3} A_{1}^{I} (\FBo )^2
         \text{J}^{I,1}_{{\rm P},-1}  
\nonumber\\&          
         - 16  m^3 \zeta_{3} A_{1}^{I} A_{2}^{I} \text{J}^{I,1}_{{\rm P},-1} - 8
          m^3 \zeta_{3} {A_{1}^{I}}^2 \text{J}^{I,2}_{{\rm P},-1}   
          - 8  m^3 \zeta_{3} \beta_{1} A_{1}^{I} \FBo  
          - 8  m^3 \zeta_{3} \beta_{0} A_{2}^{I} \FBo  - 16  m^3 \zeta_{3} \beta_{0} A_{1}^{I} \FBtw  - \frac{32}{5}  m^4
         \zeta_{2}^2 A_{1}^{I} (\FBo )^3         
 \nonumber\\&          
         - 24  m^3 \zeta_{3} \beta_{0} A_{1}^{I} \FBo  \text{J}^{I,1}_{{\rm P},-1} - 16  m^3 \zeta_{3} \beta_{0}^2 A_{1}^{I} \text{J}^{I,1}_{{\rm P},-1}- \frac{48}{5}  m^4 \zeta_{2}^2 \beta_{0} {A_{1}^{I}}^2 \text{J}^{I,1}_{{\rm P},-1} - 96  m^5 \zeta_{5} {A_{1}^{I}}^3 \text{J}^{I,1}_{{\rm P},-1}+ 64  m^5 \zeta_{2}
         \zeta_{3} {A_{1}^{I}}^3 \text{J}^{I,1}_{{\rm P},-1}
 \nonumber\\&         
         - \frac{16}{5}  m^4 \zeta_{2}^2 {A_{1}^{I}}^2 \FBo  \text{J}^{I,1}_{{\rm P},-1} \bigg]

\end{align}
%\end{small}
%\end{widetezt}
\end{small}
\hspace*{-3mm}
where we define, $\bm{\gamma_{{\rm P},i}^{I}}=\left(f_{i}^I+B_{i}^{I}\delta_{\rm{P,J}}\right)$ and $\mathcal{D}^{k}_{Q}$ is a shorthand notation for $\mathcal{D}^{k}_{Q}(z)$, see Eq.~\eqref{calLdef}. 
The constants $\text{J}^{I,i}_{{\rm P},-1}$ are given by
\begin{align}
   \text{J}^{I,1}_{{\rm P},-1}  &= 
      2{\chi^{I}_{{\rm P},1}} \, ,
\nonumber\\
   \text{J}^{I,2}_{{\rm P},-1}  &=  
      {\chi^{I}_{{\rm P},2}} + 2 \left(\chi^{I}_{{\rm P},1}\right)^2 - 2  m^2 \zeta_{2} \Big(\FBo \Big)^2 - 8  m^3 \zeta_{3}  \FBo  A_{1}^{I} - \frac{4}{5}
       m^4 \zeta_{2}^2 {A_{1}^{I}}^2 \, ,
\nonumber\\
   \text{J}^{I,3}_{{\rm P},-1}  &= 
      \frac{2}{3} {\chi^{I}_{{\rm P},3}} + 2 {\chi^{I}_{{\rm P},1}} {\chi^{I}_{{\rm P},2}} + \frac{4}{3} \left(\chi^{I}_{{\rm P},1}\right)^3 - 4  m^2 \zeta_{2} \FBo  \FBtw - 4
       m^2 \zeta_{2} \Big(\FBo \Big)^2 {\chi^{I}_{{\rm P},1}} - 8  m^2 \zeta_{2} \beta_{0} \FBo  {\chi^{I}_{{\rm P},1}} - 8   m^3 \zeta_{3}
       \FBtw A_{1}^{I} 
\nonumber\\&       
       - 8  m^3 \zeta_{3} \FBo  A_{2}^{I} - 16  m^3 \zeta_{3} \FBo  A_{1}^{I} {\chi^{I}_{{\rm P},1}} -        \frac{8}{3}
       m^3 \zeta_{3} \Big(\FBo \Big)^3 - 16  m^3 \zeta_{3} \beta_{0} A_{1}^{I} {\chi^{I}_{{\rm P},1}} - 8  m^3 \zeta_{3}      \beta_{0}
      \Big(\FBo \Big)^2 - \frac{8}{5}  m^4 \zeta_{2}^2 A_{1}^{I} A_{2}^{I} 
\nonumber\\&      
      - \frac{8}{5}  m^4 \zeta_{2}^2 {A_{1}^{I}}^2       {\chi^{I}_{{\rm P},1}}      
      - \frac{8}{5}
       m^4 \zeta_{2}^2 \Big(\FBo \Big)^2 A_{1}^{I} - 8  m^4 \zeta_{2}^2 \beta_{0} \FBo  A_{1}^{I} - 96  m^5
      \zeta_{5} \FBo  {A_{1}^{I}}^2 - 64  m^5 \zeta_{5} \beta_{0} {A_{1}^{I}}^2 + 64  m^5 \zeta_{2} \zeta_{3} \FBo
      {A_{1}^{I}}^2 
\nonumber\\&      
      + 32  m^5 \zeta_{2} \zeta_{3} \beta_{0} {A_{1}^{I}}^2 + \frac{160}{3}  m^6 \zeta_{3}^2 {A_{1}^{I}}^3 -    \frac{928}{ 105}  m^6 \zeta_{2}^3 {A_{1}^{I}}^3\, .
\end{align}
Here, $\chi^{I}_{{\rm P},i}$ are defined in Eq.~\eqref{ChiP}. 

\section{Soft and Jet functions of quarks and gluons} 
In the following, we present the soft and jet functions at the fourth order in $a_s$.
We introduce the color coefficients $C_A = N_c$ and $C_F = {(N_c^2-1)}/{2 N_c}$, which are the quadratic Casimirs of the $SU(N_c)$ gauge group 
and $n_f$, which is the number of the active quark flavors. 
We also define the following  color coefficients, including quartic Casimirs,
\begin{align}
 \frac{d^{abc} d_{abc}}{N_F} &= \frac{(N_c^2-1)(N_c^2-4)}{16N_c^2}, \,
 \frac{d^{abcd}_F d^{abcd}_F}{N_F} = \frac{(N_c^2-1)(N_c^4-6N_c^2+18)}{96 N_c^3}, \,
 \frac{d^{abcd}_F d^{abcd}_A}{N_F} = \frac{(N_c^2-1)(N_c^2+6)}{48}, \,\nonumber\\
 \frac{d^{abcd}_A d^{abcd}_A}{N_A} &= \frac{N_c^2(N_c^2+36)}{24}, \,
 \frac{d^{abcd}_F d^{abcd}_A}{N_A} = \frac{N_c(N_c^2+6)}{48}, \,
  \frac{d^{abcd}_F d^{abcd}_F}{N_A} = \frac{(N_c^4-6N_c^2+18)}{96 N_c^2}, \,
 N_F = N_c,\, N_A = (N_c^2-1), \,
% N_4= \frac{(N_c^2-4)}{N_c} \,
\end{align}
In addition, $n_{fv}$ is defined as the charge weighted sum of quark flavors i.e. $\big(\sum_{q}e_{q}\big)/e_q$. In the following equations, the constant,  $b^{q}_{4,FA} =-998.02\pm  0.02$, \cite{Kniehl:2025ttz,Moch:2023tdj}.
%see 
%https://arxiv.org/pdf/2010.02980

% Maximally non-Abelian ($\mathcal{MNA}$):
% \begin{align}
% C_F  &\leftrightarrow C_A\, ,\nonumber\\
%  \frac{d^{abcd}_F d^{abcd}_A}{N_F}  &\leftrightarrow \frac{d^{abcd}_A d^{abcd}_A}{N_A}\, , \nonumber\\
%  n_f\frac{d^{abcd}_F d^{abcd}_F}{N_F}  &\leftrightarrow  n_f\frac{d^{abcd}_F d^{abcd}_A}{N_A} \, 
% \end{align}
% see https://arxiv.org/pdf/1805.09638
%and https://arxiv.org/pdf/1909.00697
%%%%%%%%%%%%%%%%%%%%%%%%%%%%%%%
The quark soft function for the DY process is given by
\begin{align}
\mathbb{J}_{{\rm S, SV}}^{q,4}&=
%%%%%%%%%%%%%%%%%%%%%%%%%%%%%
%            SV
%%%%%%%%%%%%%%%%%%%%%%%%%%%%%
   \bm{\mathcal{D}^{7}_{Q}} \Bigg[
        C_F^4   \bigg\{ \frac{32}{3} \bigg\}\Bigg] 
+
   \bm{\mathcal{D}^{6}_{Q}} \Bigg[
        C_F^3 n_f   \bigg\{ \frac{112}{9} \bigg\}
       - C_F^3 C_A   \bigg\{   \frac{616}{9} \bigg\}\Bigg] 
+
   \bm{\mathcal{D}^{5}_{Q}} \Bigg[
        C_F^2 n_f^2   \bigg\{ \frac{128}{27} \bigg\}
       - C_F^2 C_A n_f   \bigg\{   \frac{1408}{27} \bigg\}
\nonumber\\&
       + C_F^2 C_A^2   \bigg\{ \frac{3872}{27} \bigg\}
       - C_F^3 n_f   \bigg\{  \frac{160}{3} \bigg\}
       + C_F^3 C_A   \bigg\{ \frac{1072}{3} - 96 \zeta_2 \bigg\}
       - C_F^4   \bigg\{   864 \zeta_2 \bigg\}\Bigg] 
+
   \bm{\mathcal{D}^{4}_{Q}} \Bigg[
        C_F n_f^3   \bigg\{ \frac{16}{27} \bigg\}
       - C_F C_A n_f^2   \bigg\{ \frac{88}{9} \bigg\}
\nonumber\\&
       + C_F C_A^2 n_f   \bigg\{ \frac{484}{9} \bigg\}
       - C_F C_A^3   \bigg\{   \frac{2662}{27} \bigg\}
       - C_F^2 n_f^2   \bigg\{   \frac{800}{27} \bigg\}
       + C_F^2 C_A n_f   \bigg\{ \frac{10960}{27} - \frac{160}{3} \zeta_2 \bigg\}
       - C_F^2 C_A^2   \bigg\{   \frac{33560}{27} - \frac{880}{3} \zeta_2 \bigg\}
\nonumber\\&       
       + C_F^3 n_f   \bigg\{ \frac{2960}{27} - \frac{6560}{9} \zeta_2 \bigg\}
       - C_F^3 C_A   \bigg\{   \frac{16160}{27} - 560 \zeta_3 - \frac{36080}{9} \zeta_2 \bigg\}
       + C_F^4   \bigg\{ \frac{17920}{3} \zeta_3 \bigg\}\Bigg] 
+
   \bm{\mathcal{D}^{3}_{Q}} \Bigg[
       - C_F n_f^3   \bigg\{   \frac{320}{81} \bigg\}
\nonumber\\&
       + C_F C_A n_f^2   \bigg\{ \frac{2288}{27} - \frac{64}{9} \zeta_2 \bigg\}
       - C_F C_A^2 n_f   \bigg\{   \frac{14648}{27} - \frac{704}{9} \zeta_2 \bigg\}
       + C_F C_A^3   \bigg\{ \frac{87296}{81} - \frac{1936}{9} \zeta_2 \bigg\}
       - C_F^2 C_A n_f   \bigg\{   \frac{318664}{243} 
\nonumber\\&       
       - \frac{896}{9} \zeta_3 - \frac{61376}{27} \zeta_2 \bigg\}
       + C_F^2 n_f^2   \bigg\{ \frac{20144}{243} - \frac{5056}{27} \zeta_2 \bigg\}
       + C_F^2 C_A^2   \bigg\{ \frac{1101824}{243} - \frac{17600}{9} \zeta_3 - \frac{191536}{27} \zeta_2 + \frac{1728}{5} 
         \zeta_2^2 \bigg\}
\nonumber\\&
       - C_F^3 n_f   \bigg\{   \frac{29008}{81} - \frac{12736}{3} \zeta_3 - \frac{18400}{9} \zeta_2 \bigg\}
       + C_F^3 C_A   \bigg\{ \frac{38848}{81} - \frac{65824}{3} \zeta_3 - \frac{123280}{9} \zeta_2 + 3520 \zeta_2^2
          \bigg\}
       + C_F^4   \bigg\{ 2336 \zeta_2^2 \bigg\}\Bigg] 
\nonumber\\&
+
   \bm{\mathcal{D}^{2}_{Q}} \Bigg[
        C_F n_f^3   \bigg\{ \frac{800}{81} - \frac{128}{9} \zeta_2 \bigg\}
       - C_F C_A n_f^2   \bigg\{   \frac{7894}{27} -\frac{2432}{9} \zeta_2 \bigg\}
       + C_F C_A^2 n_f   \bigg\{ \frac{20554}{9} - 352 \zeta_3 - \frac{16000}{9} \zeta_2 + \frac{352}{5} \zeta_2^2 \bigg\}
\nonumber\\&
       - C_F C_A^3   \bigg\{   \frac{412880}{81} - 1936 \zeta_3 - \frac{34720}{9} \zeta_2 + \frac{1936}{5} 
         \zeta_2^2 \bigg\}
       - C_F^2 n_f^2   \bigg\{   \frac{46816}{243} - \frac{21088}{27} \zeta_3 - \frac{18800}{27} \zeta_2 \bigg\}
\nonumber\\&
       + C_F^2 C_A n_f   \bigg\{ \frac{609392}{243} - \frac{256064}{27} \zeta_3 - \frac{88096}{9} \zeta_2 + \frac{18976}{15} \zeta_2^2 \bigg\}
       - C_F^2 C_A^2  \bigg\{   \frac{1944580}{243} + 2304 \zeta_5 - \frac{835960}{27} \zeta_3 - \frac{833212}{27} \zeta_2 
\nonumber\\&       
       + 1376 \zeta_2 \zeta_3 + \frac{98032}{15} \zeta_2^2 \bigg\}
       + C_F^3 n_f   \bigg\{ \frac{6808}{9} - \frac{26816}{3} \zeta_3 - \frac{22640}{9} \zeta_2 + \frac{19408}{15} \zeta_2^2
          \bigg\}
       + C_F^3 C_A   \bigg\{ \frac{171520}{3} \zeta_3 + \frac{122816}{9} \zeta_2 
\nonumber\\&       
       - 28128 \zeta_2 \zeta_3 - 
         \frac{22616}{3} \zeta_2^2 \bigg\}
       + C_F^4   \bigg\{ 172032 \zeta_5 - 138240 \zeta_2 \zeta_3 \bigg\}\Bigg] 
+
   \bm{\mathcal{D}^{1}_{Q}} \Bigg[
        \frac{d^{abcd}_F d^{abcd}_A}{N_F}   \bigg\{ \frac{7040}{3} \zeta_5 + \frac{256}{3} \zeta_3 - 768 \zeta_3^2 - 256 \zeta_2 
\nonumber\\&               
        - 
         \frac{15872}{35} \zeta_2^3 \bigg\}
       - n_f \frac{d^{abcd}_F d^{abcd}_F}{N_F}   \bigg\{   \frac{2560}{3} \zeta_5 + \frac{512}{3} \zeta_3 - 512 \zeta_2 \bigg\}
       + C_F C_A n_f^2   \bigg\{ \frac{116090}{243} - \frac{4640}{9} \zeta_3 - \frac{28960}{27} \zeta_2  + \frac{1024}{15} 
         \zeta_2^2 \bigg\}
\nonumber\\&              
       - C_F n_f^3   \bigg\{   \frac{8000}{729} - \frac{256}{9} \zeta_3 - \frac{1280}{27} \zeta_2 \bigg\}
       - C_F C_A^2 n_f   \bigg\{   \frac{1142774}{243} + \frac{2720}{9} \zeta_5 - \frac{14000}{3} \zeta_3 - \frac{203888}{27} \zeta_2 
       - 64 \zeta_2 \zeta_3 
\nonumber\\&              
+ \frac{4928}{5} \zeta_2^2 \bigg\}
       + C_F C_A^3   \bigg\{ \frac{9040634}{729} + \frac{30800}{9} \zeta_5 - \frac{144656}{9} \zeta_3 - 32 \zeta_3^2
          - \frac{151664}{9} \zeta_2 + 1056 \zeta_2 \zeta_3 + \frac{10912}{3} \zeta_2^2        
          - \frac{40064}{105} \zeta_2^3 \bigg\}
\nonumber\\&          
       + C_F^2 n_f^2   \bigg\{ \frac{2728228}{6561} - \frac{522112}{243} \zeta_3 - \frac{26080}{27} \zeta_2 + 
         \frac{27712}{135} \zeta_2^2 \bigg\}
       - C_F^2 C_A n_f   \bigg\{   \frac{26126858}{6561} - \frac{832}{3} \zeta_5 - \frac{766912}{27} \zeta_3 
\nonumber\\&                 
       - 
         \frac{11362256}{729} \zeta_2 + \frac{43904}{9} \zeta_2 \zeta_3 + \frac{600032}{135} \zeta_2^2 \bigg\}
       + C_F^2 C_A^2   \bigg\{ \frac{44312020}{6561} - \frac{7744}{9} \zeta_5 - \frac{20876816}{243} \zeta_3 + 
         \frac{22688}{9} \zeta_3^2
\nonumber\\&                   
         - \frac{38738840}{729} \zeta_2 + \frac{372416}{9} \zeta_2 \zeta_3 + \frac{2799328}{135} \zeta_2^2 - \frac{3060224}{945} \zeta_2^3 \bigg\}
       - C_F^3 n_f   \bigg\{   \frac{140020}{243} - \frac{512128}{9} \zeta_5 - \frac{623008}{81} \zeta_3 
\nonumber\\&        
       - \frac{327008}{81} \zeta_2 + \frac{446656}{9} \zeta_2 \zeta_3 + \frac{66368}{45} \zeta_2^2 \bigg\}
       - C_F^3 C_A   \bigg\{   315392 \zeta_5 + \frac{1034240}{27} \zeta_3 - 35840 \zeta_3^2 +
         \frac{427328}{81} \zeta_2 - \frac{2312992}{9} \zeta_2 \zeta_3 
\nonumber\\&          
         - \frac{92192}{9} \zeta_2^2 + 992 
         \zeta_2^3 \bigg\}
       + C_F^4   \bigg\{ \frac{573440}{3} \zeta_3^2 - \frac{88672}{3} \zeta_2^3 \bigg\}\Bigg] 
+
   \bm{\mathcal{D}^{0}_{Q}} \Bigg[
       - \frac{d^{abcd}_F d^{abcd}_A}{N_F}   \bigg\{   384 - 4 {\color{red} b_{4,FA}} + 6968 \zeta_7 - \frac{3680}{9} \zeta_5 
\nonumber\\&        
       - \frac{15616}{9} \zeta_3 - \frac{6688}{3} \zeta_3^2 - \frac{4352}{3} \zeta_2 + 2048 \zeta_2 \zeta_5 - 
         3584 \zeta_2 \zeta_3 + \frac{448}{15} \zeta_2^2 - \frac{1472}{5} \zeta_2^2 \zeta_3 + \frac{55616}{315} 
         \zeta_2^3 \bigg\}
\nonumber\\& 
       + n_f \frac{d^{abcd}_F d^{abcd}_F}{N_F}   \bigg\{ \frac{3200}{9} \zeta_5 - \frac{1280}{9} \zeta_3 + \frac{640}{3} \zeta_3^2 - 512 
         \zeta_2 + \frac{128}{5} \zeta_2^2 + \frac{2560}{21} \zeta_2^3 \bigg\}
       + C_F n_f^3   \bigg\{ \frac{10432}{2187} - \frac{3680}{81} \zeta_3 - \frac{3200}{81} \zeta_2 
\nonumber\\&        
       + \frac{224}{45} \zeta_2^2 \bigg\}
       - C_F C_A n_f^2   \bigg\{   \frac{898033}{2916} - \frac{608}{3} \zeta_5 - \frac{87280}{81} \zeta_3 - \frac{293528}{243} \zeta_2 + \frac{608}{9} \zeta_2 \zeta_3 + \frac{3488}{15} \zeta_2^2 \bigg\}
       + C_F C_A^2 n_f   \bigg\{ \frac{11551831}{2916} 
\nonumber\\&        
       - \frac{7064}{27} \zeta_5 - \frac{829304}{81} \zeta_3 - \frac{4552}{9} \zeta_3^2 - \frac{2400868}{243} \zeta_2 + \frac{23440}{9} \zeta_2 \zeta_3 + \frac{108896}{45} 
         \zeta_2^2 - \frac{5872}{21} \zeta_2^3 \bigg\}
       - C_F C_A^3   \bigg\{   \frac{28290079}{2187}
\nonumber\\&        
       + \frac{1}{6} {\color{red} b_{4,FA}} - \frac{11071}{3} \zeta_7 + 
         \frac{149980}{27} \zeta_5 - \frac{288544}{9} \zeta_3 + \frac{14828}{9} \zeta_3^2 - \frac{5746982}{243} 
         \zeta_2 - \frac{2752}{3} \zeta_2 \zeta_5 + \frac{120968}{9} \zeta_2 \zeta_3 + \frac{301208}{45} \zeta_2^2
\nonumber\\&           
          - \frac{8456}{15} \zeta_2^2 \zeta_3 - \frac{1009888}{945} \zeta_2^3 \bigg\}
       - C_F^2 n_f^2   \bigg\{   \frac{1064003}{2187} - \frac{33056}{9} \zeta_5 - \frac{385552}{243} \zeta_3 - 
         \frac{547648}{729} \zeta_2 + \frac{27904}{9} \zeta_2 \zeta_3 + \frac{5504}{27} \zeta_2^2 \bigg\}
\nonumber\\& 
       + C_F^2 C_A n_f   \bigg\{ \frac{18977957}{4374} - \frac{368272}{9} \zeta_5 - \frac{5770604}{243} \zeta_3 + 
         \frac{20176}{9} \zeta_3^2 - \frac{7250738}{729} \zeta_2 + \frac{1105328}{27} \zeta_2 \zeta_3 + 
         \frac{1532816}{405} \zeta_2^2 
\nonumber\\&          
         - \frac{46912}{105} \zeta_2^3 \bigg\}
       - C_F^2 C_A^2  \bigg\{   \frac{3923648}{2187} - \frac{991232}{9} \zeta_5 - \frac{18072640}{243} \zeta_3
          + \frac{282832}{9} \zeta_3^2 - \frac{21822028}{729} \zeta_2 - 8448 \zeta_2 \zeta_5 
\nonumber\\&           
          + \frac{3869768}{
         27} \zeta_2 \zeta_3 + \frac{1216048}{81} \zeta_2^2 - \frac{151904}{15} \zeta_2^2 \zeta_3 - \frac{13024}{
         15} \zeta_2^3 \bigg\}
       - C_F^3 n_f   \bigg\{  \frac{21037}{54} + \frac{119680}{3} \zeta_5 + \frac{543760}{81} \zeta_3 - \frac{326048}{9}
          \zeta_3^2 
\nonumber\\&           
          + \frac{74852}{27} \zeta_2 - \frac{102592}{3} \zeta_2 \zeta_3 - \frac{73928}{135} \zeta_2^2
          + \frac{482752}{105} \zeta_2^3 \bigg\}
       + C_F^3 C_A   \bigg\{ 274432 \zeta_5 + \frac{621568}{81} \zeta_3 - \frac{1582592}{9} \zeta_3^2 - 
         73728 \zeta_2 \zeta_5 
\nonumber\\&          
         - \frac{1972480}{9} \zeta_2 \zeta_3 - \frac{80800}{27} \zeta_2^2 + 59120 
         \zeta_2^2 \zeta_3 + \frac{82016}{3} \zeta_2^3 \bigg\}
       + C_F^4   \bigg\{ 983040 \zeta_7 - 663552 \zeta_2 \zeta_5 + 37376 \zeta_2^2 \zeta_3 \bigg\}\Bigg] 
\end{align} % Done-SG
Similarly for the Higgs boson production through ggF, the corresponding gluon soft function is given by 
\begin{align}
\mathbb{J}_{{\rm S, SV}}^{g,4}&=
%%%%%%%%%%%%%%%%%%%%%%%%%%%%%
%            SV
%%%%%%%%%%%%%%%%%%%%%%%%%%%%%
   \bm{\mathcal{D}^{7}_{Q}} \Bigg[
        C_A^4   \bigg\{ \frac{32}{3} \bigg\}\Bigg] 
+
   \bm{\mathcal{D}^{6}_{Q}} \Bigg[
        C_A^3 n_f   \bigg\{ \frac{112}{9} \bigg\}
       - C_A^4   \bigg\{  \frac{616}{9} \bigg\}\Bigg] 
+
     \bm{\mathcal{D}^{5}_{Q}} \Bigg[
        C_A^2 n_f^2   \bigg\{ \frac{128}{27} \bigg\}
       - C_A^3 n_f   \bigg\{ \frac{2848}{27} \bigg\}
       + C_A^4   \bigg\{ \frac{13520}{27} 
\nonumber\\&       
       - 960 \zeta_2 \bigg\}\Bigg] 
+
   \bm{\mathcal{D}^{4}_{Q}} \Bigg[
        C_A n_f^3   \bigg\{ \frac{16}{27} \bigg\}
       - C_A^2 n_f^2   \bigg\{   \frac{1064}{27} \bigg\}
       + C_F C_A^2 n_f   \bigg\{ \frac{80}{3} \bigg\}       
       - C_A^4   \bigg\{   \frac{52382}{27} - \frac{19600}{3} \zeta_3 - \frac{38720}{9} \zeta_2 \bigg\}
\nonumber\\&
       + C_A^3 n_f   \bigg\{ \frac{1628}{3} - \frac{7040}{9} \zeta_2 \bigg\}\Bigg] 
+
   \bm{\mathcal{D}^{3}_{Q}} \Bigg[
        C_A^2 n_f^2   \bigg\{ \frac{38576}{243} - \frac{5248}{27} \zeta_2 \bigg\}   
       - C_A n_f^3   \bigg\{   \frac{320}{81} \bigg\}
       - C_A^3 n_f   \bigg\{   \frac{454360}{243} - \frac{36800}{9} \zeta_3 
\nonumber\\&        
       - \frac{118688}{27} \zeta_2 \bigg\}      
       + C_A^4   \bigg\{ \frac{1480256}{243} - \frac{215072}{9} \zeta_3 - \frac{567184}{27} \zeta_2 + \frac{31008}{5} 
         \zeta_2^2 \bigg\}
       - C_F C_A^2 n_f   \bigg\{   \frac{3080}{9} - 256 \zeta_3 \bigg\}
\nonumber\\&       
       + C_F C_A n_f^2   \bigg\{ \frac{80}{9} \bigg\}
       \Bigg] 
+
   \bm{\mathcal{D}^{2}_{Q}} \Bigg[
        C_A n_f^3   \bigg\{ \frac{800}{81} - \frac{128}{9} \zeta_2 \bigg\}
       - C_A^2 n_f^2   \bigg\{   \frac{94534}{243} -\frac{19360}{27} \zeta_3 - \frac{26096}{27} \zeta_2 \bigg\}
       + C_A^3 n_f   \bigg\{ \frac{1026326}{243} 
\nonumber\\&       
       - \frac{486464}{27} \zeta_3        
       - \frac{120832}{9} \zeta_2 + \frac{40592}{15}
          \zeta_2^2 \bigg\}
       - C_A^4   \bigg\{   \frac{3183220}{243} - 169728 \zeta_5 - \frac{2431912}{27} \zeta_3 - \frac{1305820}{27} \zeta_2 + 167744 \zeta_2 \zeta_3 
\nonumber\\&       
       + \frac{43384}{3} \zeta_2^2 \bigg\}       
       - C_F C_A n_f^2   \bigg\{   96 - 64 \zeta_3 \bigg\}
       + C_F C_A^2 n_f   \bigg\{ \frac{11956}{9} - \frac{2272}{3} \zeta_3 - 656 \zeta_2 - \frac{384}{5} \zeta_2^2 \bigg\}
       - C_F^2 C_A n_f   \bigg\{  4 \bigg\}\Bigg] 
\nonumber\\&
+
   \bm{\mathcal{D}^{1}_{Q}} \Bigg[
        \frac{d^{abcd}_A d^{abcd}_A}{N_A}   \bigg\{ \frac{7040}{3} \zeta_5 + \frac{256}{3} \zeta_3 - 768 \zeta_3^2 - 256 \zeta_2 - 
         \frac{15872}{35} \zeta_2^3 \bigg\}
       - n_f \frac{d^{abcd}_F d^{abcd}_A}{N_A}   \bigg\{   \frac{2560}{3} \zeta_5 + \frac{512}{3} \zeta_3 - 512 \zeta_2 \bigg\}
\nonumber\\&
       - C_A n_f^3   \bigg\{   \frac{8000}{729} - \frac{256}{9} \zeta_3 - \frac{1280}{27} \zeta_2 \bigg\}
       + C_A^2 n_f^2   \bigg\{ \frac{3594334}{6561} - \frac{580000}{243} \zeta_3 - \frac{52160}{27} \zeta_2 + 
         \frac{36928}{135} \zeta_2^2 \bigg\}
\nonumber\\&
       - C_A^3 n_f   \bigg\{   \frac{40745306}{6561} - \frac{512992}{9} \zeta_5 - \frac{329296}{9} \zeta_3 - 
         \frac{16745264}{729} \zeta_2 + \frac{153856}{3} \zeta_2 \zeta_3 + \frac{939488}{135} \zeta_2^2 \bigg\}
       + C_A^4   \bigg\{ \frac{125677726}{6561}  
\nonumber\\&       
      - \frac{2815472}{9} \zeta_5
       - \frac{34090688}{243} \zeta_3 + 
         \frac{2065280}{9} \zeta_3^2 - \frac{54869576}{729} \zeta_2 + \frac{898304}{3} \zeta_2 \zeta_3 + 
         \frac{4673248}{135} \zeta_2^2 - \frac{6457984}{189} \zeta_2^3 \bigg\}
\nonumber\\&   
       + C_F C_A n_f^2   \bigg\{ \frac{28004}{81}   
       - \frac{832}{3} \zeta_3 - \frac{320}{3} \zeta_2 \bigg\}
       - C_F C_A^2 n_f   \bigg\{   \frac{772258}{243} - \frac{4672}{9} \zeta_5 - \frac{306112}{81} \zeta_3 - \frac{37840}{9} \zeta_2 + \frac{9472}{3} \zeta_2 \zeta_3 
\nonumber\\&       
       - \frac{2432}{45} \zeta_2^2 \bigg\}
       + C_F^2 C_A n_f   \bigg\{ \frac{1144}{9} - 640 \zeta_5 + \frac{1184}{3} \zeta_3 \bigg\}\Bigg] 
+
   \bm{\mathcal{D}^{0}_{Q}} \Bigg[
       - \frac{d^{abcd}_A d^{abcd}_A}{N_A}   \bigg\{  384 - 4 {\color{red} b_{4,FA}} + 6968 \zeta_7 - \frac{3680}{9} \zeta_5 
\nonumber\\&       
       - \frac{15616}{9} \zeta_3 
         - \frac{6688}{3} \zeta_3^2 - \frac{4352}{3} \zeta_2 + 2048 \zeta_2 \zeta_5 - 
         3584 \zeta_2 \zeta_3 + \frac{448}{15} \zeta_2^2 - \frac{1472}{5} \zeta_2^2 \zeta_3 + \frac{55616}{315} 
         \zeta_2^3 \bigg\}
\nonumber\\&         
       + n_f \frac{d^{abcd}_F d^{abcd}_A}{N_A}   \bigg\{ \frac{3200}{9} \zeta_5 - \frac{1280}{9} \zeta_3        
       + \frac{640}{3} \zeta_3^2 - 512 
         \zeta_2 + \frac{128}{5} \zeta_2^2 + \frac{2560}{21} \zeta_2^3 \bigg\}
       + C_A n_f^3   \bigg\{ \frac{10432}{2187} - \frac{3680}{81} \zeta_3 - \frac{3200}{81} \zeta_2 
\nonumber\\&       
       + \frac{224}{45} 
         \zeta_2^2 \bigg\}
       - C_A^2 n_f^2   \bigg\{   \frac{2987987}{8748} - \frac{34592}{9} \zeta_5 - \frac{550480}{243} \zeta_3 - 
         \frac{1148296}{729} \zeta_2 + 2912 \zeta_2 \zeta_3 + \frac{61792}{135} \zeta_2^2 \bigg\}
       + C_A^3 n_f   \bigg\{ \frac{38951173}{8748} 
\nonumber\\&       
       - \frac{2194328}{27} \zeta_5 - \frac{7714856}{243} \zeta_3 + 
         \frac{297752}{9} \zeta_3^2 - \frac{12515012}{729} \zeta_2 + \frac{2015360}{27} \zeta_2 \zeta_3 + 
         \frac{2692544}{405} \zeta_2^2 - \frac{597424}{105} \zeta_2^3 \bigg\}
\nonumber\\&
       - C_A^4   \bigg\{   \frac{1193101}{81} + \frac{1}{6} {\color{red} b_{4,FA}} - \frac{2960191}{3} \zeta_7 - 
         \frac{10233380}{27} \zeta_5 - \frac{27728032}{243} \zeta_3 + \frac{1880252}{9} \zeta_3^2 -\frac{39062974}{729} \zeta_2 
\nonumber\\&         
         + \frac{2183744}{3} \zeta_2 \zeta_5 + \frac{10150112}{27} \zeta_2 \zeta_3 + 
         \frac{10003112}{405} \zeta_2^2 - \frac{321560}{3} \zeta_2^2 \zeta_3 - \frac{5533088}{189} \zeta_2^3 \bigg\}
       - C_F C_A n_f^2   \bigg\{   \frac{110059}{243} - 32 \zeta_5 
\nonumber\\&       
       - \frac{10768}{27} \zeta_3 - 384 \zeta_2
          + 256 \zeta_2 \zeta_3 - \frac{64}{3} \zeta_2^2 \bigg\}
       + C_F C_A^2 n_f   \bigg\{ \frac{1870013}{486} - \frac{2608}{3} \zeta_5 - \frac{215116}{27} \zeta_3 + 4720 
         \zeta_3^2 - \frac{147074}{27} \zeta_2 
\nonumber\\&         
         + \frac{27872}{9} \zeta_2 \zeta_3 + \frac{816}{5} \zeta_2^2 + 
         \frac{1920}{7} \zeta_2^3 \bigg\}
       - C_F^2 C_A n_f   \bigg\{ \frac{21037}{54} - \frac{3200}{3} \zeta_5 + \frac{8848}{9} \zeta_3 - 160 
         \zeta_3^2 - 16 \zeta_2 + \frac{296}{5} \zeta_2^2 - \frac{640}{7} \zeta_2^3 \bigg\}\Bigg] 
\end{align} % Done-SG
The quark jet function is given by
\begin{align}
\mathbb{J}_{{\rm J, SV}}^{q,4}&=
%%%%%%%%%%%%%%%%%%%%%%%%%%%%%
%           SV
%%%%%%%%%%%%%%%%%%%%%%%%%%%%%
   \bm{\mathcal{D}^{7}_{Q}} \Bigg[
        C_F^4   \bigg\{ \frac{16}{3} \bigg\}\Bigg] 
+
   \bm{\mathcal{D}^{6}_{Q}} \Bigg[
        C_F^3 n_f   \bigg\{ \frac{56}{9} \bigg\}
       - C_F^3 C_A   \bigg\{ \frac{308}{9} \bigg\}
       - C_F^4   \bigg\{ 28 \bigg\}\Bigg] 
+
   \bm{\mathcal{D}^{5}_{Q}} \Bigg[
        C_F^2 n_f^2   \bigg\{ \frac{64}{27} \bigg\}
       - C_F^2 C_A n_f   \bigg\{   \frac{704}{27} \bigg\}
\nonumber\\&       
       + C_F^2 C_A^2   \bigg\{ \frac{1936}{27} \bigg\}
       - C_F^3 n_f   \bigg\{ \frac{164}{3} \bigg\}
       + C_F^3 C_A   \bigg\{ \frac{998}{3} - 48 \zeta_2 \bigg\}
       + C_F^4   \bigg\{ 110 - 144 \zeta_2 \bigg\}\Bigg] 
+
   \bm{\mathcal{D}^{4}_{Q}} \Bigg[
        C_F n_f^3   \bigg\{ \frac{8}{27} \bigg\}
\nonumber\\&
       - C_F C_A n_f^2   \bigg\{  \frac{44}{9} \bigg\}
       + C_F C_A^2 n_f   \bigg\{ \frac{242}{9} \bigg\}
       - C_F C_A^3   \bigg\{  \frac{1331}{27} \bigg\}
       - C_F^2 n_f^2   \bigg\{  \frac{640}{27} \bigg\}
       + C_F^2 C_A n_f   \bigg\{ \frac{8120}{27} - \frac{80}{3} \zeta_2 \bigg\}
\nonumber\\&       
       - C_F^2 C_A^2   \bigg\{  \frac{24040}{27} - \frac{440}{3} \zeta_2 \bigg\}
       + C_F^3 n_f   \bigg\{ \frac{6550}{27} - \frac{1120}{9} \zeta_2 \bigg\}
       - C_F^3 C_A   \bigg\{  \frac{38395}{27} - 400 \zeta_3 - \frac{7240}{9} \zeta_2 \bigg\}
       - C_F^4   \bigg\{   270 
\nonumber\\&       
       - \frac{400}{3} \zeta_3 - 660 \zeta_2 \bigg\}\Bigg] 
+
   \bm{\mathcal{D}^{3}_{Q}} \Bigg[
        C_F C_A n_f^2   \bigg\{ \frac{1540}{27} - \frac{32}{9} \zeta_2 \bigg\}
       - C_F n_f^3   \bigg\{  \frac{232}{81} \bigg\}
       - C_F C_A^2 n_f   \bigg\{  \frac{9502}{27} - \frac{352}{9} \zeta_2 \bigg\}
\nonumber\\&        
       + C_F C_A^3   \bigg\{ \frac{55627}{81} - \frac{968}{9} \zeta_2 \bigg\}
       + C_F^2 n_f^2   \bigg\{ \frac{26542}{243} - \frac{320}{9} \zeta_2 \bigg\}
       - C_F^2 C_A n_f   \bigg\{  \frac{364478}{243} - \frac{1216}{9} \zeta_3 - \frac{4816}{9} \zeta_2 \bigg\}
\nonumber\\&        
       + C_F^2 C_A^2   \bigg\{ \frac{2293955}{486} - \frac{13024}{9} \zeta_3 - \frac{17960}{9} \zeta_2 + \frac{864}{5} 
         \zeta_2^2 \bigg\}
       - C_F^3 n_f   \bigg\{  \frac{64220}{81} - \frac{1856}{9} \zeta_3 - \frac{7408}{9} \zeta_2 \bigg\}
       + C_F^3 C_A   \bigg\{ \frac{310304}{81} 
\nonumber\\&        
       - \frac{13808}{9} \zeta_3 - \frac{47200}{9} \zeta_2 + \frac{2744}{5} 
         \zeta_2^2 \bigg\}
       + C_F^4   \bigg\{ \frac{1013}{2} - 592 \zeta_3 - 1920 \zeta_2 + \frac{3136}{5} \zeta_2^2 \bigg\}\Bigg] 
+
   \bm{\mathcal{D}^{2}_{Q}} \Bigg[
        C_F n_f^3   \bigg\{ \frac{940}{81} - \frac{32}{9} \zeta_2 \bigg\}
\nonumber\\& 
       - C_F C_A n_f^2   \bigg\{  \frac{7403}{27} - 16 \zeta_3 - \frac{688}{9} \zeta_2 \bigg\}
       + C_F C_A^2 n_f   \bigg\{ \frac{17189}{9} - 352 \zeta_3 - \frac{5096}{9} \zeta_2 + \frac{176}{5} \zeta_2^2 \bigg\}
       - C_F C_A^3   \bigg\{   \frac{649589}{162}
\nonumber\\&       
       - 1452 \zeta_3 - \frac{4012}{3} \zeta_2 + \frac{968}{5} \zeta_2^2
          \bigg\}
       - C_F^2 n_f^2   \bigg\{ \frac{77552}{243} - \frac{896}{27} \zeta_3 - \frac{6152}{27} \zeta_2 \bigg\}
       + C_F^2 C_A n_f   \bigg\{ \frac{1032742}{243} - \frac{34216}{27} \zeta_3 
\nonumber\\&             
       - \frac{85052}{27} \zeta_2 + \frac{716}{3} 
         \zeta_2^2 \bigg\}
       - C_F^2 C_A^2   \bigg\{  \frac{6410317}{486} + 1392 \zeta_5 - \frac{234668}{27} \zeta_3 - 10184 
         \zeta_2 + 832 \zeta_2 \zeta_3 + \frac{19114}{15} \zeta_2^2 \bigg\}
\nonumber\\&                
       + C_F^3 n_f   \bigg\{ 1823 - 848 \zeta_3 - \frac{22390}{9} \zeta_2 + \frac{1536}{5} \zeta_2^2 \bigg\}
       - C_F^3 C_A   \bigg\{   \frac{59473}{9} + 720 \zeta_5 - \frac{16672}{3} \zeta_3 - \frac{135163}{9} \zeta_2
          + 4128 \zeta_2 \zeta_3
\nonumber\\&                 
          + \frac{14454}{5} \zeta_2^2 \bigg\}  
       - C_F^4   \bigg\{  726 - 4128 \zeta_5 - 568 \zeta_3 - 3141 \zeta_2 + 672 \zeta_2 
         \zeta_3 + \frac{13824}{5} \zeta_2^2 \bigg\}\Bigg] 
+
   \bm{\mathcal{D}^{1}_{Q}} \Bigg[
        \frac{d^{abcd}_F d^{abcd}_A}{N_F}   \bigg\{ \frac{3520}{3} \zeta_5 
\nonumber\\&         
        + \frac{128}{3} \zeta_3 - 384 \zeta_3^2 - 128 \zeta_2 - 
         \frac{7936}{35} \zeta_2^3 \bigg\}
       - n_f \frac{d^{abcd}_F d^{abcd}_F}{N_F}   \bigg\{   \frac{1280}{3} \zeta_5 + \frac{256}{3} \zeta_3 - 256 \zeta_2 \bigg\}
       - C_F n_f^3   \bigg\{ \frac{17716}{729} -\frac{464}{27} \zeta_2 \bigg\}
\nonumber\\&       
       + C_F C_A n_f^2   \bigg\{ \frac{315755}{486} - \frac{688}{9} \zeta_3 - \frac{9848}{27} \zeta_2 + \frac{64}{5} 
         \zeta_2^2 \bigg\}
       - C_F C_A^2 n_f   \bigg\{  \frac{2452247}{486} + \frac{2080}{9} \zeta_5 - \frac{5752}{3} \zeta_3 - \frac{68836}{27} \zeta_2 
\nonumber\\&       
       - 32 \zeta_2 \zeta_3 + \frac{3944}{15} \zeta_2^2 \bigg\}
       + C_F C_A^3   \bigg\{ \frac{16865531}{1458} + \frac{19360}{9} \zeta_5 - \frac{78296}{9} \zeta_3 - 16 
         \zeta_3^2 - \frac{156238}{27} \zeta_2 + 528 \zeta_2 \zeta_3 + \frac{17996}{15} \zeta_2^2 
\nonumber\\&         
         - \frac{20032}{105} \zeta_2^3 \bigg\}
       + C_F^2 n_f^2   \bigg\{ \frac{8252995}{13122} - \frac{53888}{243} \zeta_3 - \frac{47368}{81} \zeta_2 + \frac{3008}{45} \zeta_2^2 \bigg\}
       - C_F^2 C_A n_f   \bigg\{   \frac{49483216}{6561} + \frac{176}{3} \zeta_5 
\nonumber\\&       
       - \frac{317228}{81} \zeta_3 - 
         \frac{6081322}{729} \zeta_2 + \frac{8896}{9} \zeta_2 \zeta_3 + \frac{3850}{3} \zeta_2^2 \bigg\}
       + C_F^2 C_A^2   \bigg\{ \frac{1074264517}{52488} + \frac{22888}{9} \zeta_5 - \frac{4451206}{243} \zeta_3 + 
         \frac{20512}{9} \zeta_3^2 
\nonumber\\&         
         - \frac{19328317}{729} \zeta_2 + \frac{74464}{9} \zeta_2 \zeta_3 + \frac{16091}{3} 
         \zeta_2^2 - \frac{465112}{945} \zeta_2^3 \bigg\}
       - C_F^3 n_f   \bigg\{   \frac{1291945}{486} - \frac{3232}{3} \zeta_5 - \frac{144092}{81} \zeta_3 - \frac{364240}{81} \zeta_2 
\nonumber\\&       
       + \frac{6784}{9} \zeta_2 \zeta_3 + \frac{214528}{135} \zeta_2^2 \bigg\}
       + C_F^3 C_A   \bigg\{ \frac{1210163}{162} - \frac{55168}{9} \zeta_5 - \frac{29992}{3} \zeta_3 + \frac{3616}{3} 
         \zeta_3^2 - \frac{1874191}{81} \zeta_2 + \frac{77296}{9} \zeta_2 \zeta_3 
\nonumber\\&         
         + \frac{1474741}{135} 
         \zeta_2^2 - \frac{11096}{45} \zeta_2^3 \bigg\}
       + C_F^4   \bigg\{ \frac{7755}{8} - 5248 \zeta_5 + 766 \zeta_3 - \frac{640}{3} \zeta_3^2 - \frac{9730}{3} 
         \zeta_2 + 1904 \zeta_2 \zeta_3 + \frac{25592}{5} \zeta_2^2 
\nonumber\\&         
         - \frac{105568}{63} \zeta_2^3 \bigg\}\Bigg] 
+
   \bm{\mathcal{D}^{0}_{Q}} \Bigg[
       - \frac{d^{abcd}_F d^{abcd}_A}{N_F}   \bigg\{   192 - {\color{red} b_{4,FA}} + 3484 \zeta_7 - \frac{1840}{9} \zeta_5 - 
         \frac{7808}{9} \zeta_3 - \frac{3344}{3} \zeta_3^2 - \frac{2176}{3} \zeta_2 + \frac{224}{15} \zeta_2^2
\nonumber\\&         
         + 1024 \zeta_2 \zeta_5 - 
         1792 \zeta_2 \zeta_3  - \frac{736}{5} \zeta_2^2 \zeta_3 + \frac{27808}{315} 
         \zeta_2^3 \bigg\}
       + n_f \frac{d^{abcd}_F d^{abcd}_F}{N_F}   \bigg\{ 192 + \frac{11680}{9} \zeta_5 + \frac{2336}{9} \zeta_3 - \frac{448}{3} \zeta_3^2
          - \frac{2656}{3} \zeta_2 
\nonumber\\&          
          - 64 \zeta_2 \zeta_3 + \frac{896}{15} \zeta_2^2 + \frac{4864}{315} \zeta_2^3
          \bigg\}
       + C_F n_f^3   \bigg\{ \frac{50558}{2187} + \frac{80}{81} \zeta_3 - \frac{1880}{81} \zeta_2 + \frac{16}{9} \zeta_2^2 \bigg\}
       - C_F C_A n_f^2   \bigg\{   \frac{3761509}{5832} - \frac{1192}{9} \zeta_5 
\nonumber\\&       
       - \frac{6092}{81} \zeta_3 - 
         \frac{131878}{243} \zeta_2 + \frac{400}{9} \zeta_2 \zeta_3 + \frac{616}{9} \zeta_2^2 \bigg\}
       + C_F C_A^2 n_f   \bigg\{ \frac{31645735}{5832} - \frac{6562}{27} \zeta_5 - \frac{239062}{81} \zeta_3 - \frac{2612}{9} \zeta_3^2 + \frac{3314}{5} \zeta_2^2
\nonumber\\&       
       - \frac{940424}{243} \zeta_2 + \frac{8684}{9} \zeta_2 \zeta_3  - \frac{86176}{945} \zeta_2^3 \bigg\}
       - C_F C_A^3   \bigg\{   \frac{59835979}{4374} + \frac{1}{24} {\color{red} b_{4,FA}} - \frac{28291}{6} \zeta_7 + 
         \frac{128510}{27} \zeta_5 - \frac{387083}{27} \zeta_3 
\nonumber\\&         
         + \frac{17182}{9} \zeta_3^2 - \frac{2052595}{243} 
         \zeta_2 + \frac{136}{3} \zeta_2 \zeta_5 + \frac{39988}{9} \zeta_2 \zeta_3 + \frac{77341}{45} \zeta_2^2 - 
         \frac{836}{3} \zeta_2^2 \zeta_3 - \frac{55864}{189} \zeta_2^3 \bigg\}
       - C_F^2 n_f^2   \bigg\{   \frac{218969}{324} + \frac{64}{9} \zeta_5 
\nonumber\\&       
       - \frac{94832}{243} \zeta_3 - \frac{473390}{729} \zeta_2 + \frac{1280}{27} \zeta_2 \zeta_3 + \frac{21512}{135} \zeta_2^2 \bigg\}
       + C_F^2 C_A n_f   \bigg\{ \frac{21042643}{2916} + \frac{2912}{9} \zeta_5 - \frac{1603534}{243} \zeta_3 + 
         \frac{8600}{9} \zeta_3^2 
\nonumber\\&         
         - \frac{6535279}{729} \zeta_2 + \frac{70040}{27} \zeta_2 \zeta_3 + \frac{193721}{81}
          \zeta_2^2 - \frac{7088}{63} \zeta_2^3 \bigg\}
       - C_F^2 C_A^2   \bigg\{   \frac{174656377}{11664} + 8610 \zeta_7 + \frac{21118}{9} \zeta_5 - 
         \frac{5372636}{243} \zeta_3 
\nonumber\\&         
         + \frac{13958}{9} \zeta_3^2 - \frac{41540449}{1458} \zeta_2 - 4424 
         \zeta_2 \zeta_5 + \frac{164912}{9} \zeta_2 \zeta_3 + \frac{1314115}{162} \zeta_2^2 - \frac{16936}{15} 
         \zeta_2^2 \zeta_3 - \frac{168584}{315} \zeta_2^3 \bigg\}
\nonumber\\&
       + C_F^3 n_f   \bigg\{ \frac{138311}{81} - \frac{4784}{3} \zeta_5 - \frac{74008}{81} \zeta_3 - \frac{1456}{9} 
         \zeta_3^2 - \frac{208411}{54} \zeta_2 + \frac{3344}{3} \zeta_2 \zeta_3 + \frac{286976}{135} \zeta_2^2
          - \frac{17968}{105} \zeta_2^3 \bigg\}
\nonumber\\&
       - C_F^3 C_A   \bigg\{   \frac{456575}{108} - 10920 \zeta_7 - \frac{20896}{3} \zeta_5 - \frac{493891}{81} 
         \zeta_3 + \frac{33068}{9} \zeta_3^2 - \frac{1529203}{108} \zeta_2 + 2016 \zeta_2 \zeta_5 + 
         \frac{78548}{9} \zeta_2 \zeta_3 
\nonumber\\&         
         + \frac{733811}{54} \zeta_2^2 - 3272 \zeta_2^2 \zeta_3 - \frac{53974}{35} \zeta_2^3 \bigg\}
       - C_F^4   \bigg\{   \frac{37577}{48} + 2040 \zeta_7 - 6672 \zeta_5 + 3192 \zeta_3 - 1504 
         \zeta_3^2 - \frac{8411}{4} \zeta_2 + 164 \zeta_2 \zeta_3
\nonumber\\&         
         + 5472 \zeta_2 \zeta_5  + \frac{17472}{5}
          \zeta_2^2 + \frac{1984}{5} \zeta_2^2 \zeta_3 - \frac{311648}{105} \zeta_2^3 \bigg\}\Bigg] 
\end{align}    % Done-SG
and the gluon jet function is given by
\begin{align}
\mathbb{J}_{{\rm J, SV}}^{g,4}&= 
%%%%%%%%%%%%%%%%%%%%%%%%%%%%%%%
%              SV
%%%%%%%%%%%%%%%%%%%%%%%%%%%%%%%
   \bm{\mathcal{D}^{7}_{Q}} \Bigg[
        C_A^4   \bigg\{ \frac{16}{3} \bigg\}\Bigg] 
+
   \bm{\mathcal{D}^{6}_{Q}} \Bigg[
        C_A^3 n_f   \bigg\{ \frac{112}{9} \bigg\}
       - C_A^4   \bigg\{   \frac{616}{9} \bigg\}\Bigg] 
+
   \bm{\mathcal{D}^{5}_{Q}} \Bigg[
        C_A^2 n_f^2   \bigg\{ \frac{304}{27} \bigg\}
       - C_A^3 n_f   \bigg\{  \frac{4304}{27} \bigg\}
       + C_A^4   \bigg\{ \frac{15628}{27} 
\nonumber\\&       
       - 192 \zeta_2 \bigg\}\Bigg] 
+
   \bm{\mathcal{D}^{4}_{Q}} \Bigg[
        C_A n_f^3   \bigg\{ \frac{128}{27} \bigg\}
       - C_A^2 n_f^2   \bigg\{   \frac{10336}{81} \bigg\}
       + C_A^3 n_f   \bigg\{ \frac{90968}{81} - \frac{2920}{9} \zeta_2 \bigg\}
       - C_A^4   \bigg\{   \frac{250288}{81} - \frac{1600}{3} \zeta_3 
\nonumber\\&       
       - \frac{16060}{9} \zeta_2 \bigg\}
       + C_F C_A^2 n_f   \bigg\{ \frac{100}{3} \bigg\}\Bigg] 
+
   \bm{\mathcal{D}^{3}_{Q}} \Bigg[
        n_f^4   \bigg\{ \frac{64}{81} \bigg\}
       - C_A n_f^3   \bigg\{   \frac{3328}{81} \bigg\}
       + C_A^2 n_f^2   \bigg\{ \frac{171832}{243} - \frac{608}{3} \zeta_2 \bigg\}
\nonumber\\&
       + C_A^3 n_f   \bigg\{  - \frac{1191980}{243} + \frac{4736}{9} \zeta_3 + \frac{25904}{9} \zeta_2 \bigg\}
       + C_A^4   \bigg\{ \frac{2843764}{243} - \frac{35552}{9} \zeta_3 - \frac{94304}{9} \zeta_2 + \frac{6744}{5} \zeta_2^2
          \bigg\}
\nonumber\\&
       + C_F C_A n_f^2   \bigg\{ \frac{344}{9} \bigg\}
       - C_F C_A^2 n_f   \bigg\{   \frac{3872}{9} - 192 \zeta_3 \bigg\}\Bigg] 
+
   \bm{\mathcal{D}^{2}_{Q}} \Bigg[
       - n_f^4   \bigg\{   \frac{320}{81} \bigg\}
       + C_A n_f^3   \bigg\{ \frac{12416}{81} - \frac{1456}{27} \zeta_2 \bigg\}
\nonumber\\&       
       - C_A^2 n_f^2   \bigg\{   \frac{545228}{243} - \frac{4160}{27} \zeta_3 - \frac{39224}{27} \zeta_2 \bigg\}
       + C_A^3 n_f   \bigg\{ \frac{3366592}{243} - \frac{103504}{27} \zeta_3 - \frac{344732}{27} \zeta_2 + \frac{22384}{15} \zeta_2^2 \bigg\}
\nonumber\\&       
       - C_A^4   \bigg\{ \frac{7346150}{243} - 2016 \zeta_5 - \frac{494912}{27} \zeta_3 - \frac{948082}{27} 
         \zeta_2 + 5632 \zeta_2 \zeta_3 + \frac{119944}{15} \zeta_2^2 \bigg\}
       - C_F C_A n_f^2   \bigg\{   \frac{3472}{9} - \frac{512}{3} \zeta_3 \bigg\}
\nonumber\\&        
       + C_F n_f^3   \bigg\{ \frac{104}{9} \bigg\}
       + C_F C_A^2 n_f   \bigg\{ \frac{20392}{9} - \frac{3424}{3} \zeta_3 - 376 \zeta_2 - \frac{192}{5} \zeta_2^2 \bigg\}
       - C_F^2 C_A n_f   \bigg\{  8 \bigg\}\Bigg] 
+
   \bm{\mathcal{D}^{1}_{Q}} \Bigg[
        \frac{d^{abcd}_A d^{abcd}_A}{N_A}   \bigg\{ \frac{3520}{3} \zeta_5 
\nonumber\\&         
        + \frac{128}{3} \zeta_3 - 384 \zeta_3^2 - 128 \zeta_2 - 
         \frac{7936}{35} \zeta_2^3 \bigg\}
       - n_f \frac{d^{abcd}_F d^{abcd}_A}{N_A}   \bigg\{   \frac{1280}{3} \zeta_5 + \frac{256}{3} \zeta_3 - 256 \zeta_2 \bigg\}
       + n_f^4   \bigg\{ \frac{1600}{243} - \frac{128}{27} \zeta_2 \bigg\}
\nonumber\\& 
       - C_A n_f^3   \bigg\{   \frac{612662}{2187} - \frac{800}{81} \zeta_3 - \frac{20048}{81} \zeta_2 \bigg\}
       + C_A^2 n_f^2   \bigg\{ \frac{26326343}{6561} - \frac{196976}{243} \zeta_3 - \frac{1033568}{243} \zeta_2 + 
         \frac{67448}{135} \zeta_2^2 \bigg\}
\nonumber\\& 
       - C_A^3 n_f   \bigg\{   \frac{156619981}{6561} - \frac{4592}{3} \zeta_5 - \frac{103672}{9} \zeta_3 - 
         \frac{21521320}{729} \zeta_2 + \frac{24416}{9} \zeta_2 \zeta_3 + \frac{956312}{135} \zeta_2^2 \bigg\}
       + C_A^4   \bigg\{ \frac{332418839}{6561} 
\nonumber\\&        
       - \frac{64856}{9} \zeta_5 - \frac{10444660}{243} \zeta_3 + \frac{29296}{9} \zeta_3^2 - \frac{51630466}{729} \zeta_2 + \frac{193424}{9} \zeta_2 \zeta_3 + \frac{1158626}{45} 
         \zeta_2^2 - \frac{2461936}{945} \zeta_2^3 \bigg\}
\nonumber\\& 
       - C_F n_f^3   \bigg\{   \frac{2012}{27} - \frac{128}{3} \zeta_3 \bigg\}
       + C_F C_A n_f^2   \bigg\{ \frac{12896}{9} - \frac{23296}{27} \zeta_3 - \frac{680}{3} \zeta_2 - \frac{128}{15} 
         \zeta_2^2 \bigg\}
       - C_F C_A^2 n_f   \bigg\{   \frac{1611359}{243} - \frac{3776}{9} \zeta_5 
\nonumber\\&        
       - \frac{340016}{81} \zeta_3 - 
         \frac{23516}{9} \zeta_2 + \frac{3584}{3} \zeta_2 \zeta_3 - \frac{3328}{45} \zeta_2^2 \bigg\}
       + C_F^2 n_f^2   \bigg\{ 8 \bigg\}
       + C_F^2 C_A n_f   \bigg\{ \frac{1342}{9} - 640 \zeta_5 + \frac{1184}{3} \zeta_3 \bigg\}\Bigg] 
\nonumber\\& 
+
   \bm{\mathcal{D}^{0}_{Q}} \Bigg[
       - \frac{d^{abcd}_A d^{abcd}_A}{N_A}   \bigg\{   \frac{928}{9} - {\color{red} b_{4,FA}} + 3484 \zeta_7 + \frac{440}{9} \zeta_5 - 
         \frac{10160}{9} \zeta_3 - \frac{3344}{3} \zeta_3^2 - \frac{992}{3} \zeta_2 + 1024 \zeta_2 \zeta_5 - 
         2064 \zeta_2 \zeta_3 
\nonumber\\&          
         - \frac{264}{5} \zeta_2^2 - \frac{736}{5} \zeta_2^2 \zeta_3 + \frac{11264}{105} 
         \zeta_2^3 \bigg\}
       - n_f \frac{d^{abcd}_F d^{abcd}_A}{N_A}   \bigg\{   \frac{224}{9} - \frac{16240}{9} \zeta_5 + \frac{1600}{9} \zeta_3 + \frac{448}{3} 
         \zeta_3^2 + 96 \zeta_2 + 608 \zeta_2 \zeta_3 
\nonumber\\&          
         + \frac{1136}{15} \zeta_2^2 - \frac{16832}{315} 
         \zeta_2^3 \bigg\}
       + n_f^2 \frac{d^{abcd}_F d^{abcd}_F}{N_A}   \bigg\{ \frac{704}{9} - \frac{512}{3} \zeta_3 \bigg\}
       - n_f^4   \bigg\{   \frac{8000}{2187} - \frac{640}{81} \zeta_2 \bigg\}
       + C_A n_f^3   \bigg\{ \frac{4003783}{19683} - \frac{14992}{729} \zeta_3 
\nonumber\\&        
       - \frac{76624}{243} \zeta_2 + \frac{17632}{405} \zeta_2^2 \bigg\}
       - C_A^2 n_f^2   \bigg\{   \frac{168019897}{52488} - \frac{8168}{27} \zeta_5 - \frac{848845}{729} \zeta_3
          - \frac{10342238}{2187} \zeta_2 + \frac{6304}{27} \zeta_2 \zeta_3 + \frac{520424}{405} \zeta_2^2 \bigg\}
\nonumber\\&           
       + C_A^3 n_f   \bigg\{ \frac{12799895}{648} - \frac{25901}{9} \zeta_5 - \frac{10443965}{729} \zeta_3 +\frac{12440}{
         27} \zeta_3^2 - \frac{64943836}{2187} \zeta_2 + \frac{191246}{27} \zeta_2 \zeta_3 + \frac{4711811}{
         405} \zeta_2^2 
\nonumber\\&          
         - \frac{3410216}{2835} \zeta_2^3 \bigg\}
       - C_A^4   \bigg\{   \frac{1681544207}{39366} + \frac{1}{24} {\color{red} b_{4,FA}} - \frac{29911}{6} \zeta_7 - 
         \frac{190741}{27} \zeta_5 - \frac{34771520}{729} \zeta_3 + \frac{169004}{27} \zeta_3^2 + \frac{9328}{3} \zeta_2 \zeta_5
\nonumber\\&          
         - \frac{143918012}{2187} \zeta_2  + \frac{965614}{27} \zeta_2 \zeta_3 + \frac{2600237}{81} 
         \zeta_2^2 - \frac{64244}{15} \zeta_2^2 \zeta_3 - \frac{17208686}{2835} \zeta_2^3 \bigg\}
       + C_F n_f^3   \bigg\{ \frac{30310}{243} - \frac{832}{9} \zeta_3 - \frac{208}{9} \zeta_2 \bigg\}
\nonumber\\&
       - C_F C_A n_f^2   \bigg\{   \frac{3151019}{1458} - \frac{3760}{27} \zeta_5 - \frac{345640}{243} \zeta_3 - 
         \frac{22810}{27} \zeta_2 + \frac{3520}{9} \zeta_2 \zeta_3 - \frac{3008}{135} \zeta_2^2 \bigg\}
       + C_F C_A^2 n_f   \bigg\{ \frac{25496083}{2916} 
\nonumber\\&       
       - \frac{27880}{27} \zeta_5 - \frac{1514092}{243} \zeta_3 + 
         592 \zeta_3^2 - \frac{135986}{27} \zeta_2 + \frac{23384}{9} \zeta_2 \zeta_3 + \frac{5468}{27} \zeta_2^2
          + \frac{16768}{315} \zeta_2^3 \bigg\}
       - C_F^2 n_f^2   \bigg\{   \frac{619}{27} + \frac{640}{3} \zeta_5 
\nonumber\\&       
       - \frac{1792}{9} \zeta_3 \bigg\}

       - C_F^2 C_A n_f   \bigg\{   \frac{13459}{36} - \frac{5120}{3} \zeta_5 + \frac{3704}{3} \zeta_3 - 80 \zeta_3^2
          - 18 \zeta_2 + \frac{148}{5} \zeta_2^2 - \frac{320}{7} \zeta_2^3 \bigg\}

       - C_F^3 n_f   \bigg\{  23 \bigg\}\Bigg] 
\end{align}
\section{${\cal \chi}^I_{{\rm P},i}$ in Eq.{\eqref{ChiP}}}
For the SV part of the inclusive DY process, we have
\begin{align}
   \chi^{q}_{{\rm S},1} &=
        C_F   \bigg\{  - 3 \zeta_2 \bigg\}\,, 
\nonumber\\
   \chi^{q}_{{\rm S},2} &=
        C_F C_A   \bigg\{ \frac{2428}{81} - \frac{374}{9} \zeta_3 - \frac{469}{9} \zeta_2 + 4 \zeta_2^2 \bigg\}
       - C_F n_f   \bigg\{   \frac{328}{81} - \frac{68}{9} \zeta_3 - \frac{70}{9} \zeta_2 \bigg\}\,, 
\nonumber\\
   \chi^{q}_{{\rm S},3} &=
        C_F C_A^2   \bigg\{ \frac{5211949}{8748} - \frac{484}{3} \zeta_5 - \frac{128966}{81} \zeta_3 + \frac{536}{3}
         \zeta_3^2 - \frac{578479}{486} \zeta_2 + 726 \zeta_2 \zeta_3 + \frac{9457}{45} \zeta_2^2 + \frac{152}{63} \zeta_2^3 \bigg\}   
\nonumber\\&
        - C_F C_A n_f   \bigg\{   \frac{412765}{4374} + 8 \zeta_5 - \frac{9856}{27} \zeta_3 - \frac{75155}{243}
         \zeta_2 + 44 \zeta_2 \zeta_3 + \frac{2528}{45} \zeta_2^2 \bigg\}
       - C_F n_f^2   \bigg\{   \frac{128}{2187} + \frac{1480}{81} \zeta_3 + \frac{404}{27} \zeta_2 
\nonumber\\&       
       - \frac{148}{45}  \zeta_2^2 \bigg\}      
       - C_F^2 n_f   \bigg\{   \frac{42727}{324} - \frac{112}{3} \zeta_5 - \frac{2284}{27} \zeta_3 - \frac{605}{6} \zeta_2
          + 88 \zeta_2 \zeta_3 - \frac{152}{15} \zeta_2^2 \bigg\}
\end{align} % Done-SG
and for Higgs boson production,
\noindent we use $\chi^{g}_{{\rm S},i}$ = $  \frac{C_A}{C_F}   \chi^{q}_{{\rm S},i}$. \\
For the quark jet function, we find
\begin{align}
   \chi^{q}_{{\rm J},1} &=
        C_F   \bigg\{ \frac{7}{2} - 3 \zeta_2 \bigg\}\,, 
\nonumber\\
   \chi^{q}_{{\rm J},2} &=
        C_F^2   \bigg\{ \frac{9}{8} - 6 \zeta_3 - \frac{41}{2} \zeta_2 + \frac{82}{5} \zeta_2^2 \bigg\}   
       - C_F n_f   \bigg\{   \frac{4057}{324} - \frac{8}{9} \zeta_3 - \frac{68}{9} \zeta_2 \bigg\}            
       + C_F C_A   \bigg\{ \frac{53129}{648} - \frac{206}{9} \zeta_3 - \frac{416}{9} \zeta_2 - \frac{17}{5} \zeta_2^2 \bigg\}\,, 
\nonumber\\
   \chi^{q}_{{\rm J},3} &=
        C_F^3   \bigg\{ \frac{979}{8} + 84 \zeta_5 + \frac{1245}{2} \zeta_3 - 136 \zeta_3^2 + \frac{247}{4} \zeta_2
          - 204 \zeta_2 \zeta_3 + 120 \zeta_2^2 - \frac{3536}{21} \zeta_2^3 \bigg\} 
        - C_F^2 n_f   \bigg\{   \frac{88195}{324} - 8 \zeta_5 - \frac{3091}{27} \zeta_3 
\nonumber\\&           
        - \frac{1541}{9} \zeta_2 +
         \frac{16}{3} \zeta_2 \zeta_3 + \frac{2666}{45} \zeta_2^2 \bigg\}
       + C_F n_f^2   \bigg\{ \frac{124903}{8748} + \frac{188}{81} \zeta_3 - \frac{466}{27} \zeta_2 + \frac{12}{5} \zeta_2^2
          \bigg\}         
       + C_F^2 C_A   \bigg\{ \frac{4499}{48} + \frac{808}{3} \zeta_5 - \frac{7151}{9} \zeta_3
\nonumber\\&
       - 28 \zeta_3^2 - \frac{25631}{36} \zeta_2 + \frac{406}{3} \zeta_2 \zeta_3 + \frac{39223}{90} \zeta_2^2 
       + \frac{1016}{15} \zeta_2^3 \bigg\}
       - C_F C_A n_f   \bigg\{   \frac{2942843}{8748} - 8 \zeta_5 - \frac{3707}{27} \zeta_3 - \frac{68324}{243}
         \zeta_2  + 32 \zeta_2 \zeta_3 
\nonumber\\&         
         + \frac{418}{15} \zeta_2^2 \bigg\}
       + C_F C_A^2   \bigg\{ \frac{50602039}{34992} - \frac{190}{3} \zeta_5 - \frac{187951}{162} \zeta_3 + \frac{764}{3}
         \zeta_3^2 - \frac{464665}{486} \zeta_2 + 394 \zeta_2 \zeta_3 + \frac{1009}{30} \zeta_2^2 + \frac{884}{63} \zeta_2^3 \bigg\}
\end{align} % Done-SG
and for the gluon jet function,
\begin{align}
   \chi^{g}_{{\rm J},1} &=
        C_A   \bigg\{ \frac{67}{18} - 3 \zeta_2 \bigg\}
       + n_f   \bigg\{  - \frac{5}{9} \bigg\}\,, 
\nonumber\\
   \chi^{g}_{{\rm J},2} &=
        C_A^2   \bigg\{ \frac{2621}{27} - \frac{308}{9} \zeta_3 - \frac{1301}{18} \zeta_2 + 13 \zeta_2^2 \bigg\}   
       - C_A n_f   \bigg\{   \frac{1610}{81} + \frac{16}{9} \zeta_3 - \frac{136}{9} \zeta_2 \bigg\}   
       + n_f^2   \bigg\{ \frac{50}{81} - \frac{2}{3} \zeta_2 \bigg\}
\nonumber\\&              
       - C_F n_f   \bigg\{   \frac{55}{6} - 8 \zeta_3 \bigg\}\,, 
\nonumber\\
   \chi^{g}_{{\rm J},3} &=
        C_A^3   \bigg\{ \frac{35083577}{17496} + 374 \zeta_5 - \frac{127919}{81} \zeta_3 + \frac{272}{3} \zeta_3^2
          - \frac{449552}{243} \zeta_2 + 352 \zeta_2 \zeta_3 + \frac{9667}{15} \zeta_2^2 - \frac{27284}{315}
         \zeta_2^3 \bigg\}
       - C_A^2 n_f   \bigg\{   \frac{9801613}{17496} 
\nonumber\\&       
       + \frac{136}{3} \zeta_5 - \frac{4351}{27} \zeta_3 - \frac{151187}{243} \zeta_2 - 24 \zeta_2 \zeta_3 + \frac{7624}{45} \zeta_2^2 \bigg\}   
       + C_A n_f^2   \bigg\{ \frac{792839}{17496} + \frac{1130}{81} \zeta_3 - \frac{1760}{27} \zeta_2 + \frac{484}{45}
         \zeta_2^2 \bigg\}
\nonumber\\&         
       - n_f^3   \bigg\{   \frac{500}{729}        
       + \frac{16}{27} \zeta_3 - \frac{20}{9} \zeta_2 \bigg\}
       + C_F n_f^2   \bigg\{ \frac{5351}{108} - \frac{116}{3} \zeta_3 - 8 \zeta_2 \bigg\}
       - C_F C_A n_f   \bigg\{   \frac{323039}{648} - \frac{292}{3} \zeta_5 - \frac{7864}{27} \zeta_3 
       - \frac{863}{6} \zeta_2 
\nonumber\\&              
       + 88 \zeta_2 \zeta_3 - \frac{152}{15} \zeta_2^2 \bigg\}
       + C_F^2 n_f   \bigg\{ \frac{143}{6} - 120 \zeta_5 + 74 \zeta_3 \bigg\}
\end{align} % Done-SG

%%%%%%%%%%%%%%%%%%%%%%%%%%%%%%%%%%%%%%%%
\section{$\mathcal{D}^{0}_{z}$ part of inclusive cross sections}
%%%% Inclusuve case %%%
For the DY process, 
\begin{align}
\Delta^{q}_{\rmS}\Big|_{\mathcal{D}^{0}_{z}} &=
       - \frac{d^{abcd}_F d^{abcd}_A}{N_F}   \bigg\{   384 - 4 {\color{red} b_{4,FA}} + 6968 \zeta_7 - \frac{3680}{9} \zeta_5 -
         \frac{15616}{9} \zeta_3 - \frac{6688}{3} \zeta_3^2 - \frac{4352}{3} \zeta_2 + 2048 \zeta_2 \zeta_5 -
         3584 \zeta_2 \zeta_3 
\nonumber\\&         
         + \frac{448}{15} \zeta_2^2 - \frac{1472}{5} \zeta_2^2 \zeta_3 + \frac{55616}{315}
         \zeta_2^3 \bigg\}
       + n_f \frac{d^{abcd}_F d^{abcd}_F}{N_F}   \bigg\{ \frac{3200}{9} \zeta_5 - \frac{1280}{9} \zeta_3 + \frac{640}{3} \zeta_3^2 - 512
         \zeta_2 + \frac{128}{5} \zeta_2^2 + \frac{2560}{21} \zeta_2^3 \bigg\}
\nonumber\\&
       + C_F n_f^3   \bigg\{ \frac{10432}{2187} - \frac{3680}{81} \zeta_3 - \frac{3200}{81} \zeta_2 + \frac{224}{45}
         \zeta_2^2 \bigg\}
       - C_F C_A n_f^2   \bigg\{   \frac{898033}{2916} - \frac{608}{3} \zeta_5 - \frac{87280}{81} \zeta_3 - \frac{293528}
         {243} \zeta_2 
\nonumber\\&         
         + \frac{608}{9} \zeta_2 \zeta_3 + \frac{3488}{15} \zeta_2^2 \bigg\}
       + C_F C_A^2 n_f   \bigg\{ \frac{11551831}{2916} - \frac{7064}{27} \zeta_5 - \frac{829304}{81} \zeta_3 - \frac{4552}{9} \zeta_3^2 - \frac{2400868}{243} \zeta_2 + \frac{23440}{9} \zeta_2 \zeta_3 
\nonumber\\&         
         + \frac{108896}{45}
         \zeta_2^2 - \frac{5872}{21} \zeta_2^3 \bigg\}
       - C_F C_A^3   \bigg\{   \frac{28290079}{2187} + \frac{1}{6} {\color{red} b_{4,FA}} - \frac{11071}{3} \zeta_7 +
         \frac{149980}{27} \zeta_5 - \frac{288544}{9} \zeta_3 + \frac{14828}{9} \zeta_3^2 
\nonumber\\&    
        - \frac{5746982}{243}
         \zeta_2 - \frac{2752}{3} \zeta_2 \zeta_5 + \frac{120968}{9} \zeta_2 \zeta_3 + \frac{301208}{45} \zeta_2^2
          - \frac{8456}{15} \zeta_2^2 \zeta_3 - \frac{1009888}{945} \zeta_2^3 \bigg\}
       
       - C_F^2 n_f^2   \bigg\{   \frac{142769}{729} - \frac{33056}{9} \zeta_5 
\nonumber\\&       
       - \frac{113456}{81} \zeta_3 + \frac{99184}{729} \zeta_2 + \frac{79360}{27} \zeta_2 \zeta_3 - \frac{9280}{27} \zeta_2^2 \bigg\}
         
       - C_F^2 C_A n_f   \bigg\{   \frac{1792393}{1458} + \frac{368272}{9} \zeta_5 + \frac{1407448}{81} \zeta_3 -
         \frac{6800}{3} \zeta_3^2 
\nonumber\\&         
         - \frac{2636774}{729} \zeta_2 - 35232 \zeta_2 \zeta_3 + \frac{370768}{81}
         \zeta_2^2 - \frac{32384}{105} \zeta_2^3 \bigg\}
       
       + C_F^2 C_A^2   \bigg\{ \frac{15086188}{729} + \frac{1046528}{9} \zeta_5 + \frac{3043898}{81} \zeta_3 -
         \frac{82592}{3} \zeta_3^2 
\nonumber\\&         
         - \frac{12535492}{729} \zeta_2 + 3072 \zeta_2 \zeta_5 - \frac{2968640}{27}
         \zeta_2 \zeta_3 + \frac{1008832}{81} \zeta_2^2 + \frac{121888}{15} \zeta_2^2 \zeta_3 - \frac{34496}{15}
          \zeta_2^3 \bigg\}
          
       - C_F^3 n_f   \bigg\{   \frac{73309}{54} - \frac{84848}{9} \zeta_3 
\nonumber\\&       
       + \frac{119680}{3} \zeta_5 - \frac{109024}{3}
         \zeta_3^2 + \frac{8248}{27} \zeta_2 - \frac{186752}{9} \zeta_2 \zeta_3 + \frac{37672}{45} \zeta_2^2 +
         \frac{497536}{105} \zeta_2^3 \bigg\}
         
       - C_F^3 C_A   \bigg\{   \frac{206444}{27} - 274432 \zeta_5 
\nonumber\\&       
       + \frac{746878}{9} \zeta_3 + \frac{484192}{3}
         \zeta_3^2 + \frac{32740}{9} \zeta_2 + 73728 \zeta_2 \zeta_5 + \frac{1011088}{9} \zeta_2 \zeta_3
          - \frac{293536}{45} \zeta_2^2 - 30400 \zeta_2^2 \zeta_3 - \frac{406912}{15} \zeta_2^3 \bigg\}
\nonumber\\&       
       + C_F^4   \bigg\{ 983040 \zeta_7 - 196608 \zeta_5 + 32704 \zeta_3 - 15360 \zeta_3^2 -
         491520 \zeta_2 \zeta_5 + 113152 \zeta_2 \zeta_3 - \frac{391168}{5} \zeta_2^2 \zeta_3 \bigg\}\, ,

\end{align}
for the Higgs boson production through ggF, 
\begin{align}
   \Delta_{\rm S}^{g}\Big|_{\mathcal{D}^{0}_{z}} &=
-      \frac{d^{abcd}_A d^{abcd}_A}{N_A}   \bigg\{   384 - 4 {\color{red} b_{4,FA}} + 6968 \zeta_7 - \frac{3680}{9} \zeta_5 -
         \frac{15616}{9} \zeta_3 - \frac{6688}{3} \zeta_3^2 - \frac{4352}{3} \zeta_2 + 2048 \zeta_2 \zeta_5 -
         3584 \zeta_2 \zeta_3 
\nonumber\\&         
         + \frac{448}{15} \zeta_2^2 - \frac{1472}{5} \zeta_2^2 \zeta_3 + \frac{55616}{315}
         \zeta_2^3 \bigg\}
       + n_f \frac{d^{abcd}_F d^{abcd}_A}{N_A}   \bigg\{ \frac{3200}{9} \zeta_5 - \frac{1280}{9} \zeta_3 + \frac{640}{3} \zeta_3^2 - 512
         \zeta_2 + \frac{128}{5} \zeta_2^2 + \frac{2560}{21} \zeta_2^3 \bigg\}
\nonumber\\&
       + C_A n_f^3   \bigg\{ \frac{10432}{2187} - \frac{3680}{81} \zeta_3 - \frac{3200}{81} \zeta_2 + \frac{224}{45}
         \zeta_2^2 \bigg\}
       - C_A^2 n_f^2   \bigg\{   \frac{1543153}{2916} - \frac{34592}{9} \zeta_5 - \frac{176624}{81} \zeta_3 -
         \frac{1171400}{729} \zeta_2 
\nonumber\\&         
         + \frac{71200}{27} \zeta_2 \zeta_3 - \frac{2336}{45} \zeta_2^2 \bigg\}
       + C_A^3 n_f   \bigg\{ \frac{18455767}{2916} - \frac{2194328}{27} \zeta_5 - \frac{3115000}{81} \zeta_3 +
         \frac{269048}{9} \zeta_3^2 - \frac{10816300}{729} \zeta_2 
\nonumber\\&         
         + 57360 \zeta_2 \zeta_3 - \frac{1029536}{405} \zeta_2^2 - \frac{514096}{105} \zeta_2^3 \bigg\}
         
       - C_A^4   \bigg\{   \frac{40463407}{2187} + \frac{1}{6} {\color{red} b_{4,FA}} - \frac{2960191}{3} \zeta_7 -
         \frac{10233380}{27} \zeta_5 - \frac{10989496}{81} \zeta_3 
\nonumber\\&         
         + \frac{1969484}{9} \zeta_3^2 - \frac{28569362}{
         729} \zeta_2 + \frac{1683776}{3} \zeta_2 \zeta_5 + \frac{7256984}{27} \zeta_2 \zeta_3 - \frac{5197928}{
         405} \zeta_2^2 + 39144 \zeta_2^2 \zeta_3 - \frac{24472096}{945} \zeta_2^3 \bigg\}
\nonumber\\&
       - C_F C_A n_f^2   \bigg\{   \frac{155083}{243} - 32 \zeta_5 - \frac{4784}{9} \zeta_3 - \frac{5600}{9} \zeta_2
          + \frac{1280}{3} \zeta_2 \zeta_3 - \frac{64}{3} \zeta_2^2 \bigg\}
       + C_F C_A^2 n_f   \bigg\{ \frac{2519645}{486} - \frac{2608}{3} \zeta_5 
\nonumber\\&       
       + 9712 \zeta_3^2
       - \frac{143036}{9} \zeta_3  - \frac{134542}{27} \zeta_2 + \frac{27808}{9} \zeta_2 \zeta_3 - \frac{1424}{5} \zeta_2^2 +
         \frac{3328}{35} \zeta_2^3 \bigg\}
       - C_F^2 C_A n_f   \bigg\{   \frac{21037}{54} - \frac{3200}{3} \zeta_5 + \frac{8848}{9} \zeta_3 
\nonumber\\&          
         - 160 \zeta_3^2 
         - 16 \zeta_2 + \frac{296}{5} \zeta_2^2 - \frac{640}{7} \zeta_2^3 \bigg\}  \, , 
\end{align}
for the DIS cross section corresponding to the structure function ${\cal F}_2$
\begin{align}
   \Delta_{\rm J}^{q}\Big|_{\mathcal{D}^{0}_{z}} &=
       - \frac{d^{abcd}_F d^{abcd}_A}{N_F}   \bigg\{   192 - {\color{red} b_{4,FA}} + 3484 \zeta_7 - \frac{1840}{9} \zeta_5 -
         \frac{7808}{9} \zeta_3 - \frac{3344}{3} \zeta_3^2 - \frac{2176}{3} \zeta_2 + 1024 \zeta_2 \zeta_5 -
         1792 \zeta_2 \zeta_3 + \frac{224}{15} \zeta_2^2 
\nonumber\\&         
         - \frac{736}{5} \zeta_2^2 \zeta_3 + \frac{27808}{315}
         \zeta_2^3 \bigg\}
       - C_F C_A n_f^2   \bigg\{   \frac{3761509}{5832} - \frac{1192}{9} \zeta_5 - \frac{6092}{81} \zeta_3 -
         \frac{131878}{243} \zeta_2 + \frac{400}{9} \zeta_2 \zeta_3 + \frac{616}{9} \zeta_2^2 \bigg\}         
\nonumber\\&         
       + n_f \frac{d^{abcd}_F d^{abcd}_F}{N_F}   \bigg\{ 192 + \frac{11680}{9} \zeta_5 + \frac{2336}{9} \zeta_3 - \frac{448}{3} \zeta_3^2
          - \frac{2656}{3} \zeta_2 - 64 \zeta_2 \zeta_3 + \frac{896}{15} \zeta_2^2           
          + \frac{4864}{315} \zeta_2^3
          \bigg\}
\nonumber\\&
       - C_F \frac{d^{abc} d_{abc}}{N_F} n_{fv}   \bigg\{   192 - 1280 \zeta_5 + 224 \zeta_3 + 480 \zeta_2 -
         \frac{96}{5} \zeta_2^2 \bigg\}
         
       + C_F n_f^3   \bigg\{ \frac{50558}{2187} + \frac{80}{81} \zeta_3 - \frac{1880}{81} \zeta_2 + \frac{16}{9} \zeta_2^2 \bigg\}
\nonumber\\&
       + C_F C_A^2 n_f   \bigg\{ \frac{31645735}{5832} - \frac{6562}{27} \zeta_5        
       - \frac{239062}{81} \zeta_3 - \frac{2612}{9} \zeta_3^2 - \frac{940424}{243} \zeta_2 + \frac{8684}{9} \zeta_2 \zeta_3 + \frac{3314}{5} \zeta_2^2 -
         \frac{86176}{945} \zeta_2^3 \bigg\}
\nonumber\\&         
       - C_F C_A^3   \bigg\{   \frac{59835979}{4374} + \frac{1}{24} {\color{red} b_{4,FA}}       
       - \frac{28291}{6} \zeta_7 +
         \frac{128510}{27} \zeta_5 - \frac{387083}{27} \zeta_3 + \frac{17182}{9} \zeta_3^2 - \frac{2052595}{243}
         \zeta_2 + \frac{136}{3} \zeta_2 \zeta_5 
\nonumber\\&          
         + \frac{39988}{9} \zeta_2 \zeta_3 + \frac{77341}{45} \zeta_2^2 -
         \frac{836}{3} \zeta_2^2 \zeta_3
         - \frac{55864}{189} \zeta_2^3 \bigg\}
       - C_F^2 n_f^2   \bigg\{   \frac{161929}{972} + \frac{64}{9} \zeta_5 - \frac{3812}{9} \zeta_3 - \frac{385300}{729}
         \zeta_2 + \frac{1376}{27} \zeta_2 \zeta_3 
\nonumber\\&         
         + \frac{19904}{135} \zeta_2^2 \bigg\}
       - C_F^2 C_A n_f   \bigg\{  \frac{255235}{324}    
       - \frac{2984}{9} \zeta_5 + \frac{71482}{27} \zeta_3 - \frac{2920}{3}
          \zeta_3^2 + \frac{5369177}{729} \zeta_2 - \frac{21952}{9} \zeta_2 \zeta_3 - \frac{915377}{405}
         \zeta_2^2 
\nonumber\\&             
         + \frac{41488}{315} \zeta_2^3 \bigg\}
       + C_F^2 C_A^2   \bigg\{ \frac{51498031}{3888}      
       - 8610 \zeta_7 + \frac{14894}{9} \zeta_5 - \frac{188162}{27}
         \zeta_3 + \frac{5966}{3} \zeta_3^2 + \frac{17325892}{729} \zeta_2 + 3960 \zeta_2 \zeta_5 
\nonumber\\&           
         - \frac{462688}{27} \zeta_2 \zeta_3 - \frac{7041313}{810} \zeta_2^2 + \frac{25736}{15} \zeta_2^2 \zeta_3          
          + \frac{40576}{45} \zeta_2^3 \bigg\}
          
       - C_F^3 n_f   \bigg\{   \frac{16067}{27} + \frac{3952}{3} \zeta_5 + \frac{15314}{9} \zeta_3 + \frac{496}{3}
         \zeta_3^2 +
         \frac{608}{7} \zeta_2^3 
\nonumber\\&       
         - \frac{33958}{27} \zeta_2
         - \frac{10432}{9} \zeta_2 \zeta_3 - \frac{38906}{27} \zeta_2^2  \bigg\}
       - C_F^3 C_A   \bigg\{   \frac{492041}{108} - 10920 \zeta_7 - \frac{32168}{3} \zeta_5 - \frac{37640}{3}
         \zeta_3 + \frac{21668}{3} \zeta_3^2 + 17850 \zeta_2 
\nonumber\\&         
         + 2256 \zeta_2 \zeta_5 - \frac{50296}{9}
         \zeta_2 \zeta_3 + \frac{441829}{54} \zeta_2^2          
         - \frac{7736}{5} \zeta_2^2 \zeta_3 - \frac{39272}{21}
         \zeta_2^3 \bigg\}
       + C_F^4   \bigg\{ \frac{37069}{48} - 2040 \zeta_7 - 4224 \zeta_5 - 940 \zeta_3 
\nonumber\\&       
       + 1888 \zeta_3^2 + 3253 \zeta_2 - 4608 \zeta_2 \zeta_5 - 5048 \zeta_2 \zeta_3          
         + \frac{20523}{5}
         \zeta_2^2 - 176 \zeta_2^2 \zeta_3 + \frac{138784}{105} \zeta_2^3 \bigg\}    \, ,
\end{align}
for scalar (Higgs boson-mediated) DIS cross section 
\begin{align}
   \Delta_{\rm J}^{g}\Big|_{\mathcal{D}^{0}_{z}} &=
   
      - \frac{d^{abcd}_A d^{abcd}_A}{N_A}   \bigg\{   \frac{928}{9} - {\color{red} b_{4,FA}} + 3484 \zeta_7 + \frac{440}{9} \zeta_5 -
         \frac{10160}{9} \zeta_3 - \frac{3344}{3} \zeta_3^2 - \frac{992}{3} \zeta_2 + 1024 \zeta_2 \zeta_5 -
         2064 \zeta_2 \zeta_3 - \frac{264}{5} \zeta_2^2 
\nonumber \\&         
         - \frac{736}{5} \zeta_2^2 \zeta_3 + \frac{11264}{105}
         \zeta_2^3 \bigg\}
       + n_f^2 \frac{d^{abcd}_F d^{abcd}_F}{N_A}   \bigg\{ \frac{704}{9} - \frac{512}{3} \zeta_3 \bigg\}        
       + C_A n_f^3   \bigg\{ \frac{1962035}{6561} + \frac{4784}{243} \zeta_3 - \frac{74080}{243} \zeta_2 + \frac{1808}{45}
          \zeta_2^2 \bigg\}             
\nonumber \\&
       - n_f \frac{d^{abcd}_F d^{abcd}_A}{N_A}   \bigg\{   \frac{224}{9} - \frac{16240}{9} \zeta_5        
       + \frac{1600}{9} \zeta_3 + \frac{448}{3}
         \zeta_3^2 + 96 \zeta_2 + 608 \zeta_2 \zeta_3 + \frac{1136}{15} \zeta_2^2 - \frac{16832}{315}
         \zeta_2^3 \bigg\}
       - C_F^3 n_f   \bigg\{  23 \bigg\}
\nonumber \\&
       - n_f^4   \bigg\{  \frac{8000}{2187} - \frac{640}{81} \zeta_2 \bigg\} 
       - C_A^2 n_f^2   \bigg\{   \frac{86853899}{17496} - \frac{9880}{27} \zeta_5 - \frac{197167}{243} \zeta_3 -
         \frac{379097}{81} \zeta_2 + \frac{3440}{27} \zeta_2 \zeta_3 + \frac{449128}{405} \zeta_2^2 \bigg\}          
\nonumber \\&         
       + C_F n_f^3   \bigg\{ \frac{18758}{81} - \frac{4480}{27} \zeta_3 - \frac{704}{27} \zeta_2 - \frac{64}{135} \zeta_2^2
          \bigg\}
       - C_A^4   \bigg\{   \frac{373955585}{6561} + \frac{1}{24} {\color{red} b_{4,FA}} - \frac{29911}{6} \zeta_7 -
         \frac{141857}{27} \zeta_5 + \frac{64724}{9} \zeta_3^2
\nonumber \\&         
         - \frac{13935524}{243} \zeta_3  - \frac{15205433}{243}
          \zeta_2 + \frac{8848}{3} \zeta_2 \zeta_5 + \frac{303958}{9} \zeta_2 \zeta_3 + \frac{3524507}{135}
         \zeta_2^2 - \frac{16828}{5} \zeta_2^2 \zeta_3 - \frac{523556}{105} \zeta_2^3 \bigg\}          
\nonumber \\&          
       + C_A^3 n_f   \bigg\{ \frac{510680389}{17496} - \frac{26077}{9} \zeta_5 - \frac{3806447}{243} \zeta_3 +
         \frac{1064}{9} \zeta_3^2 - \frac{7087627}{243} \zeta_2 + \frac{168494}{27} \zeta_2 \zeta_3 + \frac{784091}{
                 81} \zeta_2^2 - \frac{26812}{27} \zeta_2^3 \bigg\}
\nonumber \\&
       - C_F C_A n_f^2   \bigg\{   \frac{589037}{162} - \frac{688}{3} \zeta_5 - \frac{187408}{81} \zeta_3 - \frac{28336}{
               27} \zeta_2 + 512 \zeta_2 \zeta_3 - \frac{3104}{135} \zeta_2^2 \bigg\}
       + C_F C_A^2 n_f   \bigg\{ \frac{4396589}{324} - 1528 \zeta_5 
\nonumber \\&       
       - \frac{797896}{81} \zeta_3 + 1104
         \zeta_3^2 - \frac{159673}{27} \zeta_2 + \frac{28760}{9} \zeta_2 \zeta_3 + \frac{19568}{135} \zeta_2^2
          + \frac{12736}{315} \zeta_2^3 \bigg\}

       - C_F^2 n_f^2   \bigg\{   \frac{203}{9} + \frac{1280}{3} \zeta_5 - \frac{1088}{3} \zeta_3 \bigg\}
\nonumber \\&
       - C_F^2 C_A n_f   \bigg\{   \frac{67129}{108} - 2880 \zeta_5 + \frac{17624}{9} \zeta_3 - 80 \zeta_3^2
          - 16 \zeta_2 + \frac{148}{5} \zeta_2^2 - \frac{320}{7} \zeta_2^3 \bigg\} \, ,
\end{align}
for $e^{+}e^{-}$ SIA corresponding to the transverse (${\cal F}_{T}$) structure function
\begin{align}
\Delta^{q}_{\rm SIA}\Big|_{\mathcal{D}^{0}_{z}} &=
       - \frac{d^{abcd}_F d^{abcd}_A}{N_F}   \bigg\{  192 - {\color{red} b_{4,FA}} + 3484 \zeta_7 - \frac{1840}{9} \zeta_5 -
         \frac{7808}{9} \zeta_3 - \frac{3344}{3} \zeta_3^2 - \frac{2176}{3} \zeta_2 + 1024 \zeta_2 \zeta_5 -
         1792 \zeta_2 \zeta_3 
\nonumber\\ &         
         + \frac{224}{15} \zeta_2^2 - \frac{736}{5} \zeta_2^2 \zeta_3 + \frac{27808}{315}
         \zeta_2^3 \bigg\}
       - C_F C_A n_f^2   \bigg\{   \frac{3761509}{5832} - \frac{1192}{9} \zeta_5 - \frac{6092}{81} \zeta_3 -
         \frac{131878}{243} \zeta_2 + \frac{400}{9} \zeta_2 \zeta_3 
\nonumber\\ &         
       + \frac{616}{9} \zeta_2^2 \bigg\}
       + n_f \frac{d^{abcd}_F d^{abcd}_F}{N_F}   \bigg\{ 192 + \frac{11680}{9} \zeta_5 + \frac{2336}{9} \zeta_3 - \frac{448}{3} \zeta_3^2
          - \frac{2656}{3} \zeta_2 - 64 \zeta_2 \zeta_3          
          + \frac{896}{15} \zeta_2^2 + \frac{4864}{315} \zeta_2^3
          \bigg\}
\nonumber\\ &           
       - C_F \frac{d^{abc} d_{abc}}{N_F} n_{fv}   \bigg\{   192 - 1280 \zeta_5 + 224 \zeta_3 + 480 \zeta_2 -
         \frac{96}{5} \zeta_2^2 \bigg\}
       + C_F n_f^3   \bigg\{ \frac{50558}{2187}    
       + \frac{80}{81} \zeta_3       
       - \frac{1880}{81} \zeta_2 + \frac{16}{9} \zeta_2^2 \bigg\}
\nonumber\\ &
       + C_F C_A^2 n_f   \bigg\{ \frac{31645735}{5832} - \frac{6562}{27} \zeta_5 - \frac{239062}{81} \zeta_3 - \frac{2612}{
         9} \zeta_3^2 - \frac{940424}{243} \zeta_2 + \frac{8684}{9} \zeta_2 \zeta_3 + \frac{3314}{5} \zeta_2^2 -
         \frac{86176}{945} \zeta_2^3 \bigg\}
\nonumber\\ &
       - C_F C_A^3   \bigg\{  
        \frac{59835979}{4374} + \frac{1}{24} {\color{red} b_{4,FA}} 
       - \frac{28291}{6} \zeta_7 
       + \frac{128510}{27} \zeta_5 
       - \frac{387083}{27} \zeta_3 
       + \frac{17182}{9} \zeta_3^2 
       - \frac{2052595}{243} \zeta_2 
         + \frac{136}{3} \zeta_2 \zeta_5   
\nonumber\\ &          
         + \frac{39988}{9} \zeta_2 \zeta_3 
         + \frac{77341}{45} \zeta_2^2 
         - \frac{836}{3} \zeta_2^2 \zeta_3 
         - \frac{55864}{189} \zeta_2^3 \bigg\}
       - C_F^2 n_f^2   \bigg\{   \frac{161929}{972} + \frac{64}{9} \zeta_5 - \frac{3812}{9} \zeta_3 + \frac{19760}{729}
         \zeta_2 + \frac{1760}{27} \zeta_2 \zeta_3 
\nonumber\\ &         
         - \frac{976}{135} \zeta_2^2 \bigg\}
       - C_F^2 C_A n_f   \bigg\{   \frac{255235}{324} - \frac{2984}{9} \zeta_5 + \frac{71482}{27} \zeta_3 - \frac{2920}{3}
          \zeta_3^2 - \frac{580447}{729} \zeta_2 - \frac{928}{9} \zeta_2 \zeta_3 + \frac{80503}{405} \zeta_2^2
\nonumber\\ &          
        - \frac{31088}{315} \zeta_2^3 \bigg\}
       + C_F^2 C_A^2   \bigg\{ \frac{51498031}{3888} - 8610 \zeta_7 + \frac{14894}{9} \zeta_5 - \frac{188162}{27}
         \zeta_3 + \frac{5966}{3} \zeta_3^2 - \frac{2535611}{729} \zeta_2 + 1176 \zeta_2 \zeta_5 
\nonumber\\ &         
         + \frac{42992}{
         27} \zeta_2 \zeta_3          
         + \frac{271247}{810} \zeta_2^2 + \frac{776}{15} \zeta_2^2 \zeta_3 - \frac{12992}{45}
          \zeta_2^3 \bigg\}
       - C_F^3 n_f   \bigg\{   \frac{16067}{27} + \frac{3952}{3} \zeta_5 + \frac{15314}{9} \zeta_3 + \frac{496}{3}
         \zeta_3^2 - \frac{23305}{27} \zeta_2 
\nonumber\\ &         
         - \frac{8368}{9} \zeta_2 \zeta_3    
         + \frac{1798}{27} \zeta_2^2 + \frac{608}{7} \zeta_2^3 \bigg\}
       - C_F^3 C_A   \bigg\{   \frac{492041}{108} - 10920 \zeta_7 - \frac{32168}{3} \zeta_5 - \frac{37640}{3}
         \zeta_3 + \frac{21668}{3} \zeta_3^2 + \frac{11193}{2} \zeta_2 
\nonumber\\ &         
         + 3696 \zeta_2 \zeta_5          
         + \frac{29048}{9}
         \zeta_2 \zeta_3 - \frac{28739}{54} \zeta_2^2 + \frac{904}{5} \zeta_2^2 \zeta_3 - \frac{8780}{21}
         \zeta_2^3 \bigg\}
       + C_F^4   \bigg\{ \frac{37069}{48} - 2040 \zeta_7 - 4224 \zeta_5 - 940 \zeta_3 
\nonumber\\ &       
       + 1888
         \zeta_3^2 + \frac{3503}{2} \zeta_2  
         + 576 \zeta_2 \zeta_5 - 1760 \zeta_2 \zeta_3 + \frac{3963}{5}
         \zeta_2^2 - 368 \zeta_2^2 \zeta_3 + \frac{55624}{105} \zeta_2^3 \bigg\}\,, 
\end{align}  
and finally for Higgs production in bottom-quark annihilation,
\begin{align}
   \Delta_{\rm b\overline{b}H}^{b}\Big|_{\mathcal{D}^{0}_{z}} &=
       - \frac{d^{abcd}_F d^{abcd}_A}{N_F}   \bigg\{   384 - 4 {\color{red} b_{4,FA}} + 6968 \zeta_7 - \frac{3680}{9} \zeta_5 -
         \frac{15616}{9} \zeta_3 - \frac{6688}{3} \zeta_3^2 - \frac{4352}{3} \zeta_2 + 2048 \zeta_2 \zeta_5 -
         3584 \zeta_2 \zeta_3 
\nonumber\\&         
         + \frac{448}{15} \zeta_2^2 - \frac{1472}{5} \zeta_2^2 \zeta_3 + \frac{55616}{315}
         \zeta_2^3 \bigg\}
         
       + n_f \frac{d^{abcd}_F d^{abcd}_F}{N_F}   \bigg\{ \frac{3200}{9} \zeta_5 - \frac{1280}{9} \zeta_3 + \frac{640}{3} \zeta_3^2 - 512
         \zeta_2 + \frac{128}{5} \zeta_2^2 + \frac{2560}{21} \zeta_2^3 \bigg\}
\nonumber\\&
       + C_F n_f^3   \bigg\{ \frac{10432}{2187} - \frac{3680}{81} \zeta_3 - \frac{3200}{81} \zeta_2 + \frac{224}{45}
         \zeta_2^2 \bigg\}
         
       - C_F C_A n_f^2   \bigg\{  \frac{898033}{2916} - \frac{608}{3} \zeta_5 - \frac{87280}{81} \zeta_3 - \frac{293528}{243} \zeta_2 
\nonumber\\&       
       + \frac{608}{9} \zeta_2 \zeta_3 + \frac{3488}{15} \zeta_2^2 \bigg\}
       
       + C_F C_A^2 n_f   \bigg\{ \frac{11551831}{2916} - \frac{7064}{27} \zeta_5 - \frac{829304}{81} \zeta_3 - \frac{4552}{9} \zeta_3^2 - \frac{2400868}{243} \zeta_2 + \frac{23440}{9} \zeta_2 \zeta_3 
\nonumber\\&       
       + \frac{108896}{45}
         \zeta_2^2 - \frac{5872}{21} \zeta_2^3 \bigg\}
         
       - C_F C_A^3   \bigg\{   \frac{28290079}{2187} + \frac{1}{6} {\color{red} b_{4,FA}} - \frac{11071}{3} \zeta_7 +
         \frac{149980}{27} \zeta_5 - \frac{288544}{9} \zeta_3 + \frac{14828}{9} \zeta_3^2 
\nonumber\\&         
         - \frac{5746982}{243}
         \zeta_2 - \frac{2752}{3} \zeta_2 \zeta_5 + \frac{120968}{9} \zeta_2 \zeta_3 + \frac{301208}{45} \zeta_2^2
          - \frac{8456}{15} \zeta_2^2 \zeta_3 - \frac{1009888}{945} \zeta_2^3 \bigg\}
          
       - C_F^2 n_f^2   \bigg\{   \frac{309953}{729} - \frac{33056}{9} \zeta_5 
\nonumber\\&       
       - \frac{124976}{81} \zeta_3 -
         \frac{314240}{729} \zeta_2 + \frac{79360}{27} \zeta_2 \zeta_3 - \frac{6976}{27} \zeta_2^2 \bigg\}
         
       + C_F^2 C_A^2   \bigg\{ \frac{1571464}{729} + \frac{1005056}{9} \zeta_5 + \frac{5327504}{81} \zeta_3 
       - \frac{88640}{3} \zeta_3^2 
\nonumber\\&       
       + \frac{6077552}{729} \zeta_2 + 3072 \zeta_2 \zeta_5 - \frac{3124160}{27}
         \zeta_2 \zeta_3 + \frac{3124352}{405} \zeta_2^2 + \frac{121888}{15} \zeta_2^2 \zeta_3 - \frac{34496}{15} \zeta_2^3 \bigg\}         
       + C_F^2 C_A n_f   \bigg\{ \frac{4753559}{1458} 
\nonumber\\&       
       - \frac{368272}{9} \zeta_5 - \frac{1791460}{81} \zeta_3 +
         \frac{6800}{3} \zeta_3^2       
         - \frac{3133882}{729} \zeta_2 + 36064 \zeta_2 \zeta_3 - \frac{1252496}{405}
         \zeta_2^2 + \frac{32384}{105} \zeta_2^3 \bigg\}
       - C_F^3 n_f   \bigg\{   \frac{41245}{54} 
\nonumber\\&       
       + \frac{119680}{3} \zeta_5 + \frac{30608}{9} \zeta_3 - \frac{109024}{3}
         \zeta_3^2 + \frac{13760}{27} \zeta_2 - \frac{205184}{9} \zeta_2 \zeta_3 + \frac{78184}{45} \zeta_2^2 +
         \frac{497536}{105} \zeta_2^3 \bigg\}
       - C_F^3 C_A   \bigg\{   \frac{25856}{27} 
\nonumber\\&       
       - 274432 \zeta_5 + \frac{18112}{3} \zeta_3 + \frac{511840}{3}
         \zeta_3^2 + \frac{78080}{27} \zeta_2 + 73728 \zeta_2 \zeta_5 + \frac{1275328}{9} \zeta_2 \zeta_3
          - \frac{478336}{45} \zeta_2^2           
          - 30400 \zeta_2^2 \zeta_3 
\nonumber\\&          
          - \frac{406912}{15} \zeta_2^3 \bigg\}
          
       + C_F^4   \bigg\{ 983040 \zeta_7 - 49152 \zeta_5 + 4096 \zeta_3 - 15360 \zeta_3^2 -
         491520 \zeta_2 \zeta_5 + 32768 \zeta_2 \zeta_3 - \frac{391168}{5} \zeta_2^2 \zeta_3 \bigg\}\,.
\end{align}   

\section{$\delta(1-z_1)\mathcal{D}^{0}_{z_2}$ part of rapidity distributions}
%%%% rapidity case %%%
For the DY process,
\begin{align}
\Delta^{q}_{\rm DY}\Big|_{\delta_{z_1}\mathcal{D}^{0}_{z_2}} &=
         - \frac{d^{abcd}_F d^{abcd}_A}{N_F}   \bigg\{   192 - 2 {\color{red} b_{4,FA}} + 3484 \zeta_7 - \frac{1840}{9} \zeta_5 -
         \frac{7808}{9} \zeta_3 - \frac{3344}{3} \zeta_3^2 - \frac{2176}{3} \zeta_2 + 1024 \zeta_2 \zeta_5 -
         1792 \zeta_2 \zeta_3 
\nonumber\\&         
         + \frac{224}{15} \zeta_2^2 - \frac{736}{5} \zeta_2^2 \zeta_3 + \frac{27808}{315}
         \zeta_2^3 \bigg\}
         
       + n_f \frac{d^{abcd}_F d^{abcd}_F}{N_F}   \bigg\{ \frac{1600}{9} \zeta_5 - \frac{640}{9} \zeta_3 + \frac{320}{3} \zeta_3^2 - 256
         \zeta_2 + \frac{64}{5} \zeta_2^2 + \frac{1280}{21} \zeta_2^3 \bigg\}
\nonumber\\&
       + C_F n_f^3   \bigg\{ \frac{5216}{2187} + \frac{80}{81} \zeta_3 - \frac{800}{81} \zeta_2 + \frac{16}{9} \zeta_2^2 \bigg\}
       
       - C_F C_A n_f^2   \bigg\{   \frac{898033}{5832} - \frac{304}{3} \zeta_5 - \frac{2456}{81} \zeta_3 - \frac{75718}{243} \zeta_2 - \frac{80}{9} \zeta_2 \zeta_3 
\nonumber\\&       
       + \frac{3104}{45} \zeta_2^2 \bigg\}
       + C_F C_A^2 n_f   \bigg\{ \frac{11551831}{5832} - \frac{3532}{27} \zeta_5 - \frac{150988}{81} \zeta_3 - \frac{2276}{9} \zeta_3^2 - \frac{645476}{243} \zeta_2 + \frac{4328}{9} \zeta_2 \zeta_3 + \frac{29624}{45} \zeta_2^2
\nonumber\\&
          - \frac{7288}{105} \zeta_2^3 \bigg\}
          
       - C_F C_A^3   \bigg\{   \frac{28290079}{4374} + \frac{1}{12} {\color{red} b_{4,FA}} - \frac{11071}{6} \zeta_7 +
         \frac{74990}{27} \zeta_5 - \frac{258224}{27} \zeta_3 + \frac{7414}{9} \zeta_3^2 - \frac{1634851}{243} \zeta_2
\nonumber\\&          
          - \frac{1376}{3} \zeta_2 \zeta_5 + \frac{31444}{9} \zeta_2 \zeta_3 + \frac{15632}{9} \zeta_2^2 - \frac{4228}{15} \zeta_2^2 \zeta_3 - \frac{27808}{189} \zeta_2^3 \bigg\}
          
       - C_F^2 n_f^2   \bigg\{   \frac{142769}{1458} - \frac{1168}{9} \zeta_5 - \frac{8552}{27} \zeta_3 
\nonumber\\&       
       + \frac{71900}{729} \zeta_2 + \frac{6272}{27} \zeta_2 \zeta_3 - \frac{1504}{27} \zeta_2^2 \bigg\}
       
       - C_F^2 C_A n_f   \bigg\{   \frac{1792393}{2916} + \frac{15176}{9} \zeta_5 + \frac{55924}{27} \zeta_3 - 536
          \zeta_3^2 - \frac{1571227}{729} \zeta_2 
\nonumber\\&          
          - \frac{18224}{9} \zeta_2 \zeta_3 + \frac{68168}{81} \zeta_2^2
          - \frac{3872}{105} \zeta_2^3 \bigg\}
          
       + C_F^2 C_A^2   \bigg\{ \frac{7543094}{729} + \frac{58624}{9} \zeta_5 - \frac{134305}{27} \zeta_3 - 2032
         \zeta_3^2 - \frac{7337465}{729} \zeta_2 
\nonumber\\&         
         - 768 \zeta_2 \zeta_5 - \frac{72640}{27} \zeta_2 \zeta_3
          + \frac{193424}{81} \zeta_2^2 + \frac{1840}{3} \zeta_2^2 \zeta_3 - \frac{1232}{15} \zeta_2^3 \bigg\}
          
       - C_F^3 n_f  \bigg\{   \frac{73309}{108} + \frac{2240}{3} \zeta_5 - \frac{9256}{9} \zeta_3 - \frac{538}{27} \zeta_2 
\nonumber\\&          
          - 1616 \zeta_3^2 - \frac{3040}{9} \zeta_2 \zeta_3 - \frac{764}{45} \zeta_2^2 - \frac{320}{7} \zeta_2^3
          \bigg\}
          
       - C_F^3 C_A   \bigg\{   \frac{103222}{27} - 8576 \zeta_5 + \frac{57095}{9} \zeta_3 + 5008 \zeta_3^2
          - \frac{18479}{9} \zeta_2 
\nonumber\\&           
          + 2304 \zeta_2 \zeta_5 + \frac{14936}{9} \zeta_2 \zeta_3 - \frac{8968}{45}
         \zeta_2^2 - \frac{4672}{5} \zeta_2^2 \zeta_3 + \frac{704}{5} \zeta_2^3 \bigg\}
       + C_F^4   \bigg\{ 7680 \zeta_7 - 6144 \zeta_5 + 4088 \zeta_3 - 1920 \zeta_3^2 
\nonumber\\&        
       - 4608 \zeta_2 \zeta_5 + 3904 \zeta_2 \zeta_3 - \frac{6656}{5} \zeta_2^2 \zeta_3 \bigg\}\, ,
\end{align}
for Higgs boson production through ggF,
\begin{align}
\Delta^{g}_{\rm ggH}\Big|_{\delta_{z_1}\mathcal{D}^{0}_{z_2}} &=
    - \frac{d^{abcd}_A d^{abcd}_A}{N_A}   \bigg\{   192 - 2 {\color{red} b_{4,FA}} + 3484 \zeta_7 - \frac{1840}{9} \zeta_5 -
         \frac{7808}{9} \zeta_3 - \frac{3344}{3} \zeta_3^2 - \frac{2176}{3} \zeta_2 + 1024 \zeta_2 \zeta_5 -
         1792 \zeta_2 \zeta_3 
\nonumber\\&         
         + \frac{224}{15} \zeta_2^2 - \frac{736}{5} \zeta_2^2 \zeta_3 + \frac{27808}{315}
         \zeta_2^3 \bigg\}
         
       + n_f \frac{d^{abcd}_F d^{abcd}_A}{N_A}   \bigg\{ \frac{1600}{9} \zeta_5 - \frac{640}{9} \zeta_3 + \frac{320}{3} \zeta_3^2 - 256
         \zeta_2 + \frac{64}{5} \zeta_2^2 + \frac{1280}{21} \zeta_2^3 \bigg\}
\nonumber\\&         
       + C_A n_f^3   \bigg\{ \frac{5216}{2187} + \frac{80}{81} \zeta_3 - \frac{800}{81} \zeta_2 + \frac{16}{9} \zeta_2^2 \bigg\}
       
       - C_A^2 n_f^2   \bigg\{   \frac{1543153}{5832} - \frac{1936}{9} \zeta_5 - \frac{1240}{9} \zeta_3 - \frac{258130}{729} \zeta_2 + \frac{3536}{27} \zeta_2 \zeta_3 
\nonumber\\&       
       + \frac{1504}{45} \zeta_2^2 \bigg\}
       
       + C_A^3 n_f   \bigg\{ \frac{18455767}{5832} - \frac{71884}{27} \zeta_5 - \frac{157636}{27} \zeta_3 + \frac{3196}{9}
          \zeta_3^2 - \frac{1964828}{729} \zeta_2 + \frac{24232}{9} \zeta_2 \zeta_3 + \frac{46136}{405}
         \zeta_2^2 
\nonumber\\&         
         - \frac{344}{105} \zeta_2^3 \bigg\}
         
       - C_A^4   \bigg\{   \frac{40463407}{4374} + \frac{1}{12} {\color{red} b_{4,FA}} - \frac{57151}{6} \zeta_7 -
         \frac{249490}{27} \zeta_5 - \frac{2044156}{81} \zeta_3 + \frac{110374}{9} \zeta_3^2 - \frac{4061161}{729}
         \zeta_2 
\nonumber\\&         
         + \frac{21664}{3} \zeta_2 \zeta_5 + \frac{314332}{27} \zeta_2 \zeta_3 - \frac{15392}{81}
         \zeta_2^2 - \frac{2492}{5} \zeta_2^2 \zeta_3 + \frac{71632}{945} \zeta_2^3 \bigg\}
         
       - C_F C_A n_f^2   \bigg\{   \frac{155083}{486} - 16 \zeta_5 - \frac{1912}{9} \zeta_3 
\nonumber\\&       
       - \frac{1400}{9} \zeta_2
          + \frac{320}{3} \zeta_2 \zeta_3 - \frac{32}{3} \zeta_2^2 \bigg\}
          
       + C_F C_A^2 n_f   \bigg\{ \frac{2519645}{972} - \frac{1304}{3} \zeta_5 - \frac{33742}{9} \zeta_3 + 1784
         \zeta_3^2 - \frac{22559}{27} \zeta_2 
         - \frac{392}{5} \zeta_2^2
\nonumber\\&         
         + \frac{4976}{9} \zeta_2 \zeta_3  - \frac{1024}{35} \zeta_2^3 \bigg\}
         
       - C_F^2 C_A n_f   \bigg\{   \frac{21037}{108} - \frac{1600}{3} \zeta_5 + \frac{4424}{9} \zeta_3 - 80
         \zeta_3^2 - 4 \zeta_2 + \frac{148}{5} \zeta_2^2 - \frac{320}{7} \zeta_2^3 \bigg\}\, ,
\end{align}
and finally for Higgs boson production in bottom-quark annihilation
\begin{align}
\Delta^{b}_{\rm b\overline{b}H}\Big|_{\delta_{z_1}\mathcal{D}^{0}_{z_2}} &=
       - \frac{d^{abcd}_F d^{abcd}_A}{N_F}   \bigg\{   192 - 2 {\color{red} b_{4,FA}} + 3484 \zeta_7 - \frac{1840}{9} \zeta_5 -
         \frac{7808}{9} \zeta_3 - \frac{3344}{3} \zeta_3^2 - \frac{2176}{3} \zeta_2 + 1024 \zeta_2 \zeta_5 -
         1792 \zeta_2 \zeta_3 
\nonumber\\&         
         + \frac{224}{15} \zeta_2^2 - \frac{736}{5} \zeta_2^2 \zeta_3 + \frac{27808}{315}
         \zeta_2^3 \bigg\}
         
       + n_f \frac{d^{abcd}_F d^{abcd}_F}{N_F}   \bigg\{ \frac{1600}{9} \zeta_5 - \frac{640}{9} \zeta_3 + \frac{320}{3} \zeta_3^2 - 256
         \zeta_2 + \frac{64}{5} \zeta_2^2 + \frac{1280}{21} \zeta_2^3 \bigg\}
\nonumber\\&
       + C_F n_f^3   \bigg\{ \frac{5216}{2187} + \frac{80}{81} \zeta_3 - \frac{800}{81} \zeta_2 + \frac{16}{9} \zeta_2^2 \bigg\}
       
       - C_F C_A n_f^2   \bigg\{   \frac{898033}{5832} - \frac{304}{3} \zeta_5 - \frac{2456}{81} \zeta_3 - \frac{75718}{243} \zeta_2 - \frac{80}{9} \zeta_2 \zeta_3 
\nonumber\\&       
       + \frac{3104}{45} \zeta_2^2 \bigg\}
       
       + C_F C_A^2 n_f   \bigg\{ \frac{11551831}{5832} - \frac{3532}{27} \zeta_5 - \frac{150988}{81} \zeta_3 - \frac{2276}{9} \zeta_3^2 - \frac{645476}{243} \zeta_2 + \frac{4328}{9} \zeta_2 \zeta_3 + \frac{29624}{45} \zeta_2^2
\nonumber\\&          
          - \frac{7288}{105} \zeta_2^3 \bigg\}
          
       - C_F C_A^3   \bigg\{   \frac{28290079}{4374} + \frac{1}{12} {\color{red} b_{4,FA}} - \frac{11071}{6} \zeta_7 +
         \frac{74990}{27} \zeta_5 - \frac{258224}{27} \zeta_3 + \frac{7414}{9} \zeta_3^2 - \frac{1634851}{243} \zeta_2
\nonumber\\&          
          - \frac{1376}{3} \zeta_2 \zeta_5 + \frac{31444}{9} \zeta_2 \zeta_3 + \frac{15632}{9} \zeta_2^2 - \frac{4228}{15} \zeta_2^2 \zeta_3 - \frac{27808}{189} \zeta_2^3 \bigg\}
          
       - C_F^2 n_f^2   \bigg\{   \frac{309953}{1458} - \frac{1168}{9} \zeta_5 - \frac{8168}{27} \zeta_3 
\nonumber\\&       
       - \frac{43552}{729} \zeta_2 + \frac{6272}{27} \zeta_2 \zeta_3 - \frac{928}{27} \zeta_2^2 \bigg\}
       
       + C_F^2 C_A n_f   \bigg\{ \frac{4753559}{2916} - \frac{15176}{9} \zeta_5 - \frac{94582}{27} \zeta_3 + 536
         \zeta_3^2 - \frac{63677}{729} \zeta_2 
\nonumber\\&         
         + \frac{21104}{9} \zeta_2 \zeta_3 - \frac{182728}{405} \zeta_2^2
          + \frac{3872}{105} \zeta_2^3 \bigg\}
          
       + C_F^2 C_A^2   \bigg\{ \frac{785732}{729} + \frac{37888}{9} \zeta_5 + \frac{176600}{27} \zeta_3 - 3040
         \zeta_3^2 - \frac{1902716}{729} \zeta_2 
\nonumber\\&         
         - 768 \zeta_2 \zeta_5 - \frac{136144}{27} \zeta_2 \zeta_3
          + \frac{529936}{405} \zeta_2^2 + \frac{1840}{3} \zeta_2^2 \zeta_3 - \frac{1232}{15} \zeta_2^3 \bigg\}
          
       - C_F^3 n_f   \bigg\{   \frac{41245}{108} + \frac{2240}{3} \zeta_5 + \frac{7912}{9} \zeta_3 
\nonumber\\&       
       - 1616 \zeta_3^2
          - \frac{8456}{27} \zeta_2 - \frac{5344}{9} \zeta_2 \zeta_3 + \frac{5332}{45} \zeta_2^2 - \frac{320}{7}
         \zeta_2^3 \bigg\}
         
       - C_F^3 C_A   \bigg\{   \frac{12928}{27} - 8576 \zeta_5 - 928 \zeta_3 + 6160 \zeta_3^2 
\nonumber\\&        
       + \frac{6592}{27} \zeta_2 + 2304 \zeta_2 \zeta_5 + \frac{34736}{9} \zeta_2 \zeta_3 - \frac{23488}{45}
         \zeta_2^2 - \frac{4672}{5} \zeta_2^2 \zeta_3 + \frac{704}{5} \zeta_2^3 \bigg\}
         
       + C_F^4   \bigg\{ 7680 \zeta_7 - 1536 \zeta_5 + 512 \zeta_3 
\nonumber\\&        
       - 1920 \zeta_3^2 - 4608
         \zeta_2 \zeta_5 + 1536 \zeta_2 \zeta_3 - \frac{6656}{5} \zeta_2^2 \zeta_3 \bigg\}
\end{align}

\end{widetext}

%%%%%%%%%%%%%%%%%%%%%%%%%%%%%%%%%%%%%%%%
%\begin{thebibliography}{99}
\bibliography{softjet4loop}

\bibliographystyle{apsrev4-1}
\end{document}